\documentclass[dvips,moor]{informs2}              

\long\def\hide#1{}

\newboolean{showcomments}
\setboolean{showcomments}{true}
\usepackage{color}
\usepackage[usenames,dvipsnames]{xcolor}
\newcommand{\jason}[1]{  \ifthenelse{\boolean{showcomments}}
{ \textcolor{Red}{(Jason says:  #1)}} {}  }
\newcommand{\raga}[1]{\ifthenelse{\boolean{showcomments}}
{ \textcolor{Blue}{(Raga says: #1)} } {} }
\newcommand{\adam}[1]{\ifthenelse{\boolean{showcomments}}
{ \textcolor{Red}{(Adam says:  #1)}}{}}

\usepackage[lofdepth,lotdepth]{subfig}

\usepackage{adjustbox,algorithm,algorithmicx,algpseudocode,chngpage,enumitem,multirow,tabu,type1cm}

\usepackage{natbib}
 \NatBibNumeric
 \def\bibfont{\small}%
 \bibpunct[, ]{[}{]}{,}{n}{}{,}%

\usepackage[colorlinks=true,breaklinks=true,bookmarks=true,urlcolor=blue,
     citecolor=blue,linkcolor=blue,bookmarksopen=false,draft=false]{hyperref}

\def\EMAIL#1{\href{mailto:#1}{#1}}

\TheoremsNumberedThrough     
\ECRepeatTheorems

\EquationsNumberedThrough    

\MANUSCRIPTNO{MOR-2013-108} 

\begin{document}


\RUNAUTHOR{Gopalakrishnan, Marden, and Wierman}

\RUNTITLE{Potential Games are \textit{Necessary} to Ensure PNE in Cost Sharing Games}

\TITLE{Potential Games are \textit{Necessary} to Ensure\\ Pure Nash Equilibria in Cost Sharing Games}

\ARTICLEAUTHORS{%
\AUTHOR{Ragavendran Gopalakrishnan}
\AFF{Department of Computing and Mathematical Sciences, California Institute of Technology, Pasadena, CA 91125, \EMAIL{ragad3@caltech.edu}} 
\AUTHOR{Jason R. Marden}
\AFF{Department of Electrical, Computer, and Energy Engineering, University of Colorado, Boulder, CO 80309, \EMAIL{jason.marden@colorado.edu}}
\AUTHOR{Adam Wierman}
\AFF{Department of Computing and Mathematical Sciences, California Institute of Technology, Pasadena, CA 91125, \EMAIL{adamw@caltech.edu}}
} 

\ABSTRACT{%

\vspace{-0.1in}
We consider the problem of designing distribution rules to share `welfare' (cost or revenue) among individually strategic agents. There are many known distribution rules that guarantee the existence of a (pure) Nash equilibrium in this setting, e.g., the Shapley value and its weighted variants; however, a characterization of the space of distribution rules that guarantee the existence of a Nash equilibrium is unknown. Our work provides an exact characterization of this space for a specific class of scalable and separable games, which includes a variety of applications such as facility location, routing, network formation, and coverage games. Given arbitrary local welfare functions $\mathbb{W}$, we prove that a distribution rule guarantees equilibrium existence for all games (i.e., all possible sets of resources, agent action sets, etc.) if and only if it is equivalent to a generalized weighted Shapley value on some `ground' welfare functions $\mathbb{W}'$, which can be distinct from $\mathbb{W}$. However, if budget-balance is required in addition to the existence of a Nash equilibrium, then $\mathbb{W}'$ must be the same as $\mathbb{W}$. We also provide an alternate characterization of this space in terms of `generalized' marginal contributions, which is more appealing from the point of view of computational tractability. A possibly surprising consequence of our result is that, in order to guarantee equilibrium existence in all games with any fixed local welfare functions, it is \textit{necessary} to work within the class of potential games.
}%



\maketitle

%


\vspace{-0.2in}
\section{Introduction.}\label{s.introduction}

Fair division is an issue that is at the heart of social science -- how should the cost incurred (revenue generated) by a group of self-interested agents be shared among them? This central question has led to a large literature in economics over the last decades (Young \cite{Young94a}, Young \cite{Young94b}, Moulin \cite{Moulin02}), and more recently in computer science (Anshelevich et al. \cite{Anshelevich04}, Jain and Mahdian \cite{Jain07}, Moulin \cite{Moulin13}). A standard framework within which to study this question is that of \textit{cost sharing games}, in which there is a set of agents making strategic choices of which resources to utilize. Each resource generates a welfare (cost or revenue) depending on the set of agents that choose the resource. The focus is on finding distribution rules that lead to stable and/or fair allocations, which is traditionally formalized by the concept of the \textit{core} in the cooperative theory and, more recently, by the \textit{Nash equilibrium} in the noncooperative theory.

Cost sharing has traditionally been studied in the cooperative framework. Here, the problems studied typically involve a cost value $v(S)$ for each subset of players $S$, which usually stems from the optimal solution to an underlying combinatorial optimization problem.\footnote{Note that our focus is on cost sharing \textit{games} and not cost sharing \textit{mechanisms} (Feigenbaum et al. \cite{Feigenbaum01}), which additionally involve soliciting agents' exogenous private valuations of attaining the end goal. We briefly discuss the applicability of our results to cost sharing mechanisms in Section \ref{s.extend-discuss}.} A canonical example is the multicast network formation game (Granot and Huberman \cite{Granot81}), where a set of agents (consumers) $N$ wishes to connect to a common source (a broadcaster) $s$ by utilizing links of an underlying graph. Each link (resource) has a cost associated with its usage, and the total cost of all the links used needs to be split among the agents. In such a situation, any subset $S$ of agents, if they cooperate, can form a coalition, and the best they can do is to choose the links of the minimum cost spanning tree for the set of vertices $S\cup\{s\}$, and incur its cost -- denote it by $v(S)$. Here, the core consists of all possible ways of distributing $v(N)$ to the players in $N$ in such a way that it is in their best interest to fully cooperate to form the grand coalition. That is, a distribution rule $f^v:N\rightarrow\mathbb{R}$ is in the core, if $\sum_{i\in N}f^v(i)=v(N)$, and for every subset $S\subseteq N$, $\sum_{i\in S}f^v(i)\leq v(S)$. In general, the core can be empty, though for multicast games it is not.

A cooperative framework, in effect, models a `binary choice' for the agents -- opt out, or opt in and cooperate. In large distributed (and often unregulated) systems such as the Internet, agents' options are more complex as they have the opportunity to strategically choose the best action from multiple available options. Accordingly, there is an emerging focus within cost sharing games on weaker notions of stability such as Nash equilibria. This focus is driven by applications such as network-cost sharing (Anshelevich et al. \cite{Anshelevich04}, Chen et al. \cite{Chen08}) where individually strategic behavior is commonly assumed.

Our previous example of multicast games also provides a useful illustration of the noncooperative cost sharing framework.  Multicast games were first modeled as noncooperative games in Chekuri et al. \cite{Chekuri07}, whose model also generalized facility location games, an important class of problems in operations research. The principal difference from the cooperative model is that here, the global cost share of an agent stems from \textit{local} distribution rules which specify how the local cost (cost of each link) is split between the agents using that link. Accordingly, an agent's total cost share is simply the sum of its cost shares across all the links it uses. In addition, each agent can choose between potentially several link combinations that connect to the source. A pure Nash equilibrium corresponds to a choice of links by each agent such that each agent incurs the least possible cost given the links chosen by the other agents. Similarly to the fact that the core might be empty in the cooperative model, a pure Nash equilibrium may not exist in general, but for multicast games it does (Chekuri et al. \cite{Chekuri07}).

Existing literature on noncooperative cost sharing games focuses on designing distribution rules that guarantee equilibrium existence and studying the `efficiency' of the resulting equilibria. Perhaps, the most famous such distribution rule is the Shapley value (Shapley \cite{Shapley53b}), which is budget-balanced, guarantees the existence of a Nash equilibrium in any game, and for some classes of games such as convex games, is always in the core. Generalizations of the Shapley value, e.g., weighted and generalized weighted Shapley values (Shapley \cite{Shapley53a}), exhibit many of the same properties.

In addition to guaranteeing equilibrium existence, it is also of paramount importance that these equilibria be `efficient'. That is, they should result in a system cost (usually, the total cost incurred by all the agents) that is within a small factor of the optimum. For example, in the noncooperative multicast game, which (effectively) uses the Shapley value distribution rule, a Nash equilibrium choice of links by the agents may not collectively result in the minimum spanning tree for $N\cup\{s\}$.

With these goals in mind, researchers have recently sought to provide characterizations of the class of (local) distribution rules that guarantee equilibrium existence. The first step toward this goal was made in Chen et al. \cite{Chen10}, which proves that the only \textit{budget-balanced} distribution rules that guarantee equilibrium existence in all cost sharing games are generalized weighted Shapley value distribution rules. Following on Chen et al. \cite{Chen10}, Marden and Wierman \cite{Marden13b} provides the parallel characterization in the context of revenue sharing games. Though these characterizations seem general, they are actually just worst-case characterizations. In particular, the proofs in Chen et al. \cite{Chen10} and Marden and Wierman \cite{Marden13b} consist of exhibiting a specific `worst-case' welfare function which requires that generalized weighted Shapley value distribution rules be used. Thus, characterizing the space of distribution rules (not necessarily budget-balanced) for specific local welfare functions remains an important open problem. In practice, it is exactly this issue that is important: when designing a distribution rule, one \textit{knows} the specific local welfare functions for the situation, wherein there may be distribution rules other than generalized weighted Shapley values that also guarantee the existence of an equilibrium.

\vspace{-0.1in}
\subsection*{Our contribution.}

In this article, we provide a complete characterization of the space of distribution rules (not necessarily budget-balanced) that guarantee the existence of a pure Nash equilibrium (which we will henceforth refer to as just an equilibrium) \textit{for any specific local welfare functions}. The principal contributions of this article are as follows.
\begin{enumerate}[label=\arabic*.]
\item Our main result (Theorem \ref{mainresult}) states that all games conditioned on any fixed local welfare functions possess an equilibrium if and only if the distribution rules are equivalent to generalized weighted Shapley value distribution rules on some `ground' welfare functions. This shows, perhaps surprisingly, that the results in Chen et al. \cite{Chen10} and Marden and Wierman \cite{Marden13b} hold much more generally. In particular, it is neither the existence of some worst-case welfare function, nor the restriction of budget-balance, which limits the design of distribution rules to generalized weighted Shapley values.
\item Our second result (Theorem \ref{altresult}) provides an alternative characterization of the set of distribution rules that guarantee equilibrium existence. In particular, it states that all games conditioned on any fixed local welfare functions possess an equilibrium if and only if the distribution rules are equivalent to generalized weighted marginal contribution distribution rules on some `ground' welfare functions. This result is actually a consequence of a connection between Shapley values and marginal contributions, namely that they can be viewed as equivalent given a transformation connecting their ground welfare functions (Proposition \ref{prop:svmcequiv}).
\end{enumerate}
These characterizations provide two alternative approaches for the problem of designing distribution rules, with different design tradeoffs, e.g., between budget-balance and tractability. More specifically, a design through generalized weighted Shapley values provides direct control over how close to budget-balanced the distribution rule will be; however, computing these distribution rules often requires computing exponentially many marginal contributions (Matsui and Matsui \cite{Matsui00}, Conitzer and Sandholm \cite{Conitzer04}). On the other hand, a design through generalized weighted marginal contributions requires computing only one marginal contribution; however, it is more difficult to provide bounds on the degree of budget-balance.

Another important consequence of our characterizations is that potential games are \textit{necessary} to guarantee the existence of an equilibrium in all games with fixed local welfare functions, since generalized weighted Shapley value and generalized weighted marginal contribution distribution rules result in (`weighted') potential games (Hart and Mas-Colell \cite{Hart89}, Ui \cite{Ui00}). This is particularly surprising, since the class of potential games is a relatively small subset of the class of games that possess an equilibrium (Sandholm \cite{Sandholm10}), and our characterizations imply that such a relaxation in game structure would offer no advantage in guaranteeing equilibria.

In addition to the implications of the characterizations themselves, their proofs develop tools for analyzing cost sharing games which could be useful for related models, such as cost sharing mechanisms.  The proofs consist of a sequence of counterexamples that establish novel necessary conditions for distribution rules to guarantee the existence of an equilibrium. Within this analysis, new tools for studying distribution rules using their basis representation (see Section \ref{s.basisreps}) are developed, including an inclusion-exclusion framework that is crucial for our proof. Additionally, the proofs expose a relationship between Shapley value and marginal contribution distribution rules, leading to a novel closed form expression for the potential function of the resulting games.

\vspace{-0.1in}
\section{Model.}\label{s.model}

In this work we consider a simple, but general, model of a welfare (cost or revenue) sharing game, where there is a set of self-interested agents/players $N = \{1,\ldots ,n\}$ ($n>1$) that each select a collection of resources from a set $R = \{r_1, \ldots, r_m\}$ ($m>1$). That is, each agent $i \in N$ is capable of selecting potentially multiple resources in $R$; therefore, we say that agent $i$ has an action set $\mathcal{A}_i \subseteq  2^R$. The resulting action profile, or (joint) allocation, is a tuple $a = \left(a_1, \ldots, a_n\right) \in \mathcal{A}$ where the set of all possible allocations is denoted by $\mathcal{A}= \mathcal{A}_1\times \ldots \times \mathcal{A}_n$. We occasionally denote an action profile $a$ by $(a_i, a_{-i})$ where $a_{-i}\in\mathcal{A}_{-i}$ denotes the actions of all agents except agent $i$.

Each allocation generates a welfare, $\mathcal{W}(a)$, which needs to be shared among the agents. In this work, we assume $\mathcal{W}(a)$ is \textit{(linearly) separable} across resources, i.e.,

\vspace{-0.1in}
\begin{equation*}
\mathcal{W}(a) = \sum_{r \in R} W_r\left(\{a\}_r\right),
\end{equation*}

\vspace{-0.025in}
\noindent where $\{a\}_r = \left\{i \in N: r \in a_i \right\}$ is the set of agents that are allocated to resource $r$ in $a$, and $W_r: 2^N \rightarrow \mathbb{R}$ is the local welfare function at resource $r$. This is a standard assumption (Anshelevich et al. \cite{Anshelevich04}, Chekuri et al. \cite{Chekuri07}, Chen et al. \cite{Chen10}, Marden and Wierman \cite{Marden13a}), and is quite general. Note that we incorporate both revenue and cost sharing games, since we allow for the local welfare functions $W_r$ to be either positive or negative.

The manner in which the welfare is shared among the agents determines the utility function $U_i:\mathcal{A} \rightarrow \mathbb{R}$ that agent $i$ seeks to maximize. Because the welfare is assumed to be separable, it is natural that the utility functions should follow suit. Separability corresponds to welfare garnered from each resource being distributed among only the agents allocated to that resource, which is most often appropriate, e.g., in revenue and cost sharing. This results in

\vspace{-0.1in}
\begin{equation*}
U_i(a)= \sum_{r\in a_i}f^r\left(i,\{a\}_r\right),
\end{equation*}

\vspace{-0.025in}
\noindent where $f^r: N \times 2^N \rightarrow \mathbb{R}$ is the local distribution rule at resource $r$, i.e., $f^r(i,S)$ is the portion of the local welfare $W_r$ that is allocated to agent $i\in S$ when sharing with $S$. In addition, we assume that resources with identical local welfare functions have identical distribution rules, i.e., for any two resources $r,r'\in R$,

\vspace{-0.25in}
\begin{equation*}
W_r = W_{r'}\ \Longrightarrow\ f^r = f^{r'}.
\end{equation*}

\vspace{-0.025in}
\noindent In light of this assumption, for the rest of this article, we write $f^{W_r}$ instead of $f^r$. For completeness, we define $f^{W_r}(i,S):=0$ whenever $i\notin S$. A distribution rule $f^{W_r}$ is said to be budget-balanced if, for any player set $S\subseteq N$, $\sum_{i\in S}f^{W_r}(i,S)=W_r(S).$

We represent a welfare sharing game as $G = \left(N, R, \left\{\mathcal{A}_i\right\}_{i\in N}, \left\{f^{W_r}\right\}_{r\in R}, \left\{W_r\right\}_{r\in R}\right)$, and the design of $f^{W_r}$ is the focus of this article. \textit{When there is only one local welfare function, i.e., when $W_r=W$ for all $r\in R$, we drop the subscripts and denote the local welfare function and its corresponding distribution rule by $W$ and $f^W$ respectively.}

The primary goals when designing the distribution rules $f^{W_r}$ are to guarantee (i) equilibrium existence, and (ii) equilibrium efficiency. Our focus in this work is entirely on (i) and we consider pure Nash equilibria; however it should be noted that other equilibrium concepts are also of interest (Adlakha et al. \cite{Adlakha09}, Su and van der Schaar \cite{Su09}, Marden \cite{Marden12b}). Recall that a \textit{(pure Nash) equilibrium} is an action profile $a^* \in \mathcal{A}$ such that

\vspace{-0.15in}
\begin{equation*}
\left(\forall\ i\in N\right)\quad U_i(a^*_i,a^*_{-i})\ = \ \max_{a_i \in \mathcal{A}_i} U_i(a_i,a^*_{-i}).
\end{equation*}

\subsection{Examples of distribution rules.}\label{s.examplerules}

Existing literature on cost sharing games predominantly focuses on the design and analysis of specific distribution rules.  As such, there are a wide variety of distribution rules that are known to guarantee the existence of an equilibrium.  Table \ref{table:distr} summarizes several well-known distribution rules (both budget-balanced and non-budget-balanced) from existing literature on cost sharing, and we discuss their salient features in the following.

\begin{table}
\centering
\caption{Example distribution rules}
\label{table:distr}
\adjustbox{width=\columnwidth}{
\begin{footnotesize}
{\tabulinesep=2mm
\begin{tabu}{|>{\centering\arraybackslash}m{3.5cm}|c|>{\centering}m{9cm}|}
\hline
NAME & PARAMETER & FORMULA\\
\hline
\hline
Equal share                                 & None & $f^W_{EQ}(i,S)=\frac{W(S)}{|S|}$\\
\hline
Proportional share                          & $\DS\substack{\boldsymbol{\omega}=(\omega_1,\ldots,\omega_n)\mcr \mcr \mcr \text{where}\;\omega_i>0\mcr \mcr \text{for all}\;1\leq i\leq n}$ & $f^W_{PR}[\boldsymbol{\omega}](i,S)=\frac{\omega_i}{\sum_{j\in S}\omega_j}W(S)$\\
\hline
Shapley value                               & \multirow{2}{*}{None} & $f^W_{SV}(i,S)=\sum_{T\subseteq S\backslash\{i\}}\frac{(|T|)!(|S|-|T|-1)!}{|S|!}(W(T\cup\{i\})-W(T))$\\
\cline{1-1}\cline{3-3}
Marginal contribution                       &                       & $f^W_{MC}(i,S)=W(S)-W(S-\{i\})$\\
\hline
Weighted Shapley value                      & \multirow{2}{*}{$\DS\substack{\boldsymbol{\omega}=(\omega_1,\ldots,\omega_n)\mcr \mcr \mcr \text{where}\;\omega_i>0\mcr \mcr \text{for all}\;1\leq i\leq n}$} & $f^W_{WSV}[\boldsymbol{\omega}](i,S)=\DS\sum_{T\subseteq S: i\in T}\frac{\omega_i}{\sum_{j\in T}\omega_j}\left(\sum_{R\subseteq T}(-1)^{|T|-|R|}W(R)\right)$\\
\cline{1-1}\cline{3-3}
Weighted marginal contribution              &                                                                   & $f^W_{WMC}[\boldsymbol{\omega}](i,S)=\omega_i\left(W(S)-W(S-\{i\})\right)$\\
\hline
Generalized weighted Shapley value          & \multirow{2}{*}{$\DS\substack{\omega=(\boldsymbol{\lambda},\Sigma)\mcr \mcr \boldsymbol{\lambda}=(\lambda_1,\ldots,\lambda_n)\mcr \mcr \Sigma=(S_1,\ldots,S_m)\mcr \mcr \mcr \text{where}\;\lambda_i>0\mcr \mcr \text{for all}\;1\leq i\leq n \mcr \mcr \mcr \text{and}\;S_i\cap S_j=\emptyset\mcr \mcr \text{for}\ i\not=j\mcr \mcr \mcr \text{and}\;\cup\Sigma=N}$} & $\DS\substack{f^W_{GWSV}[\omega](i,S)=\DS\sum_{T\subseteq S: i\in \overline{T}}\frac{\lambda_i}{\sum_{j\in \overline{T}}\lambda_j}\left(\sum_{R\subseteq T}(-1)^{|T|-|R|}W(R)\right)\mcr \mcr \mcr \text{where}\;\overline{T}=T\cap S_k\;\text{and}\;k=\min\left\{j|S_j\cap T\not=\emptyset\right\}}$\\
\cline{1-1}\cline{3-3}
Generalized weighted marginal contribution  & & $\DS\substack{f^W_{GWMC}[\omega](i,S)=\lambda_i\left(W(\overline{S}_k)-W(\overline{S}_k-\{i\})\right)\mcr \mcr \text{where}\;\overline{S}_k=S-\DS\bigcup_{\ell=1}^{k-1}S_{\ell}\;\text{and}\;i\in S_k}$\\
\hline
\end{tabu}}
\end{footnotesize}}
\vspace{-0.2in}
\end{table}

\vspace{-0.1in}
\subsubsection{Equal/Proportional share distribution rules.}\label{s.eqprop}
Most prior work in network cost sharing (Anshelevich et al. \cite{Anshelevich04}, Corbo and Parkes \cite{Corbo05}, Fiat et al. \cite{Fiat06}, Chekuri et al. \cite{Chekuri07}, Christodoulou et al. \cite{Christodoulou10}) deals with the equal share distribution rule, $f^W_{EQ}$, defined in Table \ref{table:distr}. Here, the welfare is divided equally among the players. The proportional share distribution rule, $f^W_{PR}[\boldsymbol{\omega}]$, is a generalization, parameterized (exogenously) by $\boldsymbol{\omega}\in\mathbb{R}^{|N|}_{++}$, a vector of strictly positive player-specific weights, and the welfare is divided among the players in proportion to their weights.

Both $f^W_{EQ}$ and $f^W_{PR}$ are budget-balanced distribution rules. However, for general welfare functions, they do not guarantee an equilibrium for all games.\footnote{When the local welfare functions $\left\{W_r\right\}$ are `anonymous', i.e., when $W_r(S)$ is purely a function of $|S|$ for all $S\subseteq N$ and $r\in R$, $\left\{f^{W_r}_{EQ}\right\}$ guarantees an equilibrium for all games. This is a consequence of it being identical to the Shapley value distribution rule (Section \ref{s.shapleyfamily}) in this case. However, the analogous property for $f^W_{PR}[\boldsymbol{\omega}]$ does not hold.}

\vspace{-0.125in}
\subsubsection{The Shapley value family of distribution rules.}\label{s.shapleyfamily}
One of the oldest and most commonly studied distribution rules in the cost sharing literature is the Shapley value (Shapley \cite{Shapley53b}). Its extensions include the \textit{weighted} Shapley value and the \textit{generalized weighted} Shapley value, as defined in Table \ref{table:distr}.

The Shapley value family of distribution rules can be interpreted as follows. For any given subset of players $S$, imagine the players of $S$ arriving one at a time to the resource, according to some order $\pi$. Each player $i$ can be thought of as contributing $W\left(P_i^\pi\cup\{i\}\right)-W\left(P_i^\pi\right)$ to the welfare $W(S)$, where $P_i^\pi$ denotes the set of players in $S$ that arrived before $i$ in $\pi$. This is the `marginal contribution' of player $i$ to the welfare, according to the order $\pi$. The Shapley value, $f^W_{SV}(i,S)$, is simply the \textit{average} marginal contribution of player $i$ to $W(S)$, under the assumption that all $|S|!$ orders are equally likely. The weighted Shapley value, $f^W_{WSV}[\boldsymbol{\omega}](i,S)$, is then a weighted average of the marginal contributions, according to a distribution with full support on all the $|S|!$ orders, determined by the parameter $\boldsymbol{\omega}\in\mathbb{R}^{|N|}_{++}$, a strictly positive vector of player weights. The (symmetric) Shapley value is recovered when all weights are equal.

The generalized weighted Shapley value, $f^W_{GWSV}[\omega]$, generalizes the weighted Shapley value to allow for the possibility of player weights being zero. It is parameterized by a \textit{weight system} given by $\omega=(\boldsymbol{\lambda},\Sigma)$, where $\boldsymbol{\lambda}\in\mathbb{R}^{|N|}_{++}$ is a vector of strictly positive player weights, and $\Sigma=\left(S_1,S_2,\ldots,S_m\right)$ is an ordered partition of the set of players $N$. Once again, players get a weighted average of their marginal contributions, but according to a distribution determined by $\boldsymbol{\lambda}$, with support only on orders that conform to $\Sigma$, i.e., for $1\leq k< \ell\leq m$, players in $S_\ell$ arrive before players in $S_k$. Note that the weighted Shapley value is recovered when $|\Sigma|=1$, i.e., when $\Sigma$ is the trivial partition, $(N)$.

The importance of the Shapley value family of distribution rules is that all distribution rules are budget-balanced, guarantee equilibrium existence in any game, and also guarantee that the resulting games are so-called `potential games' (Hart and Mas-Colell \cite{Hart89}, Ui \cite{Ui00}).\footnote{Shapley value distribution rules result in exact potential games, weighted Shapley value distribution rules result in weighted potential games, and generalized weighted Shapley value distribution rules result in a slight variation of weighted potential games (see Appendix \ref{appendix:potentialfunction} for details).} However, they have one key drawback -- computing them is often\footnote{The Shapley value has been shown to be efficiently computable in several applications (Deng and Papadimitriou \cite{Deng94}, Mishra and Rangarajan \cite{Mishra07}, Aadithya et al. \cite{Aadithya10}), where specific welfare functions and special structures on the action sets enable simplifications of the general Shapley value formula.} intractable (Matsui and Matsui \cite{Matsui00}, Conitzer and Sandholm \cite{Conitzer04}), since it requires computing the sum of exponentially many marginal contributions.\footnote{Technically, if the entire welfare function is taken as an input, then the input size is already $O(2^n)$, and Shapley values can be computed `efficiently'. However, if access to the welfare function is by means of an oracle (Liben-Nowell et al. \cite{Liben-Nowell12}), than the actual input size is still $O(n)$, and the hardness is exposed.}

\vspace{-0.1in}
\subsubsection{The marginal contribution family of distribution rules.}\label{s.mcfamily}
Another classic and commonly studied distribution rule is $f^W_{MC}$, the marginal contribution distribution rule (Wolpert and Tumer \cite{Wolpert99}), where each player's share is simply his marginal contribution to the welfare, see Table \ref{table:distr}. Clearly, $f^W_{MC}$ is not always budget-balanced. However, an equilibrium is always guaranteed to exist, and the resulting game is an exact potential game, where the potential function is the same as the welfare function. Accordingly, the marginal contribution distribution rule always guarantees that the welfare maximizing allocation is an equilibrium, i.e., the `price of stability' is one. Finally, unlike the Shapley value family of distribution rules, note that it is easy to compute, as only two calls to the welfare function are required.

Note that, it is natural to consider weighted and generalized weighted marginal contribution distribution rules which parallel those for the Shapley value described above.  These are defined formally in Table \ref{table:distr}, and they inherit the equilibrium existence and potential game properties of $f^W_{MC}$, in an analogous manner to their Shapley value counterparts. These rules have, to the best of our knowledge, not been considered previously in the literature; however, they are crucial to the characterizations provided in this article.

\subsection{Important families of cost/revenue sharing games.}

Our model for welfare sharing games generalizes several existing families of games that have received significant attention in the literature. We illustrate a few examples below, in all of which the typical distribution rule studied is the equal share or Shapley value distribution rule:
\begin{enumerate}[label=(\roman*)]
\item \textit{Multicast and facility location games} (Chekuri et al. \cite{Chekuri07}) are a special case where $N$ is the set of users, $R$ is the set of links of the underlying graph, $\mathcal{A}_i$ consists of all feasible paths from user $i$ to the source, and for all $r\in R$, $W_r=c_r W$ is the local welfare function, where $c_r$ is the cost of the link $r$, and $W$ is given by:

    \vspace{-0.2in}
    \begin{small}
    \begin{equation}
    \label{eq:timsW}
    \left(\forall\ S\subseteq N\right)\quad W(S)=
    \begin{cases}
    -1, & S\not=\emptyset\mcr
    0, & S=\emptyset
    \end{cases}
    \end{equation}
    \end{small}
\item \textit{Congestion games} (Rosenthal \cite{Rosenthal73}) are a special case where, for each $r\in R$, the local welfare function $W_r$ is `anonymous', i.e., $W_r(S)$ is purely a function of $|S|$, and is given by $|S|$ times the negative of the delay function at $r$, for all $S\subseteq N$.
\item \textit{Atomic routing games with unsplittable flows} (Roughgarden and Tardos \cite{Roughgarden02}) are a special case where $N$ is the set of source-destination pairs $(s_i,t_i)$, each of which is associated with $\rho_i$ units of flow, $R$ is the set of edges of the underlying graph, and $\mathcal{A}_i$ consists of all feasible $s_i-t_i$ paths. If $c_r(x)$ denotes the latency function on edge $r$, then $W_r$ is the negative of the cost of the total flow due to the players in $S$, i.e., $W_r(S)=-|S|c_r\left(\sum_{i\in S}\rho_i\right)$, for all $S\subseteq N$.
\item \textit{Network formation games} (Anshelevich et al. \cite{Anshelevich04}) are a special subcase of the previous case, with a suitable encoding of the players. Suppose the set of players is $N=\{0,1,\ldots,n-1\}$, and the cost of constructing each edge $r$ is $C_r(S)$ when $S\subseteq N$ is the set of players who choose that edge. Then, one possibility is to set $\rho_i=10^i$ so that $\sum_{i\in S}\rho_i$ can be decoded to obtain the set of players $S$. Therefore, $c_r$ can be defined such that for all $S\subseteq N$, $c_r\left(\sum_{i\in S}\rho_i\right)=\frac{C_r(S)}{|S|}$.
\end{enumerate}
Other notable specializations of our model that focus on the design of distribution rules are network coding (Marden and Effros \cite{Marden12}), graph coloring (Panagopoulou and Spirakis \cite{Panagopoulou08}), and coverage problems (Marden and Wierman \cite{Marden08}, Marden and Wierman \cite{Marden13a}). Designing distribution rules in our cost sharing model also has applications in distributed control (Gopalakrishnan et al. \cite{Gopalakrishnan11}).

\section{Basis representations.}\label{s.basisreps}

To gain a deeper understanding of the structural form of some of the distribution rules discussed in Section \ref{s.examplerules}, it is useful to consider their `basis' representations. Not only do these representations provide insight, they are crucial to the proofs in this paper. The basis framework we adopt was first introduced in Shapley \cite{Shapley53a} in the context of the Shapley value, and corresponds to the set of `inclusion functions'. We start by defining a basis for the local welfare functions below, and then move to introducing the basis representation of the distribution rules we introduced in Section \ref{s.examplerules}.

\vspace{-0.1in}
\subsection{A basis for welfare functions.}\label{s.basisW}

Instead of working with $W$ directly, it is often easier to represent $W$ as a linear combination of simple basis welfare functions. A natural basis, first defined in Shapley \cite{Shapley53a}, is the set of \textit{inclusion functions}. The inclusion function of a player subset $T\subseteq N$, denoted by $W^T$, is defined as:

\vspace{-0.25in}
\begin{equation}
\label{eq:Twelfare:2}
W^T(S):=
\begin{cases}
1, & \ T\subseteq S\mcr
0, & \ \mbox{otherwise}
\end{cases}
\end{equation}
In the context of cooperative game theory, inclusion functions are identified with \textit{unanimity games}. It is well-known (Shapley \cite{Shapley53a}) that the set of all inclusion functions, $\left\{W^T\ :\ T\subseteq N\right\}$, constitutes a basis for the space of all welfare functions, i.e., given any welfare function $W$, there exists a unique support set $\mathcal{T}^W\subseteq 2^N$, and a unique sequence $Q^W=\left\{q^W_T\right\}_{T\in\mathcal{T}^W}$ of non-zero weights indexed by $\mathcal{T}^W$, such that:

\vspace{-0.25in}
\begin{equation}
\label{eq:genTwelfare:2}
W=\sum_{T\in\mathcal{T}^W}q^W_TW^T
\end{equation}
We sometimes denote the welfare function $W$ by the tuple $\left(\mathcal{T}^W,Q^W\right)$.

\vspace{-0.1in}
\subsection{A basis for distribution rules.}\label{s.basisD}

The basis representation for welfare functions introduced above naturally yields a basis representation for distribution rules. To simplify notation in the following, we denote $f^{W^T}$ by $f^T$, for each $T\in\mathcal{T}^W$. That is, $f^T: N \times 2^N \rightarrow \mathbb{R}$ is a \textit{basis distribution rule} corresponding to the unanimity game $W^T$, where $f^T(i,S)$ is the portion of $W^T(S)$ allocated to agent $i\in S$ when sharing with $S$.

Given a set of basis distribution rules $\left\{f^T\ : \ T\subseteq N\right\}$, by linearity, the function $f^W$,

\vspace{-0.1in}
\begin{equation}
\label{eq:Tdistribution:2}
f^W:=\displaystyle\sum_{T\in\mathcal{T}^W}q^W_Tf^T,
\end{equation}
defines a distribution rule corresponding to the welfare function $W$. Note that if each $f^T$ is budget-balanced, meaning that for any player set $S\subseteq N$, $\sum_{i\in S}f^T(i,S)=W^T(S)$, then $f^W$ is also budget-balanced. However, unlike the basis for welfare functions, some distribution rules do not have a basis representation of the form (\ref{eq:Tdistribution:2}), e.g., equal and proportional share distribution rules (see Section \ref{s.eqprop}). But, well-known distribution rules of interest to us, like the Shapley value family of distribution rules, were originally defined in this manner. \textit{Further, our characterizations highlight that any distribution rule that guarantees equilibrium existence must have a basis representation.}

Table \ref{table:basisdistr} restates the distribution rules shown in Table \ref{table:distr} in terms of their basis representations, which, as can be seen, tend to be simpler and provide more intuition.

\begin{table}
\centering
\caption{Definition of basis distribution rules}
\label{table:basisdistr}
\adjustbox{width=\columnwidth}{
\begin{footnotesize}
{
\begin{tabu}{|>{\centering\arraybackslash}m{3.5cm}|>{\centering\arraybackslash}m{2.25cm}|>{\centering\arraybackslash}m{10cm}|}
\hline
NAME & PARAMETER & DEFINITION\\
\hline
\hline
Shapley value                               & \multirow{2}{*}{None} &
\begin{equation}
\label{eq:basisSV}
f^T_{SV}(i,S)=f^{W^T}_{EQ}(i,S)=
\begin{cases}
\frac{1}{|T|}, & \ i\in T\ \mbox{and}\ T\subseteq S\mcr
0, & \ \mbox{otherwise}
\end{cases}
\end{equation}\\
\cline{1-1}\cline{3-3}
Marginal contribution                       &  &
\begin{equation}
\label{eq:basisMC}
f^T_{MC}(i,S)=|T|f^T_{SV}(i,S)=
\begin{cases}1, & \ i\in T\ \mbox{and}\ T\subseteq S\mcr
0, & \ \mbox{otherwise}
\end{cases}
\end{equation}\\
\hline
Weighted Shapley value                      & \multirow{2}{*}{$\DS\substack{\boldsymbol{\omega}=(\omega_1,\ldots,\omega_n)\mcr \mcr \mcr \text{where}\;\omega_i>0\mcr \mcr \text{for all}\;1\leq i\leq n}$} &
\begin{equation}
\label{eq:basisWSV}
f^T_{WSV}[\boldsymbol{\omega}](i,S)=f^{W^T}_{PR}[\boldsymbol{\omega}](i,S)=
\begin{cases}
\frac{\omega_i}{\sum_{j\in T}\omega_j}, & \ i\in T\ \mbox{and}\ T\subseteq S\mcr
0, & \ \mbox{otherwise}
\end{cases}
\end{equation}\\
\cline{1-1}\cline{3-3}
Weighted marginal contribution              & &
\begin{equation*}
f^T_{WMC}[\boldsymbol{\omega}](i,S)=\left(\scriptstyle\sum_{j\in T}\omega_j\right)f^T_{WSV}[\boldsymbol{\omega}](i,S)=
    \begin{cases}
    \omega_i, & \ i\in T\ \mbox{and}\ T\subseteq S\mcr
    0, & \ \mbox{otherwise}
    \end{cases}
\end{equation*}\\
\hline
Generalized weighted Shapley value          & \multirow{2}{*}{$\DS\substack{\omega=(\boldsymbol{\lambda},\Sigma)\mcr \mcr \boldsymbol{\lambda}=(\lambda_1,\ldots,\lambda_n)\mcr \mcr \Sigma=(S_1,\ldots,S_m)\mcr \mcr \mcr \text{where}\;\lambda_i>0\mcr \mcr \text{for all}\;1\leq i\leq n \mcr \mcr \mcr \text{and}\;S_i\cap S_j=\emptyset\mcr \mcr \text{for}\ i\not=j\mcr \mcr \mcr \text{and}\;\cup\Sigma=N}$} &
\begin{equation}
\label{eq:basisGWSV}
\DS\substack{
f^T_{GWSV}[\omega](i,S)=
    \begin{cases}
    \frac{\lambda_i}{\sum_{j\in\overline{T}}\lambda_j}, & \ i\in \overline{T}\ \mbox{and}\ T\subseteq S\mcr
    0, & \ \mbox{otherwise}
    \end{cases}\mcr \mcr \mcr
\text{where}\;\overline{T}=T\cap S_k\;\text{and}\;k=\min\left\{j|S_j\cap T\not=\emptyset\right\}}
\end{equation}\\
\cline{1-1}\cline{3-3}
Generalized weighted marginal contribution  & &
\begin{equation}
\label{eq:basisGWMC}
\DS\substack{
f^T_{GWMC}[\omega](i,S)=\left(\sum_{j\in\overline{T}}\lambda_j\right)f^T_{GWSV}[\omega](i,S)=
    \begin{cases}
    \lambda_i, & \ i\in \overline{T}\ \mbox{and}\ T\subseteq S\mcr
    0, & \ \mbox{otherwise}
    \end{cases}\mcr \mcr \mcr
\text{where}\;\overline{T}=T\cap S_k\;\text{and}\;k=\min\left\{j|S_j\cap T\not=\emptyset\right\}}
\end{equation}\\
\hline
\end{tabu}}
\end{footnotesize}}
\end{table}

For example, the Shapley value distribution rule on a welfare function $W$ is quite naturally defined through its basis -- for each unanimity game $W^T$, the welfare is shared equally among the players, see (\ref{eq:basisSV}). In other words, whenever there is welfare generated (when all the players in $T$ are present), the resulting welfare is split equally among the contributing players (players in $T$).  Similarly, the weighted Shapley value, for each unanimity game $W^T$, distributes the welfare among the players in proportion to their weights, see (\ref{eq:basisWSV}).  Finally, the basis representation highlights that the generalized weighted Shapley value can be interpreted with $\Sigma$ as representing a grouping of players into priority classes, and the welfare being distributed only among the contributing players of the highest priority, in proportion to their weights, see (\ref{eq:basisGWSV}).

Interestingly, the marginal contribution distribution rule, though it was not originally defined this way, has a basis representation that highlights a core similarity to the Shapley value.  In particular, though the definitions in Table \ref{table:distr} make $f^W_{MC}$ and $f^W_{SV}$ seem radically different; from Table \ref{table:basisdistr}, their basis distribution rules, $f^T_{MC}$ and $f^T_{SV}$, are, in fact, quite intimately related, see (\ref{eq:basisSV}) and (\ref{eq:basisMC}).  We formalize this connection between the Shapley value family of distribution rules and the marginal contribution family of distribution rules in Section \ref{s.mcresult}.

\section{Results and discussion.}\label{s.result}

Our goal is to characterize the space of distribution rules that guarantee the existence of an equilibrium in welfare sharing games.  Towards this end, this paper builds on the recent works of Chen et al. \cite{Chen10} and Marden and Wierman \cite{Marden13b}, which take the first steps toward providing such a characterization. Proposition \ref{timsresult} combines the main contributions of these two papers into one statement. Let $\mathbb{W}$ denote a nonempty set of welfare functions. Let $f^\mathbb{W}=\left\{f^W\right\}_{W\in\mathbb{W}}$ denote the set of corresponding distribution rules. Let  $\mathcal{G}(N,f^\mathbb{W},\mathbb{W})$ denote the class of all welfare sharing games with player set $N$, local welfare functions $W_r\in \mathbb{W}$, and corresponding distribution rules $f^{W_r}\in f^\mathbb{W}$. We refer to $\mathbb{W}$ as the set of local welfare functions of the class $\mathcal{G}(N,f^\mathbb{W},\mathbb{W})$. Note that this class is quite general; in particular, it includes games with arbitrary resources and action sets. \textit{When there is only one local welfare function, i.e., when $\mathbb{W}=\{W\}$, we denote this class simply by $\mathcal{G}(N,f^W,W)$. Note that $\mathcal{G}(N,f^W,W)\subseteq\mathcal{G}(N,f^\mathbb{W},\mathbb{W})$ for all $W\in\mathbb{W}$.}

\vspace{-0.025in}
\begin{proposition}[Chen et al. \cite{Chen10}, Marden and Wierman \cite{Marden13b}]\label{timsresult}
There exists a local welfare function $W$ for which all games in $\mathcal{G}(N,f^W,W)$ possess a pure Nash equilibrium for a budget-balanced $f^W$ if and only if there exists a weight system $\omega$ for which $f^W$ is the generalized weighted Shapley value distribution rule, $f^W_{GWSV}[\omega]$.
\end{proposition}

\vspace{-0.05in}
Less formally, Proposition \ref{timsresult} states that if one wants to use a distribution rule that is budget-balanced and guarantees equilibrium existence for all possible welfare functions and action sets, then one is limited to the class of generalized weighted Shapley value distribution rules.\footnote{The authors of Chen et al. \cite{Chen10} and Marden and Wierman \cite{Marden13b} use the term \textit{ordered protocols} to refer to generalized weighted Shapley value distribution rules with $|\Sigma|=|N|$, i.e., where $\Sigma$ defines a total order on the set of players $N$. They state their characterizations in terms of \textit{concatenations} of positive ordered protocols, which are generalized weighted Shapley value distribution rules with an arbitrary $\Sigma$.} This result is shown by exhibiting a specific `worst-case' local welfare function $W^*$ (the one in (\ref{eq:timsW})) for which this limitation holds. In reality, when designing a distribution rule, one \textit{knows} the specific set of local welfare functions $\mathbb{W}$ for the situation, and Proposition \ref{timsresult} claims nothing in the case where it does not include $W^*$, where, in particular, there may be other budget-balanced distribution rules that guarantee equilibrium existence for all games. Recent work has shown that there are settings where this is the case (Marden and Wierman \cite{Marden13a}), at least when the agents are not allowed to choose more than one resource. In addition, the marginal contribution family of distribution rules is a non-budget-balanced class of distribution rules that guarantee equilibrium existence in all games (no matter what the local welfare functions $\mathbb{W}$), and there could potentially be others as well.

In the rest of this section, we provide two equivalent characterizations of the space of distribution rules that guarantee equilibrium existence for all games with a fixed set of local welfare functions -- one in terms of generalized weighted Shapley values and the other in terms of generalized weighted marginal contributions. We defer complete proofs to the appendices. However, we sketch an outline in Section \ref{s.proofoutline}, highlighting the proof technique and the key steps involved.

\vspace{-0.1in}
\subsection{Characterization in terms of generalized weighted Shapley values.}\label{s.svresult}
Our first characterization states that for \textit{any fixed set of local welfare functions}, even if the distribution rules are \textit{not budget-balanced}, the conclusion of Proposition \ref{timsresult} is still valid. That is, every distribution rule that guarantees the existence of an equilibrium in all games is equivalent to a generalized weighted Shapley value distribution rule:

\vspace{-0.025in}
\begin{theorem}\label{mainresult}
Given any set of local welfare functions $\mathbb{W}$, all games in $\mathcal{G}(N,f^\mathbb{W},\mathbb{W})$ possess a pure Nash equilibrium if and only if there exists a weight system $\omega$, and a mapping $g_{SV}$ that maps each local welfare function $W\in\mathbb{W}$ to a corresponding ground welfare function $g_{SV}(W)$ such that its distribution rule $f^W\in f^\mathbb{W}$ is equivalent to the generalized weighted Shapley value distribution rule, $f^{W'}_{GWSV}[\omega]$, where $W'=g_{SV}(W)$ is the actual welfare that is distributed\footnote{Note that $W'=W$ if and only if $f^W$ is budget-balanced.} by $f^W$, defined as,

\vspace{-0.1in}
\begin{small}
\begin{equation}
\label{eq:definitionW'}
(\forall S\subseteq N)\quad W'(S) = \displaystyle\sum_{i\in S}f^W(i,S).
\end{equation}
\end{small}
\end{theorem}

Refer to Appendix \ref{appendix:mainresult} for the complete proof, and Section \ref{s.proofoutline} for an outline. While Proposition \ref{timsresult} states that there exists a local welfare function for which any budget-balanced distribution rule is required to be equivalent to a generalized weighted Shapley value (on that welfare function) in order to guarantee equilibrium existence, Theorem \ref{mainresult} states a much stronger result that, \textit{for any set of local welfare functions}, the corresponding distribution rules must be equivalent to generalized weighted Shapley values on some ground welfare functions to guarantee equilibrium existence. This holds true even when the distribution rules are not budget-balanced. Proving Theorem \ref{mainresult} requires working with arbitrary local welfare functions, which is a clear distinction from the proof of Proposition \ref{timsresult}, which exhibits a specific local welfare function, showing the result for that case.

From Theorem \ref{mainresult}, it follows that designing distribution rules to ensure the existence of an equilibrium merely requires selecting a weight system $\omega=\left(\boldsymbol{\lambda},\Sigma\right)$ and a ground welfare function $W'$ for each local welfare function $W\in\mathbb{W}$ (this defines the mapping $g_{SV}$), and then applying the distribution rules $\left\{f^{W'}_{GWSV}[\omega]\right\}_{W\in\mathbb{W}}$. Budget-balance, if required, can be directly controlled through proper choice of $W'$, since $\{W'\}$ are the actual welfares distributed. For example, if exact budget-balance is desired, then $W'=W$ for all $W\in\mathbb{W}$. Notions of approximate budget-balance (Roughgarden and Sundararajan \cite{Roughgarden09}) can be similarly accommodated by keeping $W'$ `close' to $W$.

An important implication of Theorem \ref{mainresult} is that if one hopes to use a distribution rule that always guarantees equilibrium existence in games with any fixed set of local welfare functions, then one is limited to working within the class of `potential games'.  This is perhaps surprising since a priori, potential games are often thought to be a small, special class of games (Sandholm \cite{Sandholm10}).\footnote{In spite of this limitation, it is useful to point out that there are many well understood learning dynamics which guarantee equilibrium convergence in potential games (Blume \cite{Blume93}, Marden et al. \cite{Marden09a}, Marden and Shamma \cite{Marden12a}).} More specifically, generalized weighted Shapley value distribution rules result in a slight variation of weighted potential games (Hart and Mas-Colell \cite{Hart89}, Ui \cite{Ui00}),\footnote{See Definition \ref{defn:piecewiseWPG} in Appendix \ref{appendix:potentialfunction}.} whose potential function can be computed recursively as:

\vspace{-0.2in}
\begin{equation*}
\boldsymbol{\Phi}[\omega](a) = \sum_{r\in R} \boldsymbol{\phi}_r[\omega](\{a\}_r),
\end{equation*}

\vspace{-0.05in}
\noindent where $\boldsymbol{\phi}_r[\omega]\ :\ 2^N\rightarrow\mathbb{R}^m$ is the local potential function at resource $r$ (we denote the $k$th element of this vector by $\left(\phi_r[\omega]\right)_k$), and for any $1\leq k\leq m$ and any subset $S\subseteq N$,

\vspace{-0.1in}
\begin{equation}
\label{eq:GWSVpotential}
\left(\phi_r[\omega]\right)_k(S) = \frac{1}{\sum_{i\in S}\lambda_i}\left(W'_r(\overline{S}_{m-k+1}) + \sum_{i\in S}\lambda_i\left(\phi_r[\omega]\right)_k(S-\{i\})\right),
\end{equation}
where $W'_r=g_{SV}(W_r)$ and $\overline{S}_k=S-\cup_{\ell=1}^{k-1}S_\ell$. Refer to Appendix \ref{appendix:potentialfunction} for the proof.

Theorem \ref{mainresult} also has some negative implications. First, the limitation to generalized weighted Shapley value distribution rules means that one is forced to use distribution rules which may require computing exponentially many marginal contributions, as discussed in Section \ref{s.examplerules}. Second, if one desires budget-balance, then there are efficiency limits for games in $\mathcal{G}(N,f^\mathbb{W}_{WSV},\mathbb{W})$. In particular, there exists a submodular welfare function $W$ such that, for any weight vector $\boldsymbol{\omega}$, there exists a game in $\mathcal{G}{(N,f^W_{WSV}[\boldsymbol{\omega}],W)}$ where the best equilibrium has welfare that is a multiplicative factor of two worse than the optimal welfare (Marden and Wierman \cite{Marden13b}).

\subsection{Characterization in terms of generalized weighted marginal contributions.}\label{s.mcresult}

Our second characterization is in terms of the marginal contribution family of distribution rules. The key to obtaining this contribution is the connection between the marginal contribution and Shapley value distribution rules that we observed in Section \ref{s.basisreps}.  We formalize this in the following proposition. Refer to Appendix \ref{appendix:prop:svmcequiv} for the proof.

\begin{proposition}
\label{prop:svmcequiv}
For any two welfare functions $W'=\left(\mathcal{T}',Q'\right)$ and $W''=\left(\mathcal{T}'',Q''\right)$, and any weight system $\omega=\left(\boldsymbol{\lambda},\Sigma\right)$,
\begin{small}
\begin{equation}
\label{eq:svmcequiv}
f^{W'}_{GWSV}[\omega]=f^{W''}_{GWMC}[\omega]\quad\Longleftrightarrow\quad\mathcal{T}'=\mathcal{T}''\quad\text{and}\quad\left(\forall\ T\in\mathcal{T}'\right)\;q'_T=\left(\sum_{j\in\overline{T}}\lambda_j\right)q''_T.
\end{equation}
\end{small}
\end{proposition}

Informally, Proposition \ref{prop:svmcequiv} says that generalized weighted Shapley values and generalized weighted marginal contributions are equivalent, except with respect to different ground welfare functions whose relationship is through their basis coefficients, as indicated in (\ref{eq:svmcequiv}). This proposition immediately leads to the following equivalent statement of Theorem \ref{mainresult}.

\begin{theorem}\label{altresult}
Given any set of local welfare functions $\mathbb{W}$, all games in $\mathcal{G}(N,f^\mathbb{W},\mathbb{W})$ possess a pure Nash equilibrium if and only if there exists a weight system $\omega$, and a mapping $g_{MC}$ that maps each local welfare function $W\in\mathbb{W}$ to a corresponding ground welfare function $g_{MC}(W)$ such that its distribution rule $f^W\in f^\mathbb{W}$ is equivalent to the generalized weighted marginal contribution distribution rule, $f^{W''}_{GWMC}[\omega]$, where $W''=g_{MC}(W)$ is defined as,
\begin{equation}
\label{eq:definitionW''}
W'' = h(g_{SV}(W)),
\end{equation}
where $h$ denotes the mapping that maps $W'$ to $W''$ according to (\ref{eq:svmcequiv}).
\end{theorem}

Importantly, Theorem \ref{altresult} provides an alternate way of designing distribution rules that guarantee equilibrium existence. The advantage of this alternate design is that marginal contributions are much easier to compute than the Shapley value, which requires computing exponentially many marginal contributions. However, it is much more difficult to control the budget-balance of marginal contribution distribution rules.  Specifically, $\{W''\}$ are not the actual welfares distributed, and so there is no direct control over budget-balance as was the case for generalized weighted Shapley value distribution rules. Instead, it is necessary to start with desired welfares $\{W'\}$ to be distributed (equivalently, the desired mapping $g_{SV}$) and then perform a `preprocessing' step of transforming it into the ground welfare functions $\{W''\}$ using (\ref{eq:definitionW''}), which requires exponentially many calls to each $W'$.  However, this is truly a preprocessing step, and thus only needs to be performed once.

Another simplification that Theorem \ref{altresult} provides when compared to Theorem \ref{mainresult} is in terms of the potential function.  In particular, in light of Proposition \ref{prop:svmcequiv}, the distribution rules $f^{W'}_{GWSV}[\omega]$ and $f^{W''}_{GWMC}[\omega]$, where $W''=h(W')$, result in the same `weighted' potential game with the same potential function $\boldsymbol{\Phi}[\omega]$.  However, in terms of $W''$, there is a clear closed-form expression for the local potential function at resource $r$, $\boldsymbol{\phi}_r[\omega]\ :\ 2^N\rightarrow\mathbb{R}^m$. For any $1\leq k\leq m$ and any subset $S\subseteq N$,
\begin{equation*}
\left(\phi_r[\omega]\right)_k(S) = W''_r(\overline{S}_{m-k+1}),
\end{equation*}
where $W''_r=g_{MC}(W_r)$ and $\overline{S}_k=S-\cup_{\ell=1}^{k-1}S_\ell$. In other words, we have,
\begin{equation*}
\left(\forall\ S\subseteq N\right)\quad\boldsymbol{\phi}_r[\omega](S) = \left(W''_r(\overline{S}_m), W''_r(\overline{S}_{m-1}), \ldots, W''_r(\overline{S}_1)\right).
\end{equation*}
Refer to Appendix \ref{appendix:potentialfunction} for the proof. Having a simple closed form potential function is helpful for many reasons. For example, it aids in the analysis of learning dynamics and in characterizing efficiency bounds through the well-known potential function method (Tardos and Wexler \cite{Tardos07}).

\vspace{-0.1in}
\subsection{Limitations and extensions.}
\label{s.extend-discuss}

It is important to highlight that our characterizations in Theorems \ref{mainresult} and \ref{altresult} crucially depend on the fact that an equilibrium must be guaranteed in \textit{all games}, i.e., for all possibilities of resources, action sets, and choice of local welfare functions from $\mathbb{W}$. (This is the same for the characterizations given in the previous work in Chen et al. \cite{Chen10} and Marden and Wierman \cite{Marden13b}.) If this requirement is relaxed it may be possible to find situations where distribution rules that are not equivalent to generalized weighted Shapley values can guarantee equilibrium existence. For example, Marden and Wierman \cite{Marden13a} gives such a rule for a coverage game where players can select only one resource at a time. A challenging open problem is to determine the structure on the action sets that is necessary for the characterizations in Theorems \ref{mainresult} and \ref{altresult} to hold.

A second remark is that our entire focus has been on characterizing distribution rules that guarantee equilibrium existence. However, guaranteeing efficient equilibria is also an important goal for distribution rules. The characterizations in Theorems \ref{mainresult} and \ref{altresult} provide important new tools to optimize the efficiency, e.g., the price of anarchy, of distribution rules for general cost sharing and revenue sharing games through proper choice of the weight system and ground welfare functions. An important open problem in this direction is to understand the resulting tradeoffs between budget-balance and efficiency.

Finally, it is important to remember that our focus has been on cost sharing \emph{games}; however it is natural to ask if similar characterizations can be obtained for cost sharing \emph{mechanisms} (Moulin and Shenker \cite{Moulin01}, Dobzinski et al. \cite{Dobzinski08}, Immorlica and Pountourakis \cite{Immorlica12}, Johari and Tsitsiklis \cite{Johari04}, Yang and Hajek \cite{Yang07}, Moulin \cite{Moulin10}). More specifically, the model considered in this paper extends immediately to situations where players have independent heterogeneous valuations over actions, by adding more welfare functions to $\mathbb{W}$.\footnote{To see this, consider a welfare sharing game $G\in\mathcal{G}(N,f^\mathbb{W},\mathbb{W})$. Let the action set of player $i$ be $\mathcal{A}_i=\{a_1,\ldots,a_\ell\}$, and suppose he values action $a_j$ at $u_j$, for $1\leq j\leq \ell$. Then, we modify $G$ to $G'$ by adding $\ell$ more resources $r_1,\ldots,r_\ell$ to $R$, setting $W_{r_j}(S)=\begin{cases}u_j, & i\in S\\ 0, & i\notin S\end{cases}$ and $f^{W_{r_j}}=f^{W_{r_j}}_{SV}$ for $1\leq j\leq \ell$, and augmenting each action in $\mathcal{A}_i$ with its corresponding resource, so that $a_j\rightarrow a_j\cup\{r_j\}$. Then, $G'\in\mathcal{G}(N,f^{\mathbb{W}'},\mathbb{W}')$, where $\mathbb{W}'=\mathbb{W}\DS\cup_{j=1}^\ell \{W_{r_j}\}$ and $f^{\mathbb{W}'}=f^\mathbb{W}\DS\cup_{j=1}^\ell \{f^{W_{r_j}}\}$. Notice that all games in $\mathcal{G}(N,f^\mathbb{W},\mathbb{W})$ have an equilibrium if and only if all games in $\mathcal{G}(N,f^{\mathbb{W}'},\mathbb{W}')$ have an equilibrium.} However, in cost sharing mechanisms, player valuations are private, which adds a challenging wrinkle to this translation. Thus, extending our characterizations to the setting of cost sharing mechanisms is a difficult, but important, open problem.

\vspace{-0.1in}
\section{Prior work in noncooperative cost sharing games.}
\label{s.priorwork}

As noted previously, the first steps toward characterizing the space of distribution rules that guarantee equilibrium existence were provided in Chen et al. \cite{Chen10} and Marden and Wierman \cite{Marden13b}. Prior to that, almost all the literature in cost sharing games (Anshelevich et al. \cite{Anshelevich04}, Corbo and Parkes \cite{Corbo05}, Fiat et al. \cite{Fiat06}, Chekuri et al. \cite{Chekuri07}, Christodoulou et al. \cite{Christodoulou10}) considered a fixed distribution rule that guarantees equilibrium existence, namely equal share (dubbed the `fair cost allocation rule', equivalent to the Shapley value in these settings), and the focus was directed towards characterizing the efficiency of equilibria.

A recent example of work in this direction is von Falkenhausen and Harks \cite{vonFalkenhausen13}, which considers games where the action sets (strategy spaces) of the agents are either singletons or bases of a matroid defined on the ground set of resources. For such games, the authors focus on designing (possibly non-separable, non-scalable) distribution rules that result in efficient equilibria. They tackle the question of equilibrium existence with a novel characterization of the set of possible equilibria independent of the distribution rule, and then exhibit a family of distribution rules that result in any given equilibrium in this set. Our goal is fundamentally different from theirs, in that we seek to characterize distribution rules that guarantee equilibrium existence for a \textit{class of games}, whereas they directly characterize the best and worst achievable equilibria of a \textit{given game}.

An alternative approach for distribution rule design is studied in Anshelevich et al. \cite{Anshelevich03}, Hoefer and Krysta \cite{Hoefer05}. The authors consider a fundamentally different model of a cost sharing game where agents not only choose resources, but also indicate their demands for the shares of the resulting welfare at these resources. Their model essentially defers the choice of the distribution rule to the agents. In such settings, they prove that an equilibrium may not exist in general.

\section{Proof sketch of Theorem \ref{mainresult}.}\label{s.proofoutline}

We now sketch an outline of the proof of Theorem \ref{mainresult} for the special case where there is just one local welfare function $W$, i.e., $\mathbb{W}=\{W\}$, highlighting the key stages. For an independent, self-contained account of the complete proof, refer to Appendix \ref{appendix:mainresult}.

First, note that we only need to prove one direction since it is known that for any weight system $\omega$ and any two welfare functions $W,W'$, all games in $\mathcal{G}(N,f^{W'}_{GWSV}[\omega],W)$ have an equilibrium (Hart and Mas-Colell \cite{Hart89}).\footnote{Notice that $W$ has no role to play as far as equilibrium existence of games $G\in\mathcal{G}(N,f^{W'}_{GWSV}[\omega],W)$ is concerned, since it does not affect player utilities. This observation will prove crucial later.} Thus, our focus is solely on proving that for distribution rules $f^W$ that are not generalized weighted Shapley values on some ground welfare function, there exists $G\in\mathcal{G}(N,f^W,W)$ with no equilibrium.

The general proof technique is as follows. First, we present a quick reduction to characterizing only budget-balanced distribution rules $f^W$ that guarantees the existence of an equilibrium for all games in $\mathcal{G}(N,f^W,W)$. Then, we establish several necessary conditions for a budget-balanced distribution rule $f^W$ that guarantees the existence of an equilibrium for all games in $\mathcal{G}(N,f^W,W)$, which effectively eliminate all but generalized weighted Shapley values on $W$, giving us our desired result. We establish these conditions by a series of counterexamples which amount to choosing a resource set $R$ and the action sets $\left\{\mathcal{A}_i\right\}_{i\in N}$, for which failure to satisfy a necessary condition would lead to nonexistence of an equilibrium.

Throughout, we work with the basis representation of the welfare function $W$ that was introduced in Section \ref{s.basisW}. Since we are dealing with only one welfare function $W$, we drop the superscripts from $\mathcal{T}^W$, $Q^W$, and $q^W_T$ in order to simplify notation. It is useful to think of the sets in $\mathcal{T}$ as being `coalitions' of players that contribute to the welfare function $W$ (also referred to as contributing coalitions), and the corresponding coefficients in $Q$ as being their respective contributions. Also, for simplicity, we normalize $W$ by setting $W(\emptyset)=0$ and therefore, $\emptyset\notin\mathcal{T}$. Before proceeding, we introduce some notation below:
\begin{enumerate}
\item For any subset $S\subseteq N$, $\mathcal{T}(S)$ denotes the set of contributing coalitions in $S$:

\vspace{-0.1in}
    \begin{equation*}
    \mathcal{T}(S)=\left\{T\in \mathcal{T}\ |\ T\subseteq S\right\}
    \end{equation*}

\vspace{-0.025in}
\item For any subset $S\subseteq N$, $N(S)$ denotes the set of contributing players in $S$:

\vspace{-0.1in}
    \begin{equation*}
    N(S)=\bigcup\mathcal{T}(S)
    \end{equation*}

\vspace{-0.025in}
\item For any two players $i,j\in N$, $\mathcal{T}_{ij}$ denotes the set of all coalitions containing $i$ and $j$:

\vspace{-0.1in}
    \begin{equation*}
    \mathcal{T}_{ij}=\left\{T\in \mathcal{T}\ |\ \{i,j\}\subseteq T\right\}
    \end{equation*}

\vspace{-0.025in}
\item Let $\mathcal{B}\subseteq 2^{N}$ denote any collection of subsets of a set $N$. Then the relation $\subseteq$ induces a partial order on $\mathcal{B}$. $\mathcal{B}^{\min}$ denotes the set of minimal elements of the poset $\left(\mathcal{B},\subseteq\right)$:

\vspace{-0.1in}
    \begin{equation*}
    \mathcal{B}^{\min}=\left\{B\in \mathcal{B}\ |\ \left(\nexists B'\in \mathcal{B}\right)\ s.t.\ B'\subsetneq B\right\}
    \end{equation*}
\end{enumerate}

\vfill

\begin{example}
\label{example:1:po}
\textit{Let $N = \{i,j,k,\ell\}$ be the set of players. Table \ref{table:2a:po} defines a $W:2^N\rightarrow\mathbb{R}$, as well as five different distribution rules for $W$. Table \ref{table:2b:po} shows the basis representation of $W$, and Table \ref{table:2c:po} illustrates the notation defined above for $W$. Throughout the proof sketch, we periodically revisit these distribution rules to illustrate the key ideas.}
\end{example}

\vfill
The proof is divided into five sections -- each section incrementally builds on the structure imposed on the distribution rule $f$ by previous sections.

\subsection{Reduction to budget-balanced distribution rules.}
First, we reduce the problem of characterizing \textit{all} distribution rules $f^W$ that guarantee equilibrium existence for all $G\in\mathcal{G}(N,f^W,W)$ to characterizing only \textit{budget-balanced} distribution rules $f^W$ that guarantee equilibrium existence for all $G\in\mathcal{G}(N,f^W,W)$:

\newpage
\begin{proposition}\label{mainprop:1}
For all welfare functions $W$, a distribution rule $f^W$ guarantees the existence of an equilibrium for all games in $\mathcal{G}(N,f^W,W)$ if and only if it guarantees the existence of an equilibrium for all games in $\mathcal{G}(N,f^W,W')$, where, for all subsets $S\subseteq N$, $W'(S):=\sum_{i\in S}f^W(i,S)$.
\end{proposition}

\proof{Proof.} This proposition is actually a subtlety of our notation. For games $G\in\mathcal{G}(N,f^W,W)$, the welfare function $W$ does not directly affect strategic behavior (it only does so through the distribution rule $f^W$). Therefore, in terms of strategic behavior and equilibrium existence, the classes $\mathcal{G}(N,f^W,W)$ and $\mathcal{G}(N,f^W,W')$ are identical, for any two welfare functions $W,W'$. Therefore, a distribution rule $f^W$ guarantees equilibrium existence for all games in $\mathcal{G}(N,f^W,W)$ if and only if it guarantees equilibrium existence for all games in $\mathcal{G}(N,f^W,W')$. To complete the proof, simply pick $W'$ to be the actual welfare distributed by $f^W$, as defined in (\ref{eq:definitionW'}).\hfill\Halmos
\endproof

Notice that $f^W$ is a budget-balanced distribution rule for the actual welfare it distributes, namely $W'$ as defined in (\ref{eq:definitionW'}). Hence, it is sufficient to prove that for budget-balanced distribution rules $f^W$ that are not generalized weighted Shapley values, there exists a game in $\mathcal{G}(N,f^W,W)$ for which no equilibrium exists.

\vfill
\begin{example}
\label{example:2:po}
\textit{Note that $f^W_1$ through $f^W_4$ are budget-balanced, whereas $f^W_5$, which is the marginal contribution distribution rule $f^W_{MC}$, is not. Let $W'$, shown in Table \ref{table:2a:po}, be the actual welfare distributed by $f^W_5$, as defined in (\ref{eq:definitionW'}). Then $f^W_5$ is a budget-balanced distribution rule for $W'$. In fact, it is the Shapley value distribution rule $f^{W'}_{SV}$.}
\end{example}

\vfill
\begin{table}[b]
\centering
\caption{Tables for Examples 6.x}
\subfloat[][{Definitions of welfare functions and distribution rules\label{table:2a:po}}]{
\adjustbox{width=\columnwidth}{
\tabulinesep=1mm
\begin{tabu}{|c||c|c|c|c|c|c|c|c|c|c|c|c|c|c|c|c|}
\hline
$S\subseteq N$ & $\emptyset$ & $\{i\}$ & $\{j\}$ & $\{k\}$ & $\{\ell\}$ & $\{i,j\}$ & $\{i,k\}$ & $\{i,\ell\}$ & $\{j,k\}$ & $\{j,\ell\}$ & $\{k,\ell\}$ & $\{i,j,k\}$ & $\{i,j,\ell\}$ & $\{i,k,\ell\}$ & $\{j,k,\ell\}$ & $\{i,j,k,\ell\}$\\
\hline
\hline
$W(S)$ & $0$ & $5$ & $3$ & $0$ & $3$ & $6$ & $2$ & $8$ & $0$ & $6$ & $3$ & $0$ & $7$ & $5$ & $3$ & $1$\\
\hline
$f^W_1(\cdot,S)$ & $-$ & $(5)$ & $(3)$ & $(0)$ & $(3)$ & $(3,3)$ & $(1,1)$ & $(4,4)$ & $(0,0)$ & $(3,3)$ & $(\frac{3}{2},\frac{3}{2})$ & $(0,0,0)$ & $(\frac{7}{3},\frac{7}{3},\frac{7}{3})$ & $(\frac{5}{3},\frac{5}{3},\frac{5}{3})$ & $(1,1,1)$ & $(\frac{1}{4},\frac{1}{4},\frac{1}{4},\frac{1}{4})$\\
\hline
$f^W_2(\cdot,S)$ & $-$ & $(5)$ & $(3)$ & $(0)$ & $(3)$ & $(4,2)$ & $(4,-2)$ & $(5,3)$ & $(2,-2)$ & $(3,3)$ & $(0,3)$ & $(3,1,-4)$ & $(3,1,3)$ & $(4,-2,3)$ & $(2,-2,3)$ & $(2,0,-4,3)$\\
\hline
$f^W_3(\cdot,S)$ & $-$ & $(5)$ & $(3)$ & $(0)$ & $(3)$ & $(4,2)$ & $(4,-2)$ & $(5,3)$ & $(2,-2)$ & $(3,3)$ & $(0,3)$ & $(3,1,-4)$ & $(\frac{10}{3},\frac{2}{3},3)$ & $(4,-2,3)$ & $(2,-2,3)$ & $(\frac{7}{3},-\frac{1}{3},-4,3)$\\
\hline
$f^W_4(\cdot,S)$ & $-$ & $(5)$ & $(3)$ & $(0)$ & $(3)$ & $(4,2)$ & $(4,-2)$ & $(5,3)$ & $(1,-1)$ & $(3,3)$ & $(0,3)$ & $(3,0,-3)$ & $(3,1,3)$ & $(4,-2,3)$ & $(1,-1,3)$ & $(2,-1,-3,3)$\\
\hline
$f^W_5(\cdot,S)$ & $-$ & $(5)$ & $(3)$ & $(0)$ & $(3)$ & $(3,1)$ & $(2,-3)$ & $(5,3)$ & $(0,-3)$ & $(3,3)$ & $(0,3)$ & $(0,-2,-6)$ & $(1,-1,1)$ & $(2,-3,3)$ & $(0,-3,3)$ & $(-2,-4,-6,1)$\\
\hline
$W'(S)$ & $0$ & $5$ & $3$ & $0$ & $3$ & $4$ & $-1$ & $8$ & $-3$ & $6$ & $3$ & $-8$ & $1$ & $2$ & $0$ & $-11$\\
\hline
\end{tabu}}}\\
\subfloat[][{Basis representation of $W$\label{table:2b:po}}]{
\adjustbox{width=0.27\columnwidth}{
\tabulinesep=1mm
\begin{tabu}{|c|c|}
\hline
Coalition $T\in\mathcal{T}$ & Contribution $q_T\in Q$\\
\hline
\hline
$\{i\}$        & $5$\\
\hline
$\{j\}$        & $3$\\
\hline
$\{\ell\}$     & $3$\\
\hline
$\{i,j\}$      & $-2$\\
\hline
$\{i,k\}$      & $-3$\\
\hline
$\{j,k\}$      & $-3$\\
\hline
$\{i,j,\ell\}$ & $-2$\\
\hline
\end{tabu}}}
\subfloat[][Notation\label{table:2c:po}]{
\adjustbox{width=0.2\columnwidth}{
\tabulinesep=1mm
\begin{tabu}{|c|c|}
\hline
Symbol & Value\\
\hline
\hline
$\mathcal{T}(\{i,\ell\})$ & $\{\{i\},\{\ell\}\}$\\
\hline
$\mathcal{T}(\{j,k\})$    & $\{\{j\},\{j,k\}\}$\\
\hline
$N(\{i,k\})$              & $\{i,k\}$\\
\hline
$\mathcal{T}_{ij}$        & $\{\{i,j\},\{i,j,\ell\}\}$\\
\hline
$\mathcal{T}_{ij}^{\min}$ & $\{\{i,j\}\}$\\
\hline
\end{tabu}}}
\subfloat[][{Basis distribution rules computed by recursion (\ref{eq:lemma:4:2})\label{table:2d:po}}]{
\adjustbox{width=0.5\columnwidth}{
\tabulinesep=1mm
\begin{tabu}{|c||c|c|c|c|c|}
\hline
Coalition $T\in\mathcal{T}$ & $f^T_1(\cdot,T)$ & $f^T_2(\cdot,T)$ & $f^T_3(\cdot,T)$ & $f^T_4(\cdot,T)$ & $f^T_5(\cdot,T)$\\
\hline
\hline
$\{i\}$        & $(1)$                                   & $(1)$                         & $(1)$                         & $(1)$                         & $(1)$\\
\hline
$\{j\}$        & $(1)$                                   & $(1)$                         & $(1)$                         & $(1)$                         & $(1)$\\
\hline
$\{\ell\}$     & $(1)$                                   & $(1)$                         & $(1)$                         & $(1)$                         & $(1)$\\
\hline
$\{i,j\}$      & $(1,0)$                                 & $(\frac{1}{2},\frac{1}{2})$   & $(\frac{1}{2},\frac{1}{2})$   & $(\frac{1}{2},\frac{1}{2})$   & $(1,1)$\\
\hline
$\{i,k\}$      & $(\frac{4}{3},-\frac{1}{3})$            & $(\frac{1}{3},\frac{2}{3})$   & $(\frac{1}{3},\frac{2}{3})$   & $(\frac{1}{3},\frac{2}{3})$   & $(1,1)$\\
\hline
$\{j,k\}$      & $(1,0)$                                 & $(\frac{1}{3},\frac{2}{3})$   & $(\frac{1}{3},\frac{2}{3})$   & $(\frac{2}{3},\frac{1}{3})$   & $(1,1)$\\
\hline
$\{i,j,\ell\}$ & $(\frac{1}{3},\frac{1}{3},\frac{1}{3})$ & $(\frac{1}{2},\frac{1}{2},0)$ & $(\frac{1}{3},\frac{2}{3},0)$ & $(\frac{1}{2},\frac{1}{2},0)$ & $(1,1,1)$\\
\hline
\end{tabu}}}
\end{table}

\subsection{Three necessary conditions.}
The second step of the proof is to establish that for every subset $S\subseteq N$ of players, any budget-balanced distribution rule $f^W$ must distribute the welfare $W(S)$ only among contributing players, and do so as if the noncontributing players were absent:

\begin{proposition}\label{mainprop:2}
If $f^W$ is a budget-balanced distribution rule that guarantees the existence of an equilibrium in all games $G\in\mathcal{G}(N,f^W,W)$, then,
\begin{equation*}
(\forall\ S\subseteq N)\quad(\forall\ i\in S)\quad f^W(i,S)=f^W(i,N(S))
\end{equation*}
\end{proposition}

\newpage
Section \ref{s.necessaryconditions} is devoted to the proof, which consists of incrementally establishing the following necessary conditions, for any subset $S\subseteq N$:
\begin{enumerate}[label=(\alph*)]
\item If no contributing coalition is formed in $S$, then $f^W$ does not allocate any utility to the players in $S$ (Lemma \ref{lemma:1}).
\item $f^W$ distributes the welfare \textit{only} among the contributing players in $S$ (Lemma \ref{lemma:2}).
\item $f^W$ distributes the welfare among the contributing players in $S$ as if all other players were absent  (Lemma \ref{lemma:3}).
\end{enumerate}

\vspace{0.025in}
\begin{example}
\label{example:3:po}
\textit{For the welfare function $W$, $S=N(S)$ for all subsets $S$, making Proposition \ref{mainprop:2} trivial, except for the two subsets $\{k\}$ and $\{k,\ell\}$ for which $k$ is not a contributing player. Note that $f^W_2$ through $f^W_4$ allocate no welfare to $k$ in these subsets, and $\ell$ gets the same whether $k$ is present or not. But $f^W_1(k,\{k,\ell\})\not=0$ and $f^W_1(\ell,\{k,\ell\})\not=f^W_1(\ell,\{\ell\})$. Therefore, $f^W_1$, which is the equal share distribution rule $f^W_{EQ}$, violates conditions (b)-(c), and hence, Proposition \ref{mainprop:2}. So, it does not guarantee equilibrium existence in all games; see Counterexample 2 in the proof of Lemma \ref{lemma:3}.}
\end{example}

\subsection{Decomposition of the distribution rule.}
The third step of the proof establishes that $f^W$  must have a basis representation of the form (\ref{eq:Tdistribution:2}), where the basis distribution rules are generalized weighted Shapley values:

\begin{proposition}
\label{mainprop:3}
If $f$ is a budget-balanced distribution rule that guarantees the existence of an equilibrium in all games $G\in \mathcal{G}(N,f^W,W)$, then, there exists a sequence of weight systems $\Omega=\left\{\omega^T\right\}_{T\in\mathcal{T}}$ such that
\begin{equation*}
f^W=\DS\sum_{T\in\mathcal{T}}q_T f_{GWSV}^T[\omega^T]
\end{equation*}
\end{proposition}

Note that for now, the weight systems $\omega^T$ could be arbitrary, and need not be related in any way. We deal with how they should be `consistent' in the next section.

In Section \ref{s.decomposition}, we prove Proposition $\ref{mainprop:3}$ by describing a procedure to compute the basis distribution rules, $f^T$, assuming they exist, and then showing the following properties of $f^T$:
\begin{enumerate}[label=(\alph*)]
\item Each $f^T$ is a budget-balanced distribution rule for $W^T$ (Lemmas \ref{lemma:4.0}-\ref{lemma:4.1}).
\item $f^W$, and the basis distribution rules $\left\{f^T\right\}_{T\in\mathcal{T}}$, satisfy (\ref{eq:Tdistribution:2}) (Lemma \ref{lemma:4.2}).
\item Each $f^T$ is nonnegative; so, $f^T=f^T_{GWSV}[\omega^T]$ for some $\omega^T$ (Lemma \ref{lemma:4.3}).
\end{enumerate}

\vspace{0.075in}
\begin{example}
\label{example:4:po}
\textit{Table \ref{table:2d:po} shows the basis distribution rules computed by our recursive procedure in (\ref{eq:lemma:4:2}). Note that $f^T_j, 1\leq j\leq 4$, are budget-balanced distribution rules for $W^T$ (for each $T$, the shares sum up to 1). It can be verified that $f^T_2, f^T_3, f^T_4$ are nonnegative and satisfy (\ref{eq:Tdistribution:2}). Next, observe that from Table \ref{table:2a:po}, $f^W_1(i,\{i,\ell\})=4$, but from Table \ref{table:2d:po}, $\sum_{T\in\mathcal{T}}q_Tf^T_1(i,\{i,\ell\})=q_{\{i\}}f^{\{i\}}_1(i,\{i,\ell\})=5$, so $f^W_1$ violates condition (b). Also, $f^{\{i,k\}}_1(k,\{i,k\})<0$, violating condition (c). So, $f^W_1$, the equal share distribution rule, does not have a basis representation, and hence does not guarantee equilibrium existence in all games; see Counterexamples 3-4 in the proofs of Lemmas \ref{lemma:4.2} and \ref{lemma:4.3}.}
\end{example}

\subsection{Consistency of basis distribution rules.}\label{s.consistency:po}
The fourth part of the proof establishes two important consistency properties that the basis distribution rules $f^T$ must satisfy:
\begin{enumerate}[label=(\alph*)]
\item \textit{Global consistency:} If there is a pair of players $i,j$ common to two coalitions $T,T'$, then their local shares from these two coalitions must satisfy (Lemma \ref{lemma:5}): $$f^{T}(i,T)f^{T'}(j,T')=f^{T'}(i,T')f^{T}(j,T)$$
\item \textit{Cyclic consistency:} If there is a sequence of $z\geq 3$ players, $(i_1,i_2,\ldots,i_z)$ such that for each of the $z$ neighbor-pairs $\left\{\left(i_1,i_2\right),\left(i_2,i_3\right),\ldots,\left(i_z,i_1\right)\right\}$, $\exists\ T_1 \in \mathcal{T}_{i_1i_2}^{\min}, T_2 \in \mathcal{T}_{i_2i_3}^{\min}, \ldots, T_z \in \mathcal{T}_{i_zi_1}^{\min}$ and in each $T_j$, at least one of the neighbors $i_j,i_{j+1}$ gets a nonzero share, then the shares of these $z$ players must satisfy (Lemma \ref{lemma:6}): $$f^{T_1}(i_1,T_1)f^{T_2}(i_2,T_2)\cdots f^{T_z}(i_z,T_z) = f^{T_1}(i_2,T_1)f^{T_2}(i_3,T_2)\cdots f^{T_z}(i_1,T_z)$$
\end{enumerate}

Section \ref{s.consistency} is devoted to the proofs. Since $f^T=f^T_{GWSV}[\omega^T]$ for some $\omega^T$, the above translate into consistency conditions on the sequence of weight systems $\Omega=\left\{\omega^T\right\}_{T\in\mathcal{T}}$ (Corollaries \ref{corollary:5} and \ref{corollary:6} respectively). These conditions are generalizations of those used to prove Proposition \ref{timsresult} in Chen et al. \cite{Chen10} and Marden and Wierman \cite{Marden13b} -- the welfare function used, see (\ref{eq:timsW}), is such that $\mathcal{T}=2^N\backslash\{\emptyset\}$, which is `rich' enough to further simplify the above consistency conditions. In such cases, the distribution rule $f^W$ is fully determined by `pairwise shares' of the form $f^W(i,\{i,j\})$.

\vspace{0.075in}
\begin{example}
\label{example:5:po}
\textit{Among the three budget-balanced distribution rules that have a basis representation, namely $f^W_2$, $f^W_3$ and $f^W_4$, only $f^W_2$ satisfies both consistency conditions. $f^W_3$ fails the global consistency test, since $f_3^{\{i,j\}}(i,\{i,j\})f_3^{\{i,j,\ell\}}(j,\{i,j,\ell\})\not=f_3^{\{i,j,\ell\}}(i,\{i,j,\ell\})f_3^{\{i,j\}}(j,\{i,j\})$, and so, does not guarantee equilibrium existence in all games; see Counterexample 5(a) in the proof of Lemma \ref{lemma:5}. Similarly, $f^W_4$ fails the cyclic consistency test, since $f_4^{\{i,j\}}(i,\{i,j\})f_4^{\{j,k\}}(j,\{j,k\})f_4^{\{i,k\}}(k,\{i,k\})\not=f_4^{\{i,j\}}(j,\{i,j\})f_4^{\{j,k\}}(k,\{j,k\})f_4^{\{i,k\}}(i,\{i,k\})$, and hence does not guarantee equilibrium existence in all games; see Counterexample 6 in the proof of Lemma \ref{lemma:6}.}
\end{example}

\subsection{Existence of a universal weight system.}
The last step of the proof is to show that there exists a universal weight system $\omega^*=\left(\boldsymbol{\lambda}^*,\Sigma^*\right)$ that is equivalent to all the weight systems in $\Omega=\left\{\omega^T\right\}_{T\in\mathcal{T}}$. That is, replacing $\omega^T$ with $\omega^*$ for any coalition $T$ does not change the distribution rule $f_{GWSV}^T[\omega^T]$:

\begin{proposition}
\label{mainprop:4}
If $f^W=\sum_{T\in\mathcal{T}}q_T f^T_{GWSV}[\omega^T]$ is a budget-balanced distribution rule that guarantees the existence of an equilibrium in all games $G\in\mathcal{G}(N,f^W,W)$, then, there exists a weight system $\omega^*$ such that,
\begin{equation*}
(\forall\ T\in \mathcal{T})\quad f^T_{GWSV}[\omega^T]=f^T_{GWSV}[\omega^*]
\end{equation*}
\end{proposition}

In Section \ref{s.universal}, we prove this proposition by explicitly constructing $\omega^*$, given a sequence of weight systems $\Omega=\left\{\omega^T\right\}_{T\in\mathcal{T}}$ that satisfies the consistency Corollaries \ref{corollary:5} and \ref{corollary:6}.

\vspace{0.075in}
\begin{example}
\label{example:6:po}
\textit{The only budget-balanced distribution rule to have survived all the necessary conditions is $f^W_2$. Using the construction in Section \ref{s.universal}, it can be shown that $f^W_2$ is equivalent to the generalized weighted Shapley value distribution rule $f^W_{GWSV}[\omega^*]$, where the weight system $\omega^*=\left(\boldsymbol{\lambda}^*,\Sigma^*\right)$ is given by $\boldsymbol{\lambda}^*=(\frac{1}{2},\frac{1}{2},1,a)$ where $a$ is any strictly positive number, and $\Sigma^*=\{\{i,j,k\},\{\ell\}\}$.}
\end{example}

%
%
%

\begin{APPENDICES}
\section{Proof of Theorem \ref{mainresult}.}\label{appendix:mainresult}

In this appendix, we present the complete proof of Theorem \ref{mainresult}. It is our intent that this section be self-contained and independent of the partial outline presented in Section \ref{s.proofoutline}, and therefore, may contain some redundancies.

First, note that we only need to prove one direction since it is known that for any weight system $\omega$ and any mapping $g_{SV}$, all games in $\mathcal{G}(N,\{f^{g_{SV}(W)}_{GWSV}[\omega]\}_{W\in\mathbb{W}},\mathbb{W})$ have an equilibrium (Hart and Mas-Colell \cite{Hart89}).\footnote{In fact, notice that $\mathbb{W}$ has no role to play as far as equilibrium existence of games $\mathcal{G}(N,\{f^{g_{SV}(W)}_{GWSV}[\omega]\}_{W\in\mathbb{W}},\mathbb{W})$ is concerned, since it does not \textit{directly} affect player utilities. This observation will prove crucial later.} Thus, we present the bulk of the proof -- the other direction -- proving that for distribution rules $f^\mathbb{W}$ that are not generalized weighted Shapley values on some welfare function, there exists a game in $\mathcal{G}(N,f^\mathbb{W},\mathbb{W})$ for which no equilibrium exists.

The general technique of the proof is as follows. First, we present a quick reduction to characterizing only budget-balanced distribution rules $f^\mathbb{W}$ that guarantees equilibrium existence for all games in $\mathcal{G}(N,f^\mathbb{W},\mathbb{W})$. Then, we establish several necessary conditions that these rules must satisfy. Effectively, for each $W\in\mathbb{W}$, these necessary conditions eliminate any budget-balanced distribution rule $f^W$ that is not a generalized weighted Shapley value on $W$, and hence give us our desired result. We establish each of these conditions by a series of counterexamples which amount to choosing a resource set $R$, the local welfare functions $\left\{W_r\right\}_{r\in R}$, and the associated action sets $\left\{\mathcal{A}_i\right\}_{i\in N}$, for which failure to satisfy a necessary condition would lead to nonexistence of an equilibrium.

Most counterexamples involve multiple copies of the same resource. To simplify specifying such counterexamples, we introduce a scaling coefficient $v_r\in\mathbb{Z}_{++}$ for each resource $r\in R$, which denotes the number of copies of $r$, so that we have,
\begin{equation*}
(\forall\ a\in\mathcal{A})\qquad\mathcal{W}(a) = \sum_{r \in R} v_r W_r\left(\{a\}_r\right)\qquad\text{and}\qquad U_i(a) = \sum_{r\in a_i}v_r f^r\left(i,\{a\}_r\right).
\end{equation*}
Therefore, to exhibit a counterexample, in addition to choosing $R$, $\left\{W_r\right\}_{r\in R}$ and $\left\{\mathcal{A}_i\right\}_{i\in N}$, we also choose $\left\{v_r\right\}_{r\in R}$.

Throughout, we work with the basis-representation of the welfare function that was introduced in Section \ref{s.basisW}. For each $W\in\mathbb{W}$, it is useful to think of the sets in $\mathcal{T}^W$ as being `coalitions' of players that contribute to the welfare function $W$ (also referred to as contributing coalitions), and the corresponding coefficients in $Q^W$ as being their respective contributions. Also, for simplicity, we normalize $W$ by setting $W(\emptyset)=0$ and therefore, $\emptyset\notin\mathcal{T}^W$. Before proceeding, we introduce some notation below, which is also summarized in Table \ref{table:1} for easy reference.

\subsection*{Notation.} For any subset $S\subseteq N$, let $\mathcal{T}^W(S)$ denote the set of contributing coalitions in $S$:

\vspace{-0.1in}
\begin{equation*}
\mathcal{T}^W(S)=\left\{T\in \mathcal{T}^W\ |\ T\subseteq S\right\}
\end{equation*}
Using this notation, and the definition of inclusion functions from (\ref{eq:Twelfare:2}) in (\ref{eq:genTwelfare:2}), we have an alternate way of writing $W$, namely,

\vspace{-0.1in}
\begin{equation}
\label{eq:alternateW}
W(S) = \sum_{T\in\mathcal{T}^W(S)}q^W_T
\end{equation}

For any subset $S\subseteq N$, let $N^W(S)$ denote the set of contributing players in $S$:

\vspace{-0.1in}
\begin{equation*}
N^W(S)=\bigcup\mathcal{T}^W(S)
\end{equation*}
Using this notation, and the alternate definition of $W$ from (\ref{eq:alternateW}), we have,

\vspace{-0.1in}
\begin{equation}
\label{eq:propertyW}
W(S)=W(N^W(S))
\end{equation}

For any two players $i,j\in N$, let $\mathcal{T}^W_{ij}$ denote the set of all coalitions containing $i$ and $j$:

\vspace{-0.1in}
\begin{equation}
\label{eq:notation:3}
\mathcal{T}^W_{ij}=\left\{T\in \mathcal{T}^W\ |\ \{i,j\}\subseteq T\right\}
\end{equation}

Let $\mathcal{B}\subseteq 2^{N}$ denote a collection of subsets of $N$. The relation $\subseteq$ induces a partial order on $\mathcal{B}$. Let $\mathcal{B}^{\max}$ and $\mathcal{B}^{\min}$ denote the set of maximal and minimal elements of the poset $\left(\mathcal{B},\subseteq\right)$ respectively:

\vspace{-0.1in}
\begin{equation*}
\begin{split}
\mathcal{B}^{\max}&=\left\{B\in \mathcal{B}\ |\ \left(\nexists B'\in \mathcal{B}\right)\ s.t.\ B\subsetneq B'\right\}\mcr
\mathcal{B}^{\min}&=\left\{B\in \mathcal{B}\ |\ \left(\nexists B'\in \mathcal{B}\right)\ s.t.\ B'\subsetneq B\right\}
\end{split}
\end{equation*}

\begin{table}[t]
\caption{Summary of notation\label{table:1}}
\centering
\begin{footnotesize}
{\tabulinesep=1mm
\begin{tabu}{|c|c|c|}
\hline
Symbol & Definition & Meaning\\
\hline
$\mathcal{T}^W(S)$     & $\left\{T\in \mathcal{T}^W\ |\ T\subseteq S\right\}$ & set of contributing coalitions in $S$\\
$N^W(S)$               & $\bigcup\mathcal{T}^W(S)$ & set of contributing players in $S$\\
$\mathcal{T}^W_{ij}$   & $\left\{T\in \mathcal{T}^W\ |\ \{i,j\}\subseteq T\right\}$ & set of coalitions containing both players $i$ and $j$\\
$\mathcal{B}^{\max}$   & $\left\{B\in \mathcal{B}\ |\ \left(\nexists B'\in \mathcal{B}\right)\ s.t.\ B\subsetneq B'\right\}$ & set of maximal elements of the poset $\left(\mathcal{B},\subseteq\right)$\\
$\mathcal{B}^{\min}$   & $\left\{B\in \mathcal{B}\ |\ \left(\nexists B'\in \mathcal{B}\right)\ s.t.\ B'\subsetneq B\right\}$ & set of minimal elements of the poset $\left(\mathcal{B},\subseteq\right)$\\
\hline
\end{tabu}}
\end{footnotesize}
\end{table}

\begin{example}
\label{example:1}
\textit{Let $N = \{i,j,k\}$ be the set of players, and $W:2^N\rightarrow\mathbb{R}$ as defined in Table \ref{table:2a}. Table \ref{table:2b} shows the basis representation of $W$, and Table \ref{table:2c} illustrates our notation for this $W$.}
\end{example}

\begin{table}[t]
\caption{Tables for Example \ref{example:1}}
\centering
\begin{footnotesize}
\subfloat[][Definition of $W$\label{table:2a}]{
\adjustbox{width=0.16\columnwidth}{
{\tabulinesep=1mm
\begin{tabu}{|c|c|}
\hline
$S$ & $W(S)$\\
\hline
$\emptyset$ & $0$\\
$\{i\}$     & $1$\\
$\{j\}$     & $2$\\
$\{k\}$     & $3$\\
$\{i,j\}$   & $3$\\
$\{j,k\}$   & $3$\\
$\{i,k\}$   & $3$\\
$\{i,j,k\}$ & $4$\\
\hline
\end{tabu}}}}
\quad
\subfloat[][Basis representation of $W$\label{table:2b}]{
{\tabulinesep=1mm
\begin{tabu}{|c|c|}
\hline
Coalition $T\in\mathcal{T}^W$ & Contribution $q^W_T\in Q^W$\\
\hline
$\{i\}$     & $1$\\
$\{j\}$     & $2$\\
$\{k\}$     & $3$\\
$\{j,k\}$   & $-2$\\
$\{i,k\}$   & $-1$\\
$\{i,j,k\}$ & $1$\\
\hline
\end{tabu}}}
\quad
\subfloat[][Illustration of notation\label{table:2c}]{
{\tabulinesep=1mm
\begin{tabu}{|c|c|}
\hline
Symbol & Value\\
\hline
$\mathcal{T}^W(\{i,j\})$                 & $\{\{i\},\{j\}\}$\\
$\mathcal{T}^W(\{j,k\})$                 & $\{\{j\},\{k\},\{j,k\}\}$\\
$N^W(\{i,k\})$                           & $\{i,k\}$\\
$\mathcal{T}^W_{ij}$                     & $\{\{i,j,k\}\}$\\
$\left(\mathcal{T}^W_{jk}\right)^{\max}$ & $\{\{i,j,k\}\}$\\
$\left(\mathcal{T}^W_{ik}\right)^{\min}$ & $\{\{i\},\{k\}\}$\\
\hline
\end{tabu}}}
\end{footnotesize}
\vspace{-0.25in}
\end{table}

\subsection*{Proof outline.} The proof is divided into five sections -- each section incrementally builds on the structure imposed on the distribution rule $f$ by previous sections:
\begin{enumerate}[label=\arabic*.]
\item \textit{Reduction to budget-balanced distribution rules.} We reduce the problem of characterizing all distribution rules $f^\mathbb{W}$ that guarantee equilibrium existence for all $G\in\mathcal{G}(N,f^\mathbb{W},\mathbb{W})$ to characterizing only \textit{budget-balanced} distribution rules $f^\mathbb{W}$ that guarantee equilibrium existence for all $G\in\mathcal{G}(N,f^\mathbb{W},\mathbb{W})$.
\item \textit{Three necessary conditions.} We establish three necessary conditions that collectively describe, for any $W\in\mathbb{W}$, for any subset $S$ of players, which players get shares of $W(S)$ and how these shares are affected by the presence of the other players.
\item \textit{Decomposition of the distribution rule.} We use these conditions to show that for each $W\in\mathbb{W}$, $f^W$ must be representable as a linear combination of \textit{generalized weighted Shapley value distribution rules} (with possibly different weight systems) on the unanimity games corresponding to the coalitions in $\mathcal{T}^W$, with corresponding coefficients from $Q^W$.
\item \textit{Consistency of basis distribution rules.} We establish two important consistency properties (one global and one cyclic) that these `basis' distribution rules should satisfy, and restate these properties in terms of their corresponding weight systems.
\item \textit{Existence of a universal weight system.} We use the two consistency conditions on the weight systems of the basis distribution rules to show that there exists a single universal weight system that can replace the weight systems of all the basis distribution rules without changing the resulting shares of any welfare function. This establishes, for each $W\in\mathbb{W}$, the equivalence of $f^W$ to a generalized weighted Shapley value on $W$ with this universal weight system.
\end{enumerate}

\vspace{-0.05in}
\subsection{Reduction to budget-balanced distribution rules.}
First, we reduce the problem of characterizing \textit{all} distribution rules $f^\mathbb{W}$ that guarantee equilibrium existence for all $G\in\mathcal{G}(N,f^\mathbb{W},\mathbb{W})$ to characterizing only \textit{budget-balanced} distribution rules $f^\mathbb{W}$ that guarantee equilibrium existence for all $G\in\mathcal{G}(N,f^\mathbb{W},\mathbb{W})$:

\begin{proposition}\label{prop:1}
Given any set of local welfare functions $\mathbb{W}$, their corresponding local distribution rules $f^\mathbb{W}$ guarantee the existence of an equilibrium for all games in $\mathcal{G}(N,f^\mathbb{W},\mathbb{W})$ if and only if they guarantee the existence of an equilibrium for all games in $\mathcal{G}(N,f^\mathbb{W},g_{SV}(\mathbb{W}))$.
\end{proposition}
\proof{Proof.} This proposition is actually a subtlety of our notation. For games $G\in\mathcal{G}(N,f^\mathbb{W},\mathbb{W})$, the welfare functions $\mathbb{W}$ do not directly affect strategic behavior (they only do so through the distribution rules $f^\mathbb{W}$). Therefore, in terms of strategic behavior and equilibrium existence, the classes $\mathcal{G}(N,f^\mathbb{W},\mathbb{W})$ and $\mathcal{G}(N,f^\mathbb{W},\mathbb{W}')$ are identical, for any two sets of welfare functions $\mathbb{W},\mathbb{W}'$. Therefore, a distribution rule $f^\mathbb{W}$ guarantees equilibrium existence for all games in $\mathcal{G}(N,f^\mathbb{W},\mathbb{W})$ if and only if it guarantees equilibrium existence for all games in $\mathcal{G}(N,f^\mathbb{W},\mathbb{W}')$. To complete the proof, simply pick $\mathbb{W}'=g_{SV}(\mathbb{W})$, the actual welfares distributed by $f^\mathbb{W}$, as defined in (\ref{eq:definitionW'}).\hfill\Halmos
\endproof

Notice that $f^\mathbb{W}$ are budget-balanced distribution rules for the actual welfares they distribute, namely $g_{SV}(\mathbb{W})$ as defined in (\ref{eq:definitionW'}). Hence, it is sufficient to prove that for budget-balanced distribution rules $f^\mathbb{W}$ that are not generalized weighted Shapley values, there exists a game in $\mathcal{G}(N,f^\mathbb{W},\mathbb{W})$ for which no equilibrium exists.

\subsection{Constraints on individual distribution rules.}
In the next two sections, we establish common constraints that each budget-balanced distribution rule $f^W\in f^\mathbb{W}$ must satisfy, in order to guarantee equilibrium existence for all games in $\mathcal{G}(N,f^\mathbb{W},\mathbb{W})$ for any given set of local welfare functions $\mathbb{W}$. To do this, we deal with one welfare function at a time -- for each $W\in\mathbb{W}$, we only focus on the corresponding distribution rule $f^W$ guaranteeing equilibrium existence for all games in the class $\mathcal{G}(N,f^W,W)$. Note that this is justified by the fact that $\mathcal{G}(N,f^W,W)\subseteq\mathcal{G}(N,f^\mathbb{W},\mathbb{W})$ for all $W\in\mathbb{W}$, and so, if $f^\mathbb{W}$ guarantees equilibrium existence for all games in $\mathcal{G}(N,f^\mathbb{W},\mathbb{W})$, then each $f^W\in f^\mathbb{W}$ must guarantee equilibrium existence for all games in $\mathcal{G}(N,f^W,W)$.

Since we are dealing with only one welfare function at a time, we drop the superscripts from $f^W$, $\mathcal{T}^W$, $Q^W$, $q^W_T$, etc. in order to simplify notation.

\subsubsection{Three necessary conditions.}\label{s.necessaryconditions}
Our goal in this section is to establish that, for every subset $S\subseteq N$ of players, any budget-balanced distribution rule $f$ must distribute the welfare $W(S)$ only among contributing players, and do so as if the noncontributing players were absent:

\begin{proposition}\label{prop:2}
If $f$ is a budget-balanced distribution rule that guarantees the existence of an equilibrium in all games $G\in\mathcal{G}(N,f,W)$, then,

\begin{equation*}
(\forall\ S\subseteq N)\quad(\forall\ i\in S)\quad f(i,S)=f(i,N(S))
\end{equation*}
\end{proposition}

We prove this proposition in incremental stages, by establishing the following necessary conditions, for any subset $S\subseteq N$:
\begin{enumerate}[label=(\alph*)]
\item If no contributing coalition is formed in $S$, then $f$ does not allocate any utility to the players in $S$. Formally, in Lemma \ref{lemma:1}, we show that if $\mathcal{T}(S)=\emptyset$, then for all players $i\in S$, $f(i,S)=0$.
\item $f$ distributes the welfare \textit{only} among the contributing players in $S$. Formally, in Lemma \ref{lemma:2}, we generalize Lemma \ref{lemma:1} by showing that for all players $i\notin N(S)$, $f(i,S)=0$.
\item $f$ distributes the welfare among the contributing players in $S$ as if all other players were absent. Formally, in Lemma \ref{lemma:3}, we show that for all players $i\in N(S)$, $f(i,S)=f(i,N(S))$.
\end{enumerate}

\begin{lemma}\label{lemma:1}
If $f$ is a budget-balanced distribution rule that guarantees the existence of an equilibrium in all games $G\in\mathcal{G}(N,f,W)$, then,
\begin{equation}
\label{eq:lemma1}
(\forall\ S\subseteq N\ s.t.\ \mathcal{T}(S)=\emptyset)\quad(\forall\ i\in S)\quad f(i,S)=0
\end{equation}
\end{lemma}

\proof{Proof.} The proof is by induction on $|S|$. The base case, where $|S|=1$ is immediate, because from budget-balance, we have that for any player $i\in N$,
\begin{equation*}
f(i,\{i\}) =
\begin{cases}
q_{\{i\}} & ,\ \{i\}\in\mathcal{T}\\
0 & ,\ \text{otherwise}
\end{cases}
\end{equation*}
Our induction hypothesis is that (\ref{eq:lemma1}) holds for all subsets $S$ of size $z$, for some $0<z<|N|$. Assuming that this is true, we show that (\ref{eq:lemma1}) holds for all subsets $S$ of size $z+1$. The proof is by contradiction, and proceeds as follows.

Assume to the contrary, that $f(i,S)\not=0$ for some $i\in S$, for some $S\subseteq N$, where $\mathcal{T}(S)=\emptyset$ and $|S|=z+1$. Since $f$ is budget-balanced, and $z+1\geq 2$, it follows that there is some $j\in S-\{i\}$ with $f(j,S)\not=0$, such that $f(i,S)\cdot f(j,S)<0$, i.e., $f(i,S)$ and $f(j,S)$ have opposite signs. Without loss of generality, assume that $f(i,S)<0$ and $f(j,S)>0$.

\begin{figure}[ht]
    \centering
    \subfloat[][The game \label{fig:2a}]{
        \includegraphics*[height=4.5cm,width=0.275\textwidth]{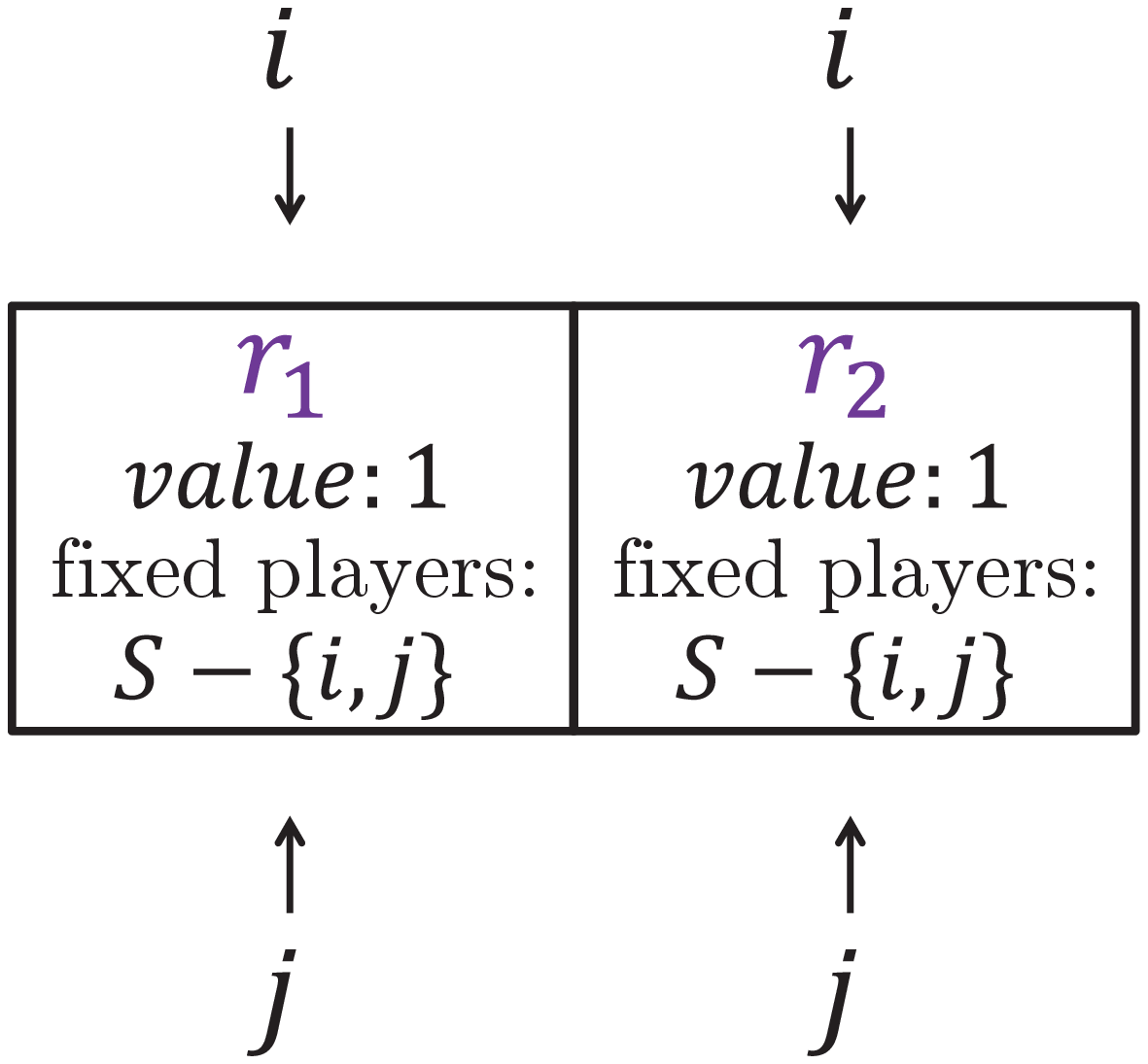}}
    \qquad
    \subfloat[][The payoff matrix\label{fig:2b}]{
        \includegraphics*[height=4.5cm,width=0.575\textwidth]{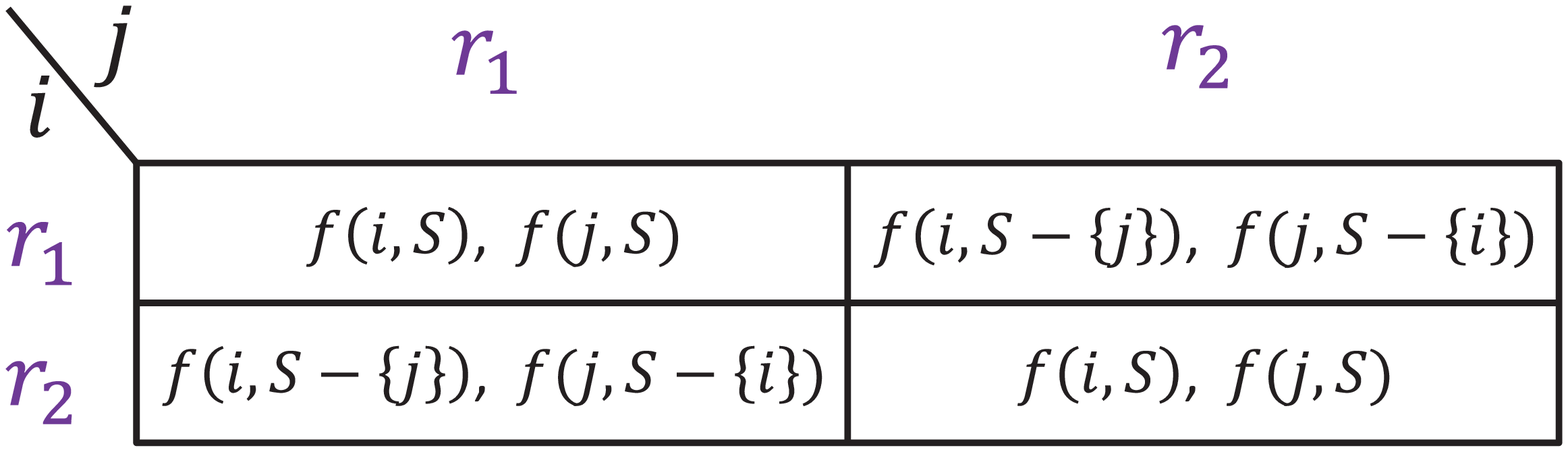}}
    \caption{Counterexample 1\label{fig:2}}
\vspace{-0.15in}
\end{figure}

\textit{Counterexample 1:} Consider the game in Figure \ref{fig:2a}, with resource set $R=\left\{r_1,r_2\right\}$ and local resource coefficients $v_{r_1}=v_{r_2}=1$. Players $i$ and $j$ have the same action sets -- they can each choose either $r_1$ or $r_2$. All other players in $S$ have a fixed action -- they choose both resources. Formally,

\vspace{-0.075in}
\begin{equation*}
\mathcal{A}_k = \begin{cases}
\{\{r_1\},\{r_2\}\} & ,\ k\in\{i,j\}\mcr
\{\{r_1,r_2\}\} & ,\ k\in S-\{i,j\}
\end{cases}
\end{equation*}
This is essentially a game between $i$ and $j$, with the payoff matrix in Figure \ref{fig:2b}.

Since $\mathcal{T}(S)=\emptyset$, it follows that $\mathcal{T}(S')=\emptyset$ for all $S'\subseteq S$. Therefore, by letting $S'=S-\{i\}$, we can apply the induction hypothesis to $S'$ to obtain $f(j,S-\{i\})=0$. Similarly, by letting $S'=S-\{j\}$, we get $f(i,S-\{j\})=0$. We now use this to show that none of the four outcomes of Counterexample 1 is an equilibrium -- this contradicts the fact that $f$ guarantees the existence of an equilibrium in all games $G\in\mathcal{G}(N,f,W)$. First, consider the outcome $\left(\{r_1\},\{r_1\}\right)$. Given that player $j$ is in $r_1$, player $i$ obtains a payoff of $f(i,S)$ in $r_1$, which, by our assumption in step (ii), is negative. By deviating to $r_2$, player $i$ would obtain a payoff of $f(i,S-\{j\})=0$, which is strictly better for player $i$. Hence, $\left(\{r_1\},\{r_1\}\right)$ is not an equilibrium. By nearly identical arguments, it can be shown that the other three outcomes are also not equilibria. This completes the inductive argument.\hfill\Halmos
\endproof

\begin{lemma}\label{lemma:2}
If $f$ is a budget-balanced distribution rule that guarantees the existence of an equilibrium in all games $G\in\mathcal{G}(N,f,W)$, then,

\vspace{-0.125in}
\begin{equation}
\label{eq:lemma2}
(\forall\ S\subseteq N)\quad(\forall\ i\in S-N(S))\quad f(i,S)=0
\end{equation}
\end{lemma}

\proof{Proof.} For $0\leq p\leq |\mathcal{T}|$ and $0\leq q\leq n$, let $\mathcal{P}_p^q$ denote the collection of all nonempty subsets $S$ for which $|\mathcal{T}(S)|=p$ and $|S-N(S)|=q$, i.e., $S$ has exactly $p$ contributing coalitions and $q$ noncontributing players in it. Then, $\mathbb{P}=\{\mathcal{P}_0^0, \mathcal{P}_0^1, \ldots, \mathcal{P}_0^n\;,\; \mathcal{P}_1^0, \mathcal{P}_1^1, \ldots, \mathcal{P}_1^n\;,\; \ldots\;,\; \mathcal{P}_{|\mathcal{T}|}^0, \mathcal{P}_{|\mathcal{T}|}^1, \ldots, \mathcal{P}_{|\mathcal{T}|}^n\}$ is an ordered partition of all nonempty subsets of $N$. Note that we have slightly abused the usage of the term `partition', since it is possible that $\mathcal{P}_p^q=\emptyset$ for some $p,q$.

We prove the lemma by induction on $\mathbb{P}=\left\{\{\mathcal{P}_p^q\}\right\}$, i.e., the tuple $(p,q)$. Our base cases are twofold:
\begin{enumerate}[label=(\roman*)]
\item When $p=0$, i.e., for any subset $S\in\bigcup_{q=0}^n\mathcal{P}_0^q$, $\mathcal{T}(S)=\emptyset$. So, (\ref{eq:lemma2}) is true from Lemma \ref{lemma:1}.
\item When $q=0$, i.e., for any subset $S\in\bigcup_{p=0}^n\mathcal{P}_p^0$, $S=N(S)$. So, (\ref{eq:lemma2}) is vacuously true.
\end{enumerate}
Our induction hypothesis is the following statement:

\vspace{-0.175in}
\begin{equation*}
\text{(\ref{eq:lemma2}) holds for all }S\in\displaystyle\bigcup_{p=0}^{z}\displaystyle\bigcup_{q=0}^{n}\mathcal{P}_p^q\displaystyle\bigcup_{q=0}^{y}\mathcal{P}_{z+1}^q,\text{ for some }0\leq z < |\mathcal{T}|,\text{ and for some }0\leq y < n.
\end{equation*}

\vspace{-0.05in}
\noindent Assuming that this is true, we prove that (\ref{eq:lemma2}) holds for all $S\in\mathcal{P}_{z+1}^{y+1}$. In other words, assuming that for all subsets $S\in\bigcup\{\mathcal{P}_0^0, \mathcal{P}_0^1, \ldots, \mathcal{P}_0^n\;,\; \ldots\;,\; \mathcal{P}_z^0, \mathcal{P}_z^1, \ldots, \mathcal{P}_z^n\;,\; \ldots\;,\; \mathcal{P}_{z+1}^0, \mathcal{P}_{z+1}^1, \ldots, \mathcal{P}_{z+1}^y\}$ we have already proved the lemma, we focus on proving the lemma for $S\in\mathcal{P}_{z+1}^{y+1}$, the next collection in $\mathbb{P}$. The proof is by contradiction, and proceeds as follows.

\newpage

Assume to the contrary, that $f(i,S)\not=0$ for some $i\in S-N(S)$, for some $S\in\mathcal{P}_{z+1}^{y+1}$. Since $z+1\geq 1$ and $y+1\geq 1$, it must be that $|S|\geq 2$, i.e., $S$ has at least two players. Also, because $i\notin N(S)$, it follows that $N(S)=N(S-\{i\})$, and so, from (\ref{eq:propertyW}), we have, $W(S)=W(S-\{i\})$. Since $f$ is budget-balanced, and $W(S)=W(S-\{i\})$, we can express $f(i,S)$ as,

\vfill
\begin{equation*}
f(i,S)=\displaystyle\sum_{k\in S-\{i\}}\left(f(k,S-\{i\})-f(k,S)\right)
\end{equation*}

\vfill
\noindent Because $f(i,S)\not=0$, it is clear that at least one of the difference terms on the right hand side is nonzero and has the same sign as $f(i,S)$. That is, there is some $j\in S-\{i\}$ such that

\vfill
\begin{equation}
\label{eq:lemma:2:1}
f(i,S)\left(f(j,S-\{i\})-f(j,S)\right)>0
\end{equation}

\vfill
\noindent Also, $f(i,S-\{j\})=0$. To see this, we consider the following two cases, where, for ease of expression, we let $S'=S-\{j\}$.

\vfill
\begin{enumerate}[label=(\roman*)]
\item If $j\in N(S)$, then $|\mathcal{T}(S')|<|\mathcal{T}(S)|=z+1$, and so $S'\in\bigcup_{p=0}^{z}\bigcup_{q=0}^{n}\mathcal{P}_p^q$.

\vfill
\item If $j\notin N(S)$, then $|\mathcal{T}(S')|=|\mathcal{T}(S)|=z+1$, and $|S'-N(S')|<|S-N(S)|=y+1$ and so $S'\in\bigcup_{q=0}^{y}\mathcal{P}_{z+1}^q$.
\end{enumerate}

\vfill
\noindent In either case, we can apply the induction hypothesis to $S-\{j\}$ to conclude that $f(i,S-\{j\})=0$, since $i\notin N(S-\{j\})$. Therefore, (\ref{eq:lemma:2:1}) can be rewritten as,

\vfill
\begin{equation*}
\left(f(i,S)-f(i,S-\{j\})\right)\left(f(j,S)-f(j,S-\{i\})\right)<0
\end{equation*}

\vfill
\noindent To complete the proof, let us first consider the case where $f(i,S)-f(i,S-\{j\})<0$ and $f(j,S)-f(j,S-\{i\})>0$. For this case, \textit{Counterexample 1} illustrated in Figure \ref{fig:2}, along with the arguments for non-existence of equilibrium therein (the proof of Lemma \ref{lemma:1}), serves as a counterexample here too. The proof for when $f(i,S)-f(i,S-\{j\})>0$ and $f(j,S)-f(j,S-\{i\})<0$ is symmetric.\hfill\Halmos
\endproof

\vfill
\begin{lemma}\label{lemma:3}
If $f$ is a budget-balanced distribution rule that guarantees the existence of an equilibrium in all games $G\in\mathcal{G}(N,f,W)$, then,

\vfill
\begin{equation}
\label{eq:lemma3}
(\forall\ S\subseteq N)\quad(\forall\ i\in N(S))\quad f(i,S)=f(i,N(S))
\end{equation}
\end{lemma}

\vfill
\proof{Proof.} Since this is a tautology when $N(S)=S$, let us assume that $N(S)\subsetneq S$. We consider two cases below.

\vfill

\noindent\textbf{Case 1:} $|N(S)|=1$. Without loss of generality, let $N(S)=\{i\}$. Since $f$ is budget-balanced, and $W(S)=W(N(S))$, we can express $f(i,N(S))$ as,

\vfill
\begin{equation*}
f(i,N(S)) = \displaystyle\sum_{k\in S}f(k,S)
\end{equation*}

\vfill
\noindent From Lemma \ref{lemma:2}, we know that $f(k,S)=0$ for all $k\in S-N(S)$. Accordingly, $f(i,S)=f(i,N(S))$.

\vfill

\noindent\textbf{Case 2:} $|N(S)|\not=1$. For $0\leq p\leq |\mathcal{T}|$, let $\mathcal{P}_p$ denote the collection of all nonempty subsets $S$ such that $|N(S)|\not=1$ and $N(S)\subsetneq S$, for which $|\mathcal{T}(S)|=p$, i.e., $S$ has exactly $p$ contributing coalitions in it. Then, $\mathbb{P}=\{\mathcal{P}_0, \mathcal{P}_1, \ldots, \mathcal{P}_{|\mathcal{T}|}\}$ is an ordered partition of all nonempty subsets $S$ such that $|N(S)|\not=1$ and $N(S)\subsetneq S$. Note that we have slightly abused the usage of the term `partition', since it is possible that $\mathcal{P}_p=\emptyset$ for some $p$.

\newpage

We prove the lemma by induction on $\mathbb{P}$. The base case, where $S\in\mathcal{P}_0$, is vacuously true, since $N(S)=\emptyset$. Our induction hypothesis is that (\ref{eq:lemma3}) holds for all subsets $S\in\bigcup_{p=0}^z\mathcal{P}_p$, for some $0\leq z<|\mathcal{T}|$. Assuming that this is true, we show that (\ref{eq:lemma3}) holds for all subsets $S\in\mathcal{P}_{z+1}$.

\vfill
Before proceeding with the proof, we point out the following observation. Since $f$ is budget-balanced, and $W(S)=W(N(S))$, we have,

\vfill
\begin{equation}
\label{eq:lemma:3:0.5}
\sum_{k\in N(S)}f(k,N(S)) = \sum_{k\in S}f(k,S) = \sum_{k\in N(S)}f(k,S)
\end{equation}

\vfill
\noindent where the second equality comes from Lemma \ref{lemma:2}, which gives us $f(k,S)=0$ for all $k\in S-N(S)$.

\vfill
The proof is by contradiction, and proceeds as follows. Assume to the contrary, that $f(k,S)\not=f(k,N(S))$ for some $k\in N(S)$, for some $S\in\mathcal{P}_{z+1}$. Since $z+1\geq 1$, and $|N(S)|\not=1$, $|N(S)|\geq 2$. Then, from (\ref{eq:lemma:3:0.5}), we can pick $i,j\in N(S)$ such that,

\vfill
\begin{equation}
\label{eq:lemma:3:1}
f(i,S)<f(i,N(S))
\end{equation}

\vfill
\begin{equation}
\label{eq:lemma:3:2}
f(j,S)>f(j,N(S))
\end{equation}

\begin{figure}[ht]
    \centering
    \subfloat[][The game\label{fig:3a}]{
        \includegraphics*[height=5cm,width=0.37\textwidth]{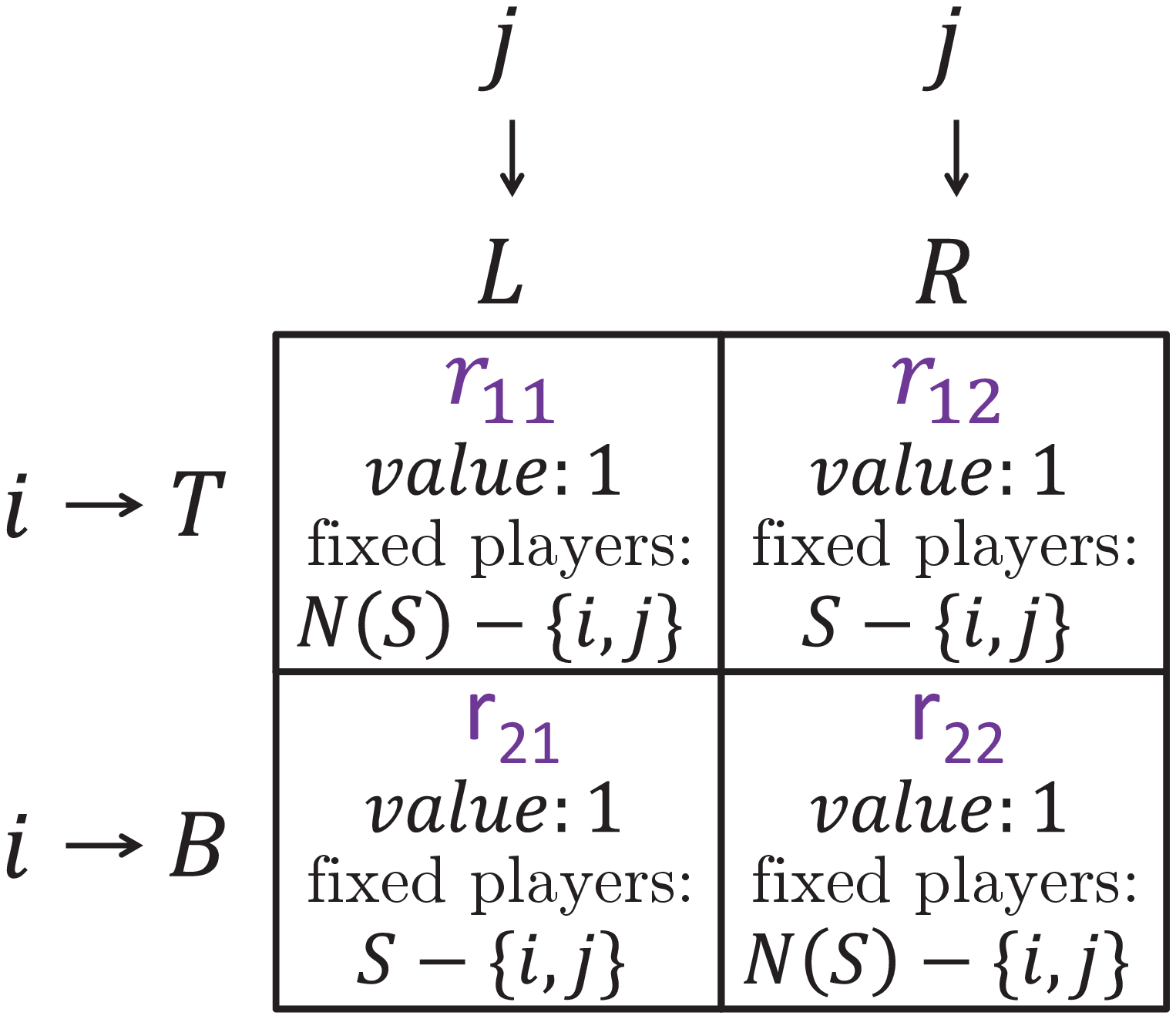}}
    \qquad
    \subfloat[][The payoff matrix\label{fig:3b}]{
        \includegraphics*[height=5cm,width=0.45\textwidth]{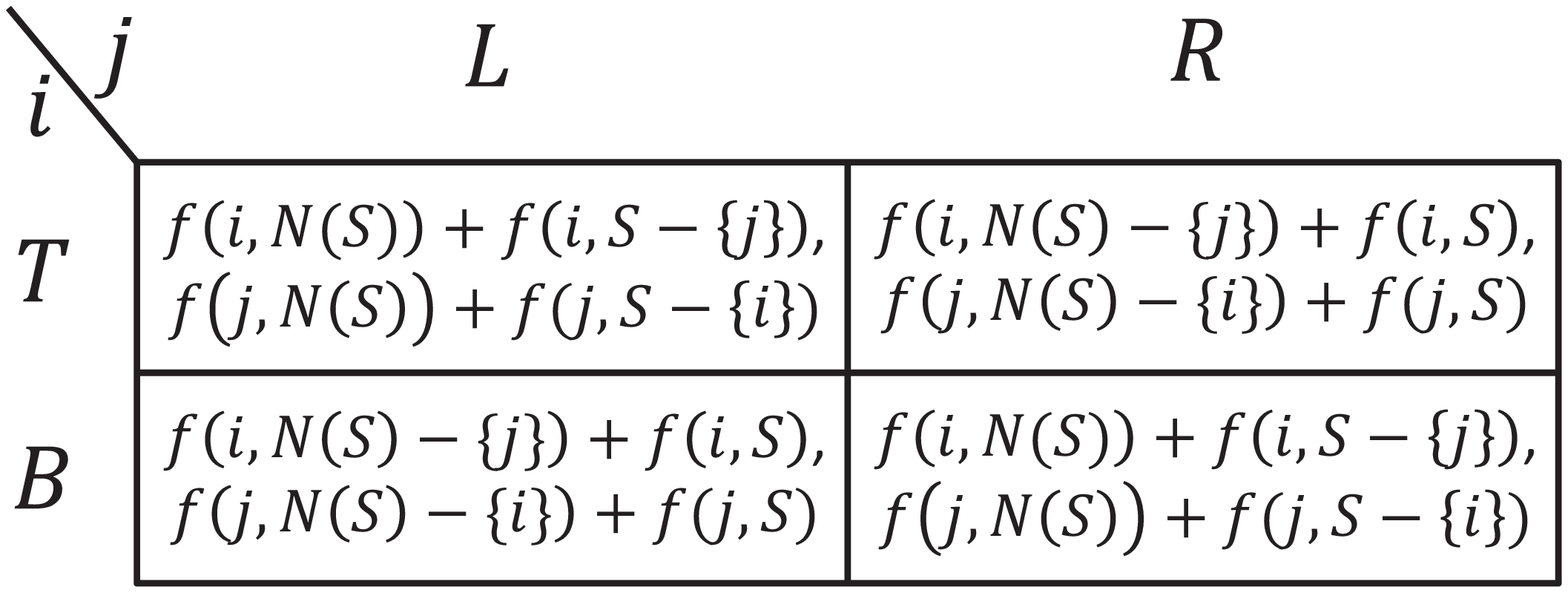}}
    \caption{Counterexample 2\label{fig:3}}
\end{figure}

\textit{Counterexample 2:} Consider the game in Figure \ref{fig:3a}, with resource set $R=\left\{r_{11},r_{12},r_{21},r_{22}\right\}$ and local resource coefficients $v_{r_{11}}=v_{r_{12}}=v_{r_{21}}=v_{r_{22}}=1$. Player $i$ is the row player and player $j$ is the column player. All other players in $N(S)$ have a fixed action -- they choose all four resources. And all players in $S-N(S)$ also have a fixed action -- they choose resources $r_{12}$ and $r_{21}$. Formally,

\vfill
\begin{equation*}
\mathcal{A}_k = \begin{cases}
\{T=\{r_{11},r_{12}\},B=\{r_{21},r_{22}\}\} & ,\ k=i\mcr
\{L=\{r_{11},r_{21}\},R=\{r_{12},r_{22}\}\} & ,\ k=j\mcr
\{\{r_{11},r_{12},r_{21},r_{22}\}\} & ,\ k\in N(S)-\{i,j\}\mcr
\{\{r_{12},r_{21}\}\} & ,\ k\in S-N(S)
\end{cases}
\end{equation*}

\vfill
\noindent This is essentially a game between players $i$ and $j$, with the payoff matrix in Figure \ref{fig:3b}. The set of joint action profiles can therefore be represented as $\mathcal{A}=\left\{TL,TR,BL,BR\right\}$.

\newpage

Because $i\in N(S)$, $|\mathcal{T}(S-\{i\})| = |\mathcal{T}(N(S)-\{i\})| < |\mathcal{T}(S)|$. Also note that $N(S-\{i\})=N(N(S)-\{i\})$. Now, consider two cases:
\begin{enumerate}[label=(\roman*)]
\item If $j\notin N(S-\{i\})$, then $j\notin N(N(S)-\{i\})$, and so, from Lemma \ref{lemma:2}, $f(j,S-\{i\})=f(j,N(S)-\{i\})=0$.
\item If $j\in N(S-\{i\})$, then $j\in N(N(S)-\{i\})$. If $N(S-\{i\})=\{j\}$, then, applying our analysis in Case 1 to $S-\{i\}$ and $N(S)-\{i\}$, we have,
    \begin{equation}
    \label{eq:lemma:3:2.5}
    \begin{split}
    f(j,S-\{i\})&=f(j,N(S-\{i\}))\mcr
    f(j,N(S)-\{i\})&=f(j,N(N(S)-\{i\}))
    \end{split}
    \end{equation}
    If $N(S-\{i\})\not=\{j\}$, then we know that $|N(S-\{i\})|\geq 2$. Accordingly, we can apply our induction hypothesis to $S-\{i\}$ and $N(S)-\{i\}$ to obtain (\ref{eq:lemma:3:2.5}).
\end{enumerate}
In either case, we have,
\begin{equation}
\label{eq:lemma:3:3}
f(j,S-\{i\}) = f(j,N(S)-\{i\})
\end{equation}
By similar arguments, we obtain,
\begin{equation}
\label{eq:lemma:3:4}
f(i,S-\{j\})=f(i,N(S)-\{j\})
\end{equation}

We use the four properties in (\ref{eq:lemma:3:1}), (\ref{eq:lemma:3:2}), (\ref{eq:lemma:3:3}), and (\ref{eq:lemma:3:4}) to show that Counterexample 2 does not possess an equilibrium, thereby contradicting the fact that $f$ guarantees the existence of an equilibrium in all games $G\in\mathcal{G}(N,f,W)$. We show this for each outcome:
\begin{enumerate}[label=(\roman*)]
\item $TL$ is not an equilibrium, since player $j$ has an incentive to deviate from $L$ to $R$:
    \begin{equation*}
    f(j,N(S)-\{i\})+f(j,S)>f(j,S-\{i\})+f(j,N(S))
    \end{equation*}
    This results from combining (\ref{eq:lemma:3:2}) and (\ref{eq:lemma:3:3}).
\item $TR$ is not an equilibrium, since player $i$ has an incentive to deviate from $T$ to $B$:
    \begin{equation*}
    f(i,S-\{j\})+f(i,N(S))>f(i,N(S)-\{j\})+f(i,S)
    \end{equation*}
    This results from combining (\ref{eq:lemma:3:1}) and (\ref{eq:lemma:3:4}).
\item $BR$ and $BL$ are also not equilibria, because in these action profiles, players $j$ and $i$ respectively have incentives to deviate -- the arguments are identical to cases (a) and (b) above, respectively.
\end{enumerate}
This completes the inductive argument.\hfill\Halmos
\endproof

\subsubsection{Decomposition of the distribution rule.}\label{s.decomposition}
Our goal in this section is to use the necessary conditions above (Proposition \ref{prop:2}) to establish that $f$ must be representable as a linear combination of generalized weighted Shapley value distribution rules (see (\ref{eq:basisGWSV}) in Table \ref{table:basisdistr}) on the unanimity games corresponding to the coalitions in $\mathcal{T}$, with corresponding coefficients from $Q$:

\begin{proposition}
\label{prop:3}
If $f$ is a budget-balanced distribution rule that guarantees the existence of an equilibrium in all games $G\in \mathcal{G}(N,f,W)$, then, there exists a sequence of weight systems $\Omega=\left\{\omega^T\right\}_{T\in\mathcal{T}}$ such that
\begin{equation*}
f=\sum_{T\in\mathcal{T}}q_T f_{GWSV}^T[\omega^T]
\end{equation*}
\end{proposition}

Note that for now, the weight systems $\omega^T$ could be arbitrary, and need not be related in any way. We deal with how they should be `consistent' later, in Section \ref{s.consistency}.

Before proceeding, we define a useful abstract mathematical object. The $\min$-partition of a finite poset $\left(\mathcal{T},\subseteq\right)$, denoted by $\mathbb{P}_{\min}(\mathcal{T})=\left\{\mathcal{P}_1, \mathcal{P}_2,\ldots, \mathcal{P}_\ell\right\}$, is an ordered partition of $\mathcal{T}$, constructed iteratively as specified in Algorithm \ref{alg:1}.

\begin{algorithm}
\caption{Construction of $\mathbb{P}_{\min}(\mathcal{T})$\label{alg:1}}
\begin{algorithmic}
\State $\mathcal{P}_1 = \mathcal{T}^{\min}$
\State $z \gets 2$
\While{$\mathcal{T}\not=\bigcup_{1\leq p < z}\mathcal{P}_p$}
    $\mathcal{P}_z = (\mathcal{T}-\bigcup_{1\leq p < z}\mathcal{P}_p)^{\min}$;
    $z \gets z+1$;
\EndWhile
\end{algorithmic}
\end{algorithm}

\begin{example}
\label{example:2}
\textit{Let $\left(\mathcal{T},\subseteq\right)$ be a poset, where $\mathcal{T}=\left\{\{i\},\{j\},\{j,k\},\{k,\ell\},\{j,\ell\},\{i,j,k\}\right\}$. Then, $$\mathbb{P}_{\min}(\mathcal{T})=\left\{\;\left\{\{i\},\{j\},\{k,\ell\}\right\}\;,\;\left\{\{j,k\},\{j,\ell\}\right\}\;,\;\left\{\{i,j,k\}\right\}\;\right\}$$}
\end{example}

\noindent\textit{Construction of basis distribution rules:} Given a budget-balanced distribution rule $f$ that guarantees the existence of an equilibrium in all games $G\in\mathcal{G}(N,f,W)$, we now show how to construct a sequence of basis distribution rules $\left\{f^T\right\}_{T\in\mathcal{T}}$ such that (\ref{eq:Tdistribution:2}) is satisfied. Let $\mathbb{P}_{\min}(\mathcal{T})=\left\{\mathcal{P}_1,\mathcal{P}_2,\ldots,\mathcal{P}_\ell\right\}$ be the $\min$-partition of the poset $\left(\mathcal{T},\subseteq\right)$, and let $f$ be a distribution rule for $W$. Starting with $z=1$, recursively define $f^T$ for each $T\in\mathcal{P}_z$ as,
\begin{equation}
\label{eq:lemma:4:2}
\left(\forall S\subseteq N\right)\ \left(\forall i\in N\right)\quad f^T(i,S)=
\begin{cases}
\frac{1}{q_T}\left(f(i,T)-\displaystyle\sum_{T'\in \mathcal{T}(T)-\{T\}}q_{T'}f^{T'}(i,S)\right) & ,\ T\subseteq S\\
0 & ,\ \mbox{otherwise}
\end{cases}
\end{equation}
At the end of this procedure, we obtain the basis distribution rules $\left\{f^T\right\}_{T\in\mathcal{T}}$. Note that it is not obvious from this construction that these basis distribution rules satisfy (\ref{eq:Tdistribution:2}), or that they are generalized weighted Shapley value distribution rules on their corresponding unanimity games. The rest of this section is devoted to showing these properties. But first, here is an example to demonstrate this recursive construction.

\vfill
\begin{example}
\label{example:3}
\textit{Consider the setting in Example \ref{example:1}, where $N=\{i,j,k\}$ is the set of players, and $W: 2^N\rightarrow\mathbb{R}$ is the welfare function defined in Table \ref{table:2a}. The basis representation of $W$ is shown in Table \ref{table:2b}. The set of coalitions is therefore given by $\mathcal{T}=\left\{\{i\},\{j\},\{k\},\{j,k\},\{i,k\},\{i,j,k\}\right\}$. For the poset $\left(\mathcal{T},\subseteq\right)$, we have,
$$\mathbb{P}_{\min}(\mathcal{T})=\left\{\;\left\{\{i\},\{j\},\{k\}\right\}\;,\;\left\{\{j,k\},\{i,k\}\right\}\;,\;\left\{\{i,j,k\}\right\}\;\right\}$$
Consider the following two distribution rules for $W$.
\begin{enumerate}[label=(\roman*)]
\item $f_{SV}$, the \textit{Shapley value} distribution rule (see Section \ref{s.shapleyfamily}).
\item $f_{EQ}$, the \textit{equal share} distribution rule (see Section \ref{s.eqprop}).
\end{enumerate}
Table \ref{table:3a} shows $f_{SV}$ and $f_{EQ}$ for this welfare function. The basis distribution rules $f^T_{SV}$ and $f^T_{EQ}$ that result from applying our construction (\ref{eq:lemma:4:2}) above are shown in Table \ref{table:3b}. For simplicity, we show only $f^T_{SV}(\cdot,T)$ and $f^T_{EQ}(\cdot,T)$.}
\end{example}

\begin{table}[ht]
\caption{Tables for Example \ref{example:3}}
\centering
\subfloat[][Definition of $f_{SV}$ and $f_{EQ}$\label{table:3a}]{
{\tabulinesep=1mm
\begin{tabu}{|c|c|c|}
\hline
$S$ & $f_{SV}(\cdot,S)$ & $f_{EQ}(\cdot,S)$\\
\hline
$\{i\}$     & $(1)$                                    & $(1)$\\
$\{j\}$     & $(2)$                                    & $(2)$\\
$\{k\}$     & $(3)$                                    & $(3)$\\
$\{i,j\}$   & $(1,2)$                                  & $(\frac{3}{2},\frac{3}{2})$\\
$\{j,k\}$   & $(1,2)$                                  & $(\frac{3}{2},\frac{3}{2})$\\
$\{i,k\}$   & $(\frac{1}{2},\frac{5}{2})$              & $(\frac{3}{2},\frac{3}{2})$\\
$\{i,j,k\}$ & $(\frac{5}{6},\frac{4}{3},\frac{11}{6})$ & $(\frac{4}{3},\frac{4}{3},\frac{4}{3})$\\
\hline
\end{tabu}}}
\qquad
\subfloat[][Basis distribution rules for $f_{SV}$ and $f_{EQ}$\label{table:3b}]{
{\tabulinesep=1mm
\begin{tabu}{|c|c|c|}
\hline
Coalition $T\in\mathcal{T}$ & $f^T_{SV}(\cdot,T)$ & $f^T_{EQ}(\cdot,T)$\\
\hline
$\{i\}$     & $(1)$                                   & $(1)$\\
$\{j\}$     & $(1)$                                   & $(1)$\\
$\{k\}$     & $(1)$                                   & $(1)$\\
$\{j,k\}$   & $(\frac{1}{2},\frac{1}{2})$             & $(\frac{1}{4},\frac{3}{4})$\\
$\{i,k\}$   & $(\frac{1}{2},\frac{1}{2})$             & $(-\frac{1}{2},\frac{3}{2})$\\
$\{i,j,k\}$ & $(\frac{1}{3},\frac{1}{3},\frac{1}{3})$ & $(-\frac{1}{6},-\frac{1}{6},\frac{4}{3})$\\
\hline
\end{tabu}}}
\end{table}

\vfill
The proof of Proposition \ref{prop:3} consists of four lemmas, as outlined below:

\vfill
\begin{enumerate}[label=(\alph*)]
\item In Lemma \ref{lemma:4.0}, we show that each $f^T$, as constructed in (\ref{eq:lemma:4:2}) mimics $f$ locally for its corresponding unanimity game $W^T$, i.e., $f^T$ satisfies Proposition \ref{prop:2} for $W=W^T$.

\vfill
\item Using this property, in Lemma \ref{lemma:4.1}, we show that each $f^T$ is a budget-balanced distribution rule for its corresponding unanimity game $W^T$.

\vfill
\item In Lemma \ref{lemma:4.2}, we show that $f$, and the basis distribution rules $\left\{f^T\right\}_{T\in\mathcal{T}}$, satisfy (\ref{eq:Tdistribution:2}), i.e., $f=\sum_{T\in\mathcal{T}}q_Tf^T$.

\vfill
\item Finally, in Lemma \ref{lemma:4.3}, we show that for each $T\in\mathcal{T}$, there exists a weight system $\omega^T$ such that $f^T=f^T_{GWSV}[\omega^T]$.
\end{enumerate}

\newpage

\begin{lemma}
\label{lemma:4.0}
Each $f^T$ as defined in (\ref{eq:lemma:4:2}) satisfies,

\vfill
\begin{equation}
\label{eq:lemma4.0}
\left(\forall S\subseteq N\right)\ \left(\forall i\in N\right)\quad f^T(i,S)=\begin{cases}
f^T(i,T) & ,\ i\in T \text{ and } T\subseteq S\mcr
0 & ,\ \text{otherwise}
\end{cases}
\end{equation}
\end{lemma}

\vfill
\proof{Proof.} The proof is by induction on $\mathbb{P}_{\min}(\mathcal{T})$. The base case, where $T\in\mathcal{P}_1$ is immediate, because from (\ref{eq:lemma:4:2}), for any $T\in\mathcal{P}_1$,

\vfill
\begin{equation*}
\left(\forall S\subseteq N\right)\ \left(\forall i\in N\right)\quad f^T(i,S)=
\begin{cases}
\frac{1}{q_T}f(i,T) & ,\ T\subseteq S\mcr
0 & ,\ \mbox{otherwise}
\end{cases}
\end{equation*}

\vfill
\noindent Our induction hypothesis is that $f^T$ satisfies (\ref{eq:lemma4.0}) for all $T\in\bigcup_{p=1}^{z}\mathcal{P}_p$, for some $1\leq z<\ell$. Assuming that this is true, we prove that $f^T$ satisfies (\ref{eq:lemma4.0}) for all $T\in\mathcal{P}_{z+1}$. To evaluate $f^T(i,S)$ for some $i\in S\subseteq N$, we consider the following three cases:

\vfill
\begin{enumerate}[label=(\roman*)]
\item $T\nsubseteq S$. In this case, from (\ref{eq:lemma:4:2}), $f^T(i,S)=0$.
\item $i\notin T\subseteq S$. Here, we know that $f(i,T)=0$ by definition. Also, for all $T'\in\mathcal{T}(T)-\{T\}$, we have, $T'\in\bigcup_{p=1}^{z}\mathcal{P}_p$ and $i\notin T'$; so, from the induction hypothesis, $f^{T'}(i,S)=0$. Therefore, evaluating (\ref{eq:lemma:4:2}), we get $f^T(i,S)=0$.

\vfill
\item $i\in T\subseteq S$. In this case, we need to show that $f^T(i,S)=f^T(i,T)$. By (\ref{eq:lemma:4:2}), we have,
    \begin{equation*}
    \begin{split}
    f^T(i,S)&=\frac{1}{q_T}\left(f(i,T)-\displaystyle\sum_{T'\in \mathcal{T}(T)-\{T\}}q_{T'}f^{T'}(i,S)\right)\mcr
    f^T(i,T)&=\frac{1}{q_T}\left(f(i,T)-\displaystyle\sum_{T'\in \mathcal{T}(T)-\{T\}}q_{T'}f^{T'}(i,T)\right)
    \end{split}
    \end{equation*}
    For each $T'\in\mathcal{T}(T)-\{T\}$, we know that $T'\in\bigcup_{p=1}^{z}\mathcal{P}_p$, and hence, from the induction hypothesis, we have $f^{T'}(i,S)=f^{T'}(i,T)$. Therefore, $f^T(i,T)=f^T(i,S)$, as desired.
\end{enumerate}

\vfill
\noindent Hence, $f^T$ satisfies (\ref{eq:lemma4.0}).\hfill\Halmos
\endproof

\newpage

\begin{lemma}
\label{lemma:4.1}
If $f$ is a budget-balanced distribution rule for $W$, then each $f^T$ as defined in (\ref{eq:lemma:4:2}) is a budget-balanced distribution rule for $W^T$, i.e.,
\begin{equation*}
\left(\forall T\in \mathcal{T}\right)\ \left(\forall S\subseteq N\right)\quad \sum_{i\in S}f^T(i,S)=W^T(S)
\end{equation*}
\end{lemma}

\proof{Proof.} Since $f^T$ is of the form (\ref{eq:lemma4.0}) from Lemma \ref{lemma:4.0}, to show (local) budget-balance, we need only show that
\begin{equation}
\label{eq:lemma4.1}
\left(\forall T\in \mathcal{T}\right)\quad \sum_{i\in T}f^T(i,T)=1
\end{equation}
Once again, the proof is by induction on $\mathbb{P}_{\min}(\mathcal{T})$. The base case, where $T\in\mathcal{P}_1$ follows from the budget-balance of $f$. Our induction hypothesis is that $f^T$ satisfies (\ref{eq:lemma4.1}) for all $T\in\cup_{p=1}^{z}\mathcal{P}_p$, for some $1\leq z<\ell$. Assuming that this is true, we prove that $f^T$ satisfies (\ref{eq:lemma4.1}) for all $T\in\mathcal{P}_{z+1}$. For any $T\in\mathcal{P}_{z+1}$, using (\ref{eq:lemma:4:2}), we have,
\begin{small}
\begin{equation*}
\begin{split}
\sum_{i\in T}f^T(i,T) &= \frac{1}{q_T}\left(\sum_{i\in T}f(i,T)-\sum_{i\in T}\sum_{T'\in \mathcal{T}(T)-\{T\}}q_{T'}f^{T'}(i,T')\right)\mcr
&= \frac{1}{q_T}\left(W(T)-\sum_{T'\in \mathcal{T}(T)-\{T\}}\sum_{i\in T'}q_{T'}f^{T'}(i,T')\right)\mcr
&= \frac{1}{q_T}\left(\sum_{T'\in\mathcal{T}(T)}q_{T'}-\sum_{T'\in \mathcal{T}(T)-\{T\}}q_{T'}\right) = \frac{1}{q_T}\left(q_T\right) = 1
\end{split}
\end{equation*}
\end{small}
where we have used the budget-balance of $f$, followed by the induction hypothesis and (\ref{eq:alternateW}). This completes the inductive argument and hence the proof.\hfill\Halmos
\endproof

\vspace{0.075in}

\begin{example}
\label{example:3.5}
\textit{Consider the decomposition of $f_{SV}$ and $f_{EQ}$ illustrated in Example \ref{example:3}. Both $f^T_{SV}$ and $f^T_{EQ}$ are locally budget-balanced for all $T\in\mathcal{T}$, that is, they satisfy (\ref{eq:lemma4.1}).}
\end{example}

\vspace{0.075in}

Before continuing with the proof, in the next lemma, we present the \textit{conditional inclusion-exclusion principle}, an important and useful property of the basis distribution rules $\left\{f^T\right\}_{T\in\mathcal{T}}$.

\begin{lemma}
\label{lemma:4.1.5}
(Conditional inclusion-exclusion principle) For any $T\in\mathcal{T}$, there exist integers $\left\{n_T(T')\right\}_{T'\in\mathcal{T}}$ such that the basis distribution rules $\left\{f^T\right\}_{T\in\mathcal{T}}$ defined in (\ref{eq:lemma:4:2}) satisfy,
\begin{equation}
\label{eq:condinclexcl}
\left(\forall i\in T\right)\quad q_T f^T(i,T) = \displaystyle\sum_{T'\in\mathcal{T}(T)}n_T(T')f(i,T')
\end{equation}
Furthermore, if $\left\{f^T\right\}_{T\in\mathcal{T}}$ satisfies
\begin{equation}
\label{eq:lemma:4.1.5}
\left(\forall S\subsetneq T\right)\ \left(\forall i\in S\right)\quad f(i,S) = \sum_{T'\in\mathcal{T}(S)}q_{T'}f^{T'}(i,T'),
\end{equation}
then,
\begin{equation}
\label{eq:condinclexcl_1}
\left(\forall i\in T\right)\ \left(\forall j\in T-\{i\}\right)\quad 0 = q_T f^T(i,T-\{j\}) = \displaystyle\sum_{T'\in\mathcal{T}(T)}n_T(T')f(i,T'-\{j\})
\end{equation}
\end{lemma}

\proof{Proof.} For any $T\in\mathcal{T}$, setting $S=T$ in (\ref{eq:lemma:4:2}), and using Lemma \ref{lemma:4.0}, we get,
\begin{equation}
\label{eq:recursion}
\left(\forall i\in T\right)\quad q_T f^T(i,T)=f(i,T)-\displaystyle\sum_{T'\in \mathcal{T}(T)-\{T\}}q_{T'}f^{T'}(i,T')
\end{equation}

\noindent It follows that by unravelling the recursion above, i.e., by repeatedly substituting for the terms $q_{T'}f^{T'}(i,T')$ that appear in the summation, we obtain (\ref{eq:condinclexcl}), where $\left\{n_T(T')\right\}_{T'\in\mathcal{T}}$ are some integers.

\newpage

Let $\mathcal{T}_i(T)=\left\{T'\in\mathcal{T}(T):i\in T'\right\}$ denote the set of coalitions contained in $T$ that contain $i$. Before proving (\ref{eq:condinclexcl_1}), we make the following observation. From (\ref{eq:condinclexcl}), we have,
\begin{equation}
\label{eq:lemma:4.1.5:1}
\begin{split}
\left(\forall i\in T\right)\quad q_T f^T(i,T) &= \sum_{T'\in\mathcal{T}(T)}n_T(T')f(i,T')\mcr
&= \sum_{T'\in\mathcal{T}_i(T)}n_T(T')\sum_{T''\in\mathcal{T}_i(T')}q_{T''}f^{T''}(i,T'')\quad\text{(from (\ref{eq:lemma:4:2}) and Lemma \ref{lemma:4.0})}\mcr
&= \sum_{T'\in\mathcal{T}_i(T)}m_{i,T}(T')q_{T'}f^{T'}(i,T')
\end{split}
\end{equation}
where $\left\{m_{i,T}(T')\right\}_{T'\in\mathcal{T}}$ are some integer coefficients. We now exploit the fact that (\ref{eq:lemma:4.1.5:1}) holds for all distribution rules $f$ to show that the unique solution for the coefficients $m_{i,T}(T')$ is given by,
\begin{equation}
\label{eq:lemma:4.1.5:2}
m_{i,T}(T')=
\begin{cases}
1 & ,\ T'=T\mcr
0 & ,\ \text{otherwise}
\end{cases}
\end{equation}
To see this, we first prove that, given $T\in\mathcal{T}$ and $i\in T$, $m_{i,T}(T'')=0$ for all $T''\in\mathcal{T}_i(T)-\{T\}$ by induction on $\mathbb{P}_{\min}(\mathcal{T}_i(T)-\{T\})$. To do this, we focus on the family of generalized weighted Shapley value distribution rules $\left\{f_{GWSV}[\omega^{S}]\right\}_{S\in\mathcal{T}(T)}$ with weight systems $\omega^{S}=(\boldsymbol{\lambda},\Sigma^{S})$, where $\boldsymbol{\lambda}=(1,1,\ldots,1)$ and $\Sigma^{S}=(N-S,S)$. By definition (see (\ref{eq:basisGWSV}) in Table \ref{table:basisdistr}), for each $T'\in\mathcal{T}_i(T)$,
\begin{equation}
\label{eq:lemma:4.1.5:3}
f^{T'}_{GWSV}[\omega^{S}](i,T') = \begin{cases}
\frac{1}{|T'|} & ,\ T'\subseteq S\mcr
0 & ,\ \mbox{otherwise}
\end{cases}
\end{equation}
\begin{enumerate}[label=(\roman*)]
\item For the base case, when $T''\in\mathcal{P}_1$, it follows from (\ref{eq:lemma:4.1.5:3}), with $S=T''$, that, for all $T'\in\mathcal{T}_i(T)$,
    \begin{equation}
    \label{eq:lemma:4.1.5:4}
    f^{T'}_{GWSV}[\omega^{T''}](i,T') = \begin{cases}
    \frac{1}{|T'|} & ,\ T' = T''\mcr
    0 & ,\ \mbox{otherwise}
    \end{cases}
    \end{equation}
    This results from the fact that since $T''\in\mathcal{P}_1$, $T'\subseteq T''$ if and only if $T'=T''$. Now, we evaluate (\ref{eq:lemma:4.1.5:1}) for the distribution rule $f_{GWSV}[\omega^{T''}]$ to get,
    \begin{equation*}
    q_T f^T_{GWSV}[\omega^{T''}](i,T) = \displaystyle\sum_{T'\in\mathcal{T}_i(T)}m_{i,T}(T')q_{T'}f^{T'}_{GWSV}[\omega^{T''}](i,T')
    \end{equation*}
    Using (\ref{eq:lemma:4.1.5:4}) to simplify the above equation, we get,
    \begin{equation*}
    0 = m_{i,T}(T'') q_{T''} \frac{1}{|T''|}
    \end{equation*}
    since, for any $T'\not=T''$, $T'\cap (N-T')\not=\emptyset$. Therefore, $m_{i,T}(T'') = 0$.
\item Our induction hypothesis is that $m_{i,T}(T'')=0$ for all $T''\in\bigcup_{p=1}^{z}\mathcal{P}_p$, for some $1\leq z<\ell$. Assuming that this is true, we prove that $m_{i,T}(T'')=0$ for all $T''\in\mathcal{P}_{z+1}$. If $T''\in\mathcal{P}_{z+1}$, it follows from (\ref{eq:lemma:4.1.5:3}), with $S=T''$, that, for all $T'\in\{T\}\bigcup_{p=z+1}^{\ell}\mathcal{P}_p$, (\ref{eq:lemma:4.1.5:4}) holds, from a similar reasoning as above. Now, we evaluate (\ref{eq:lemma:4.1.5:1}) for the distribution rule $f_{GWSV}[\omega^{T''}](\cdot,\cdot)$ to get,
    \begin{equation*}
    q_T f^T_{GWSV}[\omega^{T''}](i,T) = \sum_{T'\in\mathcal{T}_i(T)}m_{i,T}(T')q_{T'}f^{T'}_{GWSV}[\omega^{T''}](i,T')
    \end{equation*}
    By grouping together terms on the right hand side, we can rewrite this as,

\vspace{-0.15in}
    \begin{equation}
    \label{eq:local:1}
    q_T f^T_{GWSV}[\omega^{T''}](i,T) = m_{i,T}(T)q_{T}f^{T}_{GWSV}[\omega^{T''}](i,T) + \sum_{p=1}^{\ell}\sum_{T'\in\mathcal{P}_p}m_{i,T}(T')q_{T'}f^{T'}_{GWSV}[\omega^{T''}](i,T')
    \end{equation}

\vspace{-0.15in}
    Using the induction hypothesis, we get that,

\vspace{-0.1in}
    \begin{equation}
    \label{eq:local:2}
    \displaystyle\sum_{p=1}^{z}\sum_{T'\in\mathcal{P}_p}m_{i,T}(T')q_{T'}f^{T'}_{GWSV}[\omega^{T''}](i,T') = 0
    \end{equation}

\vspace{-0.025in}
    Using (\ref{eq:lemma:4.1.5:4}), we get that,

\vspace{-0.15in}
    \begin{equation}
    \label{eq:local:3}
    \begin{split}
    \displaystyle\sum_{p=z+1}^{\ell}\sum_{T'\in\mathcal{P}_p}m_{i,T}(T')q_{T'}f^{T'}_{GWSV}[\omega^{T''}](i,T') &= m_{i,T}(T'') q_{T''} \frac{1}{|T''|}\\
    f^T_{GWSV}[\omega^{T''}](i,T) &= 0
    \end{split}
    \end{equation}
    since, for all $T'\in\{T\}\bigcup_{p=z+1}^{\ell}\mathcal{P}_p$, if $T'\not=T''$, then $T'\cap (N-T')\not=\emptyset$. Therefore, using (\ref{eq:local:2}) and (\ref{eq:local:3}) in (\ref{eq:local:1}), we get $m_{i,T}(T'') = 0$.
\end{enumerate}
This completes the inductive argument. From this, it is straightforward to see that $m_{i,T}(T)=1$.

\vspace{0.05in}

We now return to proving the remainder of the lemma, that is (\ref{eq:condinclexcl_1}). The right hand side of (\ref{eq:condinclexcl_1}) can be evaluated as,

\vspace{-0.125in}
\begin{equation*}
\begin{split}
\sum_{T'\in\mathcal{T}(T)}n_T(T')f(i,T'-\{j\}) &= \sum_{T'\in\mathcal{T}(T)}n_T(T')\sum_{T''\in\mathcal{T}(T'-\{j\})}q_{T''}f^{T''}(i,T'')\quad\;\;\ \text{(from (\ref{eq:lemma:4.1.5}))}\mcr
&= \sum_{T'\in\mathcal{T}_i(T)}n_T(T')\sum_{T''\in\mathcal{T}_i(T')}q_{T''}f^{T''}(i,T''-\{j\})\;\;\;\text{(from Lemma \ref{lemma:4.0})}\mcr
&= \sum_{T'\in\mathcal{T}_i(T)}m_{i,T}(T')q_{T'}f^{T'}(i,T'-\{j\})\quad\quad\quad\;\;\;\ \ \text{(from (\ref{eq:lemma:4.1.5:1}))}\mcr
&= q_T f^T(i,T-\{j\}) = 0\quad\quad\quad\quad\quad\quad\quad\quad\quad\quad\;\text{(from (\ref{eq:lemma:4.1.5:2}))}
\end{split}
\end{equation*}

\vspace{-0.025in}
\noindent This completes the proof.\hfill\Halmos
\endproof

\begin{lemma}
\label{lemma:4.2}
If $f$ is a budget-balanced distribution rule that guarantees the equilibrium existence in all games $G\in\mathcal{G}(N,f,W)$, then the basis distribution rules $\left\{f^T\right\}_{T\in\mathcal{T}}$ defined in (\ref{eq:lemma:4:2}) satisfy

\vspace{-0.075in}
\begin{equation}
\label{eq:lemma:4:Tdistribution}
\left(\forall S\subseteq N\right)\ \left(\forall i\in S\right)\quad f(i,S) = \sum_{T\in\mathcal{T}}q_Tf^T(i,S)
\end{equation}
\end{lemma}

\vspace{-0.025in}
\proof{Proof.} From Lemma \ref{lemma:4.0}, (\ref{eq:lemma:4:Tdistribution}) is equivalent to

\vspace{-0.05in}
\begin{equation}
\label{eq:lemma:4:Tdistribution:alt}
\left(\forall S\subseteq N\right)\ \left(\forall i\in S\right)\quad f(i,S) = \sum_{T\in\mathcal{T}(S)}q_Tf^T(i,T)
\end{equation}

\vspace{-0.075in}
\noindent Let $S\subseteq N$. We consider three cases:

\vfill
\noindent\textbf{Case 1:} $S\in\mathcal{T}$. The proof is immediate here, because, rearranging the terms in (\ref{eq:lemma:4:2}), we get,
\begin{equation}
\label{eq:lemma:4:3}
f(i,S)=\sum_{T'\in \mathcal{T}(S)}q_{T'}f^{T'}(i,S)=\sum_{T'\in \mathcal{T}(S)}q_{T'}f^{T'}(i,T')
\end{equation}
where the last equality follows from Lemma \ref{lemma:4.0}.

\vfill
\noindent\textbf{Case 2:} $S\not=N(S)$. In this case, we can apply Proposition \ref{prop:2}, i.e., $f(i,S)=f(i,N(S))$, to reduce it to the following case, replacing $S$ with $N(S)$.

\vfill
\noindent\textbf{Case 3:} $S=N(S)$. In other words, $S$ is a union of one or more coalitions in $\mathcal{T}$. The remainder of the proof is devoted to this case.

\newpage
For any subset $S\subseteq N$ such that $S=N(S)$, i.e., $S$ is exactly a union of one or more coalitions in $\mathcal{T}$, we prove this lemma by induction on $|\mathcal{T}(S)|$. The base case, where $|\mathcal{T}(S)|=1$ (and hence $S\in\mathcal{T}$) is true from (\ref{eq:lemma:4:3}). Our induction hypothesis is that (\ref{eq:lemma:4:Tdistribution:alt}) holds for all subsets $S\subseteq N$ such that $S=N(S)$, with $|\mathcal{T}(S)|\leq z$ for some $1\leq z < |\mathcal{T}|$. Assuming that this is true, we prove that (\ref{eq:lemma:4:Tdistribution:alt}) holds for all subsets $S\subseteq N$ such that $S=N(S)$, with $|\mathcal{T}(S)|= z+1$. If $S\in\mathcal{T}$, then the proof is immediate from (\ref{eq:lemma:4:3}), so let us assume $S\notin\mathcal{T}$. Before proceeding with the proof, we point out the following observation. Since $f$ is budget-balanced, we have,
\begin{equation}
\label{eq:lemma:4:3.5}
\begin{split}
\sum_{i\in S}f(i,S) = W(S) &= \sum_{T\in\mathcal{T}(S)} q_T\quad\quad\quad\quad\quad\quad\quad\quad\text{(from (\ref{eq:alternateW}))}\mcr
&= \sum_{T\in\mathcal{T}(S)} q_T \sum_{i\in T} f^T(i,T)\quad\quad\;\;\;\ \text{(from Lemma \ref{lemma:4.1})}\mcr
&= \sum_{i\in S}\sum_{T\in\mathcal{T}(S)}q_T f^T(i,T)\quad\quad\;\;\;\;\,\text{(from Lemma \ref{lemma:4.0})}
\end{split}
\end{equation}

The proof is by contradiction, and proceeds as follows. Assume to the contrary, that $f(k,S) \not= \sum_{T\in\mathcal{T}(S)}q_Tf^T(k,T)$ for some $k\in S$, for some $S\subseteq N$ such that $S=N(S)$, with $|\mathcal{T}(S)|= z+1$. Since $z+1\geq 2$, and $S=N(S)$, it must be that $|S|\geq 2$, i.e., $S$ has at least two players. Then, from (\ref{eq:lemma:4:3.5}), it follows that we can pick $i,j\in S$ such that,
\begin{equation}
\label{eq:lemma:4:4}
f(i,S) > \sum_{T\in\mathcal{T}(S)}q_Tf^T(i,T)
\end{equation}
\begin{equation}
\label{eq:lemma:4:5}
f(j,S) < \sum_{T\in\mathcal{T}(S)}q_Tf^T(j,T)
\end{equation}
Because $S=N(S)$, for any $S'\subsetneq S$, $|\mathcal{T}(S')|<|\mathcal{T}(S)|$. Hence, applying the induction hypothesis,
\begin{equation}
\label{eq:lemma:4:2:indhyp}
\left(\forall S'\subsetneq S\right)\ \left(\forall i\in S'\right)\quad f(i,S') = \sum_{T\in\mathcal{T}(S')}q_T f^T(i,T)
\end{equation}
Since every coalition $T\in\mathcal{T}(S)$ is a subset of $S$, (\ref{eq:lemma:4:2:indhyp}) holds when $S$ is replaced with any $T\in\mathcal{T}(S)$. Therefore, Lemma \ref{lemma:4.1.5}, the conditional inclusion-exclusion principle, can be applied to obtain, for any coalition $T\in\mathcal{T}(S)$,
\begin{equation*}
\begin{split}
\left(\forall i\in T\right)\quad q_T f^T(i,T) &= \sum_{T'\in\mathcal{T}(T)}n_T(T')f(i,T')\mcr
\left(\forall i\in T\right)\ \left(\forall j\in T-\{i\}\right)\quad q_T f^T(i,T-\{j\}) &= \sum_{T'\in\mathcal{T}(T)}n_T(T')f(i,T'-\{j\})
\end{split}
\end{equation*}
Summing up these equations over all $T\in\mathcal{T}(S)$, we get,
\begin{equation}
\label{eq:lemma:4:2:1}
\begin{split}
\left(\forall i\in S\right)\quad \sum_{T\in\mathcal{T}(S)} q_T f^T(i,T) &= \sum_{T\in\mathcal{T}(S)}\widetilde{n}_S(T)f(i,T)\mcr
\left(\forall i\in S\right)\ \left(\forall j\in S-\{i\}\right)\quad \sum_{T\in\mathcal{T}(S)} q_T f^T(i,T-\{j\}) &= \sum_{T\in\mathcal{T}(S)}\widetilde{n}_S(T)f(i,T-\{j\})
\end{split}
\end{equation}
where the constants $\left\{\widetilde{n}_S(T)\right\}_{T\in\mathcal{S}}$ are given by,
\begin{equation*}
\widetilde{n}_S(T)=\sum_{\substack{T'\in\mathcal{T}(S)\mcr T\subseteq T'}}n_{T'}(T)
\end{equation*}

\newpage
\begin{figure}[ht]
    \FIGURE
    {\includegraphics*[height=7cm,width=7cm]{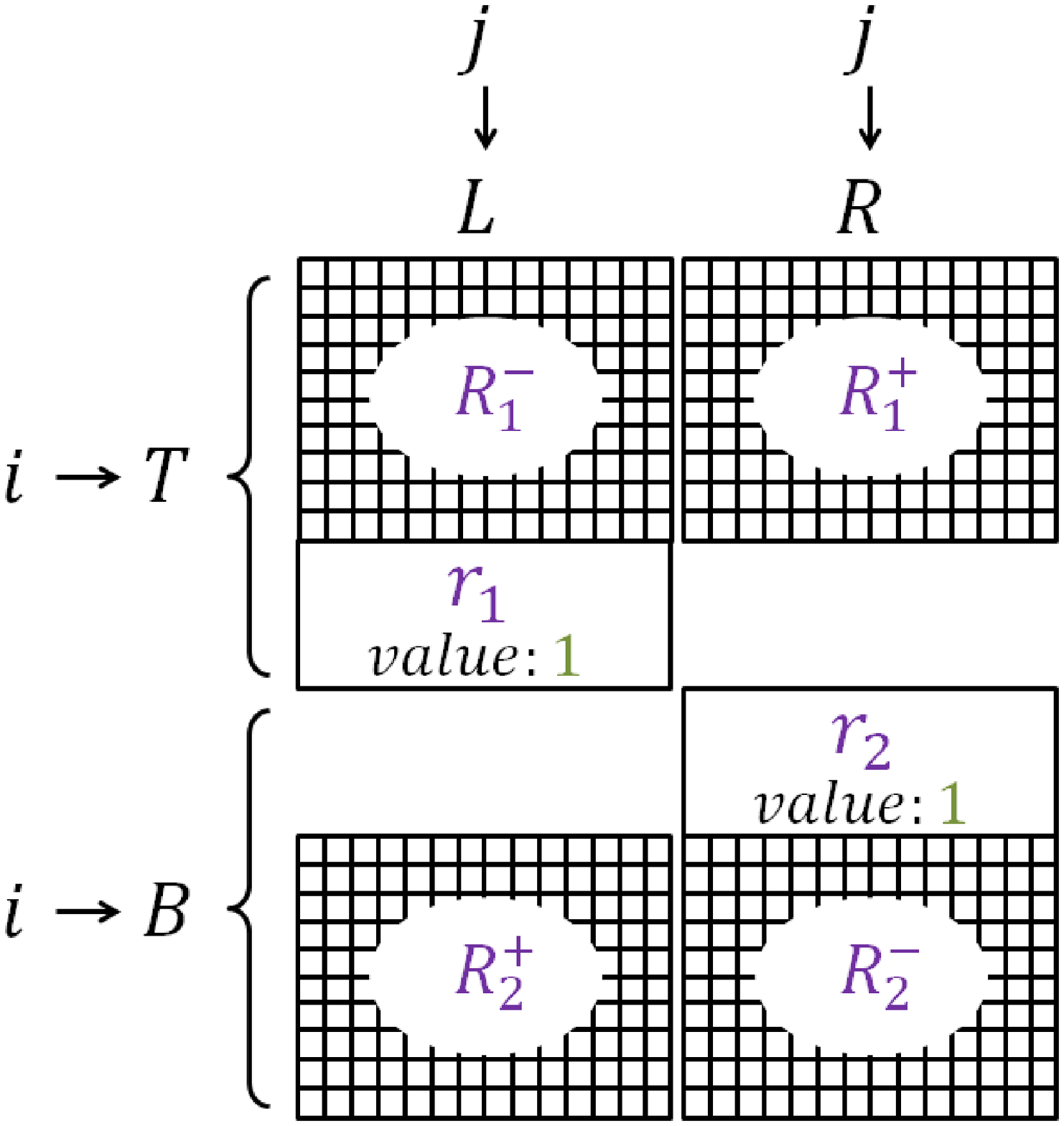}}
    {Counterexample 3\label{fig:4}}
    {}
\vspace{-0.1in}
\end{figure}

\textit{Counterexample 3.} Our goal is to exploit inequalities (\ref{eq:lemma:4:4}) and (\ref{eq:lemma:4:5}) to build a counterexample that mimics \textit{Counterexample 2} illustrated in Figure \ref{fig:3}, leading to a similar best-response cycle involving just players $i$ and $j$. Equation (\ref{eq:lemma:4:2:1}) suggests the following technique for achieving precisely this. Consider the game in Figure \ref{fig:4}, which has the same underlying $2\times 2$ box structure of \textit{Counterexample 2}. The resources in the top half are added as follows:

\vfill
\begin{enumerate}[label=(\roman*)]
\item Add a resource $r_1$ to the top left box.

\vfill
\item Add resources in $R^+_1=\left\{r^T_1:T\in\mathcal{T}(S)\text{ and }\widetilde{n}_S(T)>0\right\}$ to the top right box.

\vfill
\item Add resources in $R^-_1=\left\{r^T_1:T\in\mathcal{T}(S)\text{ and }\widetilde{n}_S(T)<0\right\}$ to the top left box.
\end{enumerate}

\vfill
\noindent Then, the bottom half is symmetrically filled up as follows.

\vfill
\begin{enumerate}[label=(\roman*)]
\item Add a resource $r_2$ to the bottom right box.

\vfill
\item Add resources in $R^+_2=\left\{r^T_2:T\in\mathcal{T}(S)\text{ and }\widetilde{n}_S(T)>0\right\}$ to the bottom left box.

\vfill
\item Add resources in $R^-_2=\left\{r^T_2:T\in\mathcal{T}(S)\text{ and }\widetilde{n}_S(T)<0\right\}$ to the bottom right box.
\end{enumerate}

\vfill
\noindent The resource set $R$ is therefore given by,

\vspace{-0.125in}
\begin{equation*}
R=\left\{r_1,r_2\right\}\cup R^+_1\cup R^-_1\cup R^+_2\cup R^-_2
\end{equation*}
The local resource coefficients are given by,

\vspace{-0.125in}
\begin{equation*}
v_{r_1}=v_{r_2}=1\quad\quad\text{and}\quad\quad (\forall T\in\mathcal{T}(S))\quad v_{r^T_1}=v_{r^T_2}=|\widetilde{n}_S(T)|
\end{equation*}
In resources $r_1$ and $r_2$, we fix players in $S-\{i,j\}$. For each $T\in\mathcal{T}(S)$, in resources $r^T_1$ and $r^T_2$, we fix players in $T-\{i,j\}$. Effectively, all players other than $i$ and $j$ have a fixed action in their action set, determined by these fixtures. The action set of player $i$ is given by $\mathcal{A}_i = \left\{T,B\right\}$, where,

\vspace{-0.1in}
\begin{equation*}
T = \left\{r^T_1 \in R^-_1 : i\in T\right\}\cup \left\{r_1\right\}\cup \left\{r^T_1 \in R^+_1 : i\in T\right\}
\end{equation*}

\vspace{-0.175in}
\begin{equation*}
B = \left\{r^T_2 \in R^+_2 : i\in T\right\}\cup \left\{r_2\right\}\cup \left\{r^T_2 \in R^-_2 : i\in T\right\}
\end{equation*}
The action set of player $j$ is given by $\mathcal{A}_j = \left\{L,R\right\}$, where,

\vspace{-0.1in}
\begin{equation*}
L = \left\{r^T_1 \in R^-_1 : j\in T\right\}\cup \left\{r_1\right\}\cup \left\{r^T_2 \in R^+_2 : j\in T\right\}
\end{equation*}

\vspace{-0.175in}
\begin{equation*}
R = \left\{r^T_1 \in R^+_1 : j\in T\right\}\cup \left\{r_2\right\}\cup \left\{r^T_2 \in R^-_2 : j\in T\right\}
\end{equation*}
This is essentially a game between players $i$ and $j$. The set of joint action profiles can therefore be represented as $\mathcal{A}=\left\{TL,TR,BL,BR\right\}$.

We use the four properties in (\ref{eq:lemma:4:4}), (\ref{eq:lemma:4:5}), (\ref{eq:lemma:4:2:indhyp}), and (\ref{eq:lemma:4:2:1}) to show that \textit{Counterexample 3} does not possess an equilibrium, thereby contradicting the fact that $f$ guarantees the existence of an equilibrium in all games $G\in\mathcal{G}(N,f,W)$. We show this for each outcome:

\newpage
\begin{enumerate}[label=(\roman*)]
\item $TL$ is not an equilibrium, since player $j$ has an incentive to deviate from $L$ to $R$. To see this, consider the utilities of player $j$ when choosing $L$ and $R$,

    \vspace{-0.125in}
    \begin{equation*}
    U_j(T,L)=-\sum_{\substack{T\in\mathcal{T}(S)\mcr \widetilde{n}_S(T)<0}}\widetilde{n}_S(T)f(j,T) + f(j,S) + \sum_{\substack{T\in\mathcal{T}(S)\mcr \widetilde{n}_S(T)>0}}\widetilde{n}_S(T)f(j,T-\{i\})
    \end{equation*}

    \vspace{-0.075in}
    \begin{equation*}
    U_j(T,R)=\sum_{\substack{T\in\mathcal{T}(S)\mcr \widetilde{n}_S(T)>0}}\widetilde{n}_S(T)f(j,T) + f(j,S-\{i\}) - \sum_{\substack{T\in\mathcal{T}(S)\mcr \widetilde{n}_S(T)<0}}\widetilde{n}_S(T)f(j,T-\{i\})
    \end{equation*}

    \vspace{-0.025in}
    \noindent The difference in utilities for $j$ between choosing $R$ and $L$ is therefore given by,

    \vspace{-0.175in}
    \begin{equation*}
    \begin{split}
    U_j(T,R&)-U_j(T,L)\mcr
    &= \left(\sum_{T\in\mathcal{T}(S)}\widetilde{n}_S(T)f(j,T) - f(j,S)\right)\mcr
    &\quad\quad\quad\quad\quad\quad\quad\quad\quad + \left(f(j,S-\{i\})-\sum_{T\in\mathcal{T}(S)}\widetilde{n}_S(T)f(j,T-\{i\})\right)\mcr
    &= \left(\sum_{T\in\mathcal{T}(S)}q_T f^T(j,T) - f(j,S)\right) + \left(f(j,S-\{i\}) - \sum_{T\in\mathcal{T}(S)}q_T f^T(j,T-\{i\})\right)\mcr
    &= \left(\sum_{T\in\mathcal{T}(S)}q_T f^T(j,T) - f(j,S)\right) + \left(f(j,S-\{i\}) - \sum_{T\in\mathcal{T}(S-\{i\})}q_T f^T(j,T)\right)\mcr
    &= \sum_{T\in\mathcal{T}(S)}q_T f^T(j,T) - f(j,S) > 0
    \end{split}
    \end{equation*}

    \vspace{-0.05in}
    \noindent This results from using (\ref{eq:lemma:4:2:1}) first, followed by (\ref{eq:lemma:4:2:indhyp}), Lemma \ref{lemma:4.0}, and then (\ref{eq:lemma:4:5}).
\item $TR$ is not an equilibrium, since player $i$ has an incentive to deviate from $T$ to $B$. The proof is along the same lines as the previous case. Using similar arguments, we get,

    \vspace{-0.125in}
    \begin{equation*}
    U_i(B,R)-U_i(T,R) = f(i,S) - \displaystyle\sum_{T\in\mathcal{T}(S)}q_T f^T(i,T)
    \end{equation*}

    \vspace{-0.15in}
    \noindent which is positive, from (\ref{eq:lemma:4:4}).
\item $BR$ and $BL$ are also not equilibria, because in these action profiles, players $j$ and $i$ respectively have incentives to deviate -- the arguments are identical to cases (a) and (b) above, respectively.
\end{enumerate}
This completes the inductive argument.\hfill\Halmos
\endproof

\vfill

\begin{example}
\label{example:4}
\textit{Consider the decomposition of $f_{SV}$ and $f_{EQ}$ into their respective basis distribution rules, as illustrated in Example \ref{example:3}. Let $S=\{i,j\}$.
\begin{enumerate}[label=(\roman*)]
\item $f_{SV}(i,S)=1$, and $\sum_{T\in\mathcal{T}(S)}q_T f^T_{SV}(i,T) = q_{\{i\}} f^{\{i\}}_{SV}(i,\{i\}) = 1$. Thus, $f_{SV}$ satisfies (\ref{eq:lemma:4:Tdistribution}).
\item $f_{EQ}(i,S)=\frac{3}{2}$, and $\sum_{T\in\mathcal{T}(S)}q_T f^T_{EQ}(i,T) = q_{\{i\}} f^{\{i\}}_{EQ}(i,\{i\}) = 1$. Therefore, $f_{EQ}$ does not satisfy (\ref{eq:lemma:4:Tdistribution}), and hence does not guarantee the existence of an equilibrium in all games $G\in\mathcal{G}(N,f,W)$.
\end{enumerate}}
\end{example}

\vfill

From Lemma \ref{lemma:4.2}, it follows that any budget-balanced distribution rule $f$ that guarantees the existence of an equilibrium in all games $G\in\mathcal{G}(N,f,W)$ satisfies (\ref{eq:lemma:4.1.5}) for all $T\in\mathcal{T}$. So, for such $f$, condition (\ref{eq:lemma:4.1.5}) can be stripped off of Lemma \ref{lemma:4.1.5}, leading to the (unconditional) \textit{inclusion-exclusion principle}, a powerful tool that we use extensively in proving several subsequent lemmas. We formally state this in the following corollary:

\newpage

\begin{corollary}
\label{corollary:4.2.5}
(Inclusion-exclusion principle) If $f$ is a budget-balanced distribution rule that guarantees the existence of an equilibrium in all games $G\in\mathcal{G}(N,f,W)$, and $\left\{f^T\right\}_{T\in\mathcal{T}}$ are the basis distribution rules defined in (\ref{eq:lemma:4:2}), then, for every $T\in\mathcal{T}$, there exist integers $\left\{n_T(T')\right\}_{T'\in\mathcal{T}}$ such that the following equations hold:

\vspace{-0.15in}
\begin{equation}
\label{eq:inclexcl}
\left(\forall i\in T\right)\quad q_T f^T(i,T) = \displaystyle\sum_{T'\in\mathcal{T}(T)}n_T(T')f(i,T')
\end{equation}

\vspace{-0.1in}
\begin{equation}
\label{eq:inclexcl_1}
\left(\forall i\in T\right)\ \left(\forall j\in T-\{i\}\right)\quad 0 = q_T f^T(i,T-\{j\}) = \displaystyle\sum_{T'\in\mathcal{T}(T)}n_T(T')f(i,T'-\{j\})
\end{equation}
\end{corollary}

\vfill

\vfill
\begin{example}
\label{example:4.5}
\textit{To illustrate the inclusion-exclusion principle, let the set of players be $N=\{i,j,k,\ell\}$, and let the set of contributing coalitions be $\mathcal{T}=\left\{\{i\},\{i,j\},\{i,k\},\{i,\ell\},\{i,j,k,\ell\}\right\}$. Then, for $T=\{i,j,k,\ell\}$, unraveling the recursion in (\ref{eq:recursion}) gives the following inclusion-exclusion formula for isolating $f^T(i,T)$, in terms of $f$:
\begin{equation*}
q_T f^T(i,T) = f(i,\{i,j,k,\ell\}) - f(i,\{i,j\}) - f(i,\{i,k\}) - f(i,\{i,\ell\}) + 2 f(i,\{i\})
\end{equation*}
The corresponding coefficients are given by,
\begin{equation*}
n_T(\{i,j,k,\ell\}) = 1 \quad\quad n_T(\{i,j\}) = n_T(\{i,k\}) = n_T(\{i,\ell\}) = -1 \quad\quad n_T(\{i\}) = 2
\end{equation*}}
\end{example}

\vfill

\vfill
\begin{lemma}
\label{lemma:4.3}
If $f$ is a budget-balanced distribution rule that guarantees the existence of an equilibrium in all games $G\in\mathcal{G}(N,f,W)$, then, for each basis distribution rule $f^T$ defined in (\ref{eq:lemma:4:2}), there exists a weight system $\omega^T$, such that,

\vspace{-0.15in}
\begin{equation}
\label{eq:lemma:4:gwsv}
f^T=f^T_{GWSV}[\omega^T]
\end{equation}
\end{lemma}

\vfill
\proof{Proof.} First, we show that each basis distribution rule $f^T$ is nonnegative. The proof is by contradiction, and proceeds as follows. Assume to the contrary, that $f^T(k,T)<0$ for some $k\in T$. From (\ref{eq:lemma4.1}), this is possible only if $|T|\geq 2$, and it follows that we can pick $i,j\in T$ such that $q_T f^T(i,T) < 0$ and $q_T f^T(j,T) > 0$.

\vfill
\textit{Counterexample 4.} Our goal is to exploit the inequalities $q_T f^T(i,T) < 0$ and $q_T f^T(j,T) > 0$ to build a counterexample that mimics \textit{Counterexample 3} illustrated in Figure \ref{fig:4}, leading to a similar best-response cycle involving just players $i$ and $j$. The inclusion-exclusion principle (Corollary \ref{corollary:4.2.5}) suggests the following technique for achieving precisely this. Consider the game in Figure \ref{fig:4.5}, which is nearly identical to \textit{Counterexample 3}, except that resources $r_1$ and $r_2$ are absent. The resources in the top half are added as follows.

\vfill
\begin{enumerate}[label=(\roman*)]
\item Add resources in $R^+_1=\left\{r^{T'}_1:T'\in\mathcal{T}(T)\text{ and }n_T(T')>0\right\}$ to the top right box.

\vfill
\item Add resources in $R^-_1=\left\{r^{T'}_1:T'\in\mathcal{T}(T)\text{ and }n_T(T')<0\right\}$ to the top left box.
\end{enumerate}

\vfill
\noindent Then, the bottom half is symmetrically filled up as follows.

\vfill
\begin{enumerate}[label=(\roman*)]
\item Add resources in $R^+_2=\left\{r^{T'}_2:T'\in\mathcal{T}(T)\text{ and }n_T(T')>0\right\}$ to the bottom left box.

\vfill
\item Add resources in $R^-_2=\left\{r^{T'}_2:T'\in\mathcal{T}(T)\text{ and }n_T(T')<0\right\}$ to the bottom right box.
\end{enumerate}

\vfill
\noindent The resource set $R$ is therefore given by,

\vspace{-0.1in}
\begin{equation*}
R=R^+_1\cup R^-_1\cup R^+_2\cup R^-_1
\end{equation*}
The local resource coefficients are given by,

\vspace{-0.1in}
\begin{equation*}
(\forall T'\in\mathcal{T}(T))\quad v_{r^{T'}_1}=v_{r^{T'}_2}=|n_T(T')|
\end{equation*}
For each $T'\in\mathcal{T}(T)$, in resources $r^{T'}_1$ and $r^{T'}_2$, we fix players in $T'-\{i,j\}$. Effectively, all players other than $i$ and $j$ have a fixed action in their action set, determined by these fixtures.

\newpage
\noindent The action set of player $i$ is given by $\mathcal{A}_i = \left\{T,B\right\}$, where,

\vspace{-0.1in}
\begin{equation*}
T = \left\{r^{T'}_1 \in R^-_1 : i\in T'\right\}\cup \left\{r^{T'}_1 \in R^+_1 : i\in T'\right\}
\end{equation*}

\vspace{-0.1in}
\begin{equation*}
B = \left\{r^{T'}_2 \in R^+_2 : i\in T'\right\}\cup \left\{r^{T'}_2 \in R^-_2 : i\in T'\right\}
\end{equation*}
The action set of player $j$ is given by $\mathcal{A}_j = \left\{L,R\right\}$, where,

\vspace{-0.1in}
\begin{equation*}
L = \left\{r^{T'}_1 \in R^-_1 : j\in T'\right\}\cup \left\{r^{T'}_2 \in R^+_2 : j\in T'\right\}
\end{equation*}

\vspace{-0.1in}
\begin{equation*}
R = \left\{r^{T'}_2 \in R^+_1 : j\in T'\right\}\cup \left\{r^{T'}_1 \in R^-_2 : j\in T'\right\}
\end{equation*}
This is essentially a game between players $i$ and $j$. The set of joint action profiles can therefore be represented as $\mathcal{A}=\left\{TL,TR,BL,BR\right\}$.

\begin{figure}[t]
    \FIGURE
    {\includegraphics*[height=5cm,width=6.5cm]{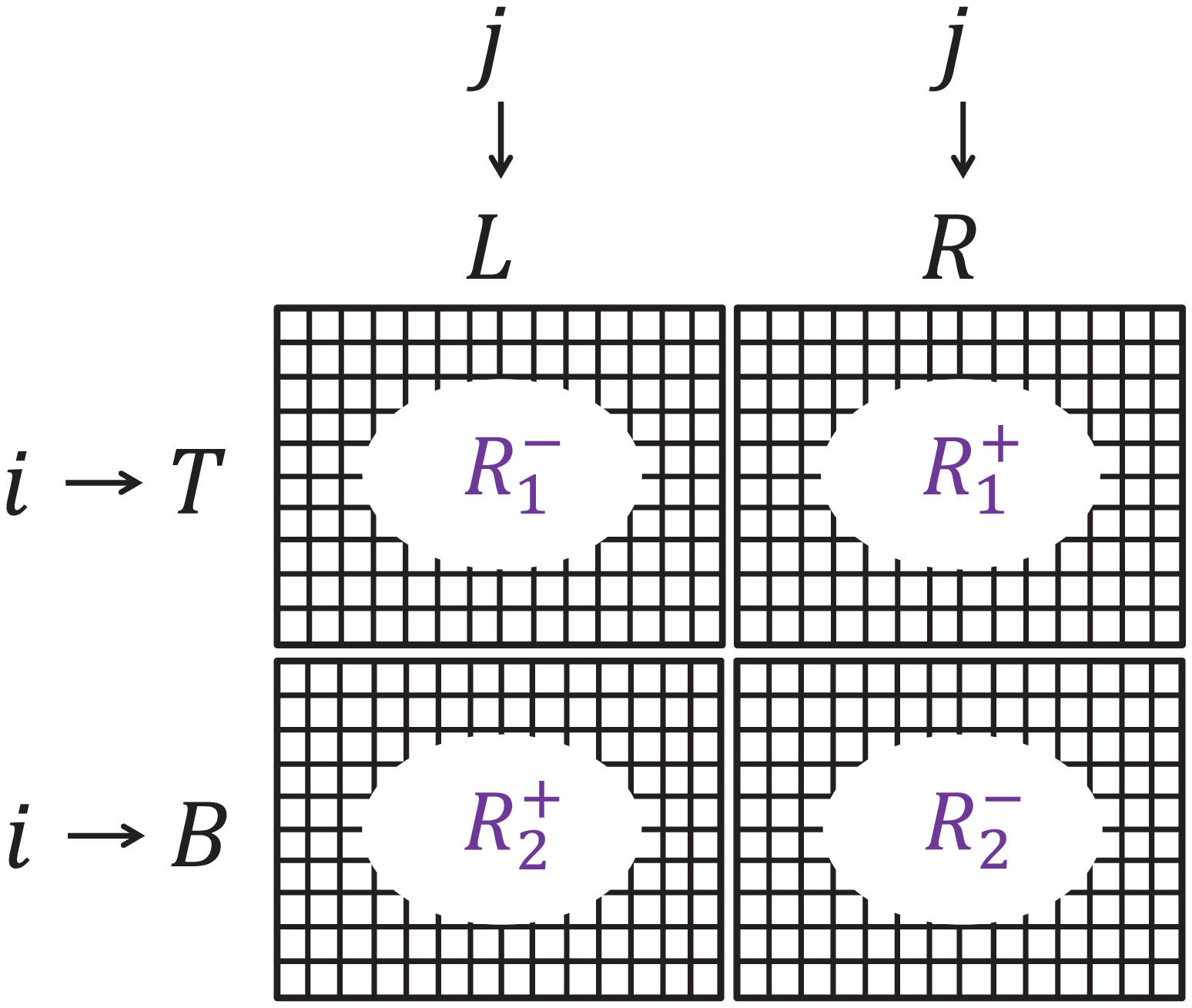}}
    {Counterexample 4\label{fig:4.5}}
    {}
\vspace{-0.1in}
\end{figure}

We use the inclusion-exclusion principle (Corollary \ref{corollary:4.2.5}) to show that Counterexample 4 does not possess an equilibrium, thereby contradicting the fact that $f$ guarantees the existence of an equilibrium in all games $G\in\mathcal{G}(N,f,W)$. We show this for each outcome:
\begin{enumerate}[label=(\roman*)]
\item $TL$ is not an equilibrium, because player $j$ has an incentive to deviate from $L$ to $R$. To see this, consider the utilities of player $j$ when choosing $L$ and $R$,

    \vspace{-0.1in}
    \begin{equation*}
    U_j(T,L)=-\sum_{\substack{T'\in\mathcal{T}(T)\mcr n_T(T')<0}}n_T(T')f(j,T') + \sum_{\substack{T'\in\mathcal{T}(T)\mcr n_T(T')>0}}n_T(T')f(j,T'-\{i\})
    \end{equation*}

    \vspace{-0.05in}
    \begin{equation*}
    U_j(T,R)=\sum_{\substack{T'\in\mathcal{T}(T)\mcr n_T(T')>0}}n_T(T')f(j,T') - \sum_{\substack{T'\in\mathcal{T}(T)\mcr n_T(T')<0}}n_T(T')f(j,T'-\{i\})
    \end{equation*}
    The difference in utilities for $j$ between choosing $R$ and $L$ is therefore given by,

    \vspace{-0.15in}
    \begin{equation*}
    \begin{split}
    U_j(T,R)-U_j(T,L) &= \sum_{T'\in\mathcal{T}(T)}n_T(T')f(j,T') - \sum_{T'\in\mathcal{T}(T)}n_T(T')f(j,T'-\{i\})\mcr
    &= \sum_{T'\in\mathcal{T}(T)}n_T(T')f(j,T')\quad\quad\quad\text{(from (\ref{eq:inclexcl_1}))}\mcr
    &= q_T f^T(j,T) > 0\quad\quad\quad\quad\quad\quad\text{(from (\ref{eq:inclexcl}))}
    \end{split}
    \end{equation*}

\vspace{-0.025in}
\item $TR$ is not an equilibrium, because player $i$ has an incentive to deviate from $T$ to $B$. The proof resembles the previous case. By using similar arguments, we get,

    \vspace{-0.125in}
    \begin{equation*}
    U_i(B,R)-U_i(T,R) = q_T f^T(i,T) < 0
    \end{equation*}

\vspace{-0.025in}
\item $BR$ and $BL$ are also not equilibria, because in these action profiles, players $j$ and $i$ respectively have incentives to deviate -- the arguments are identical to cases (a) and (b) above, respectively.
\end{enumerate}
This completes the inductive argument.

\newpage

Now, since each $f^T$ is nonnegative, budget-balanced, and satisfies Lemma \ref{lemma:4.0}, it is completely specified by $|T|$ nonnegative values, $\{f^T(i,T):i\in T\}$, that sum to 1. Let $\Sigma^T=\left(S_1^T,S_2^T\right)$ be an ordered partition of $T$, where $S_1^T=\left\{i\ :\ f^T(i,T)>0\right\}$, and $S_2^T=T-S_1^T$. Define a weight vector $\boldsymbol{\lambda}^T$ as follows:

\vspace{-0.25in}
\begin{equation*}
\lambda_i^T=\begin{cases}
f^T(i,T) & ,\ i\in S_1^T\mcr
\text{arbitrary positive value} & ,\ i\in S_2^T
\end{cases}
\end{equation*}
Then, it follows that $f^T$ satisfies (\ref{eq:lemma:4:gwsv}) with weight system $\omega^T=\left(\boldsymbol{\lambda}^T,\Sigma^T\right)$ constructed above. (See (\ref{eq:basisGWSV}) in Table \ref{table:basisdistr} to recall the definition of the generalized weighted Shapley value distribution rule.) This completes the proof.\hfill\Halmos
\endproof

\vspace{0.05in}
\begin{example}
\label{example:5}
\textit{Consider the decomposition of $f_{SV}$ and $f_{EQ}$ into their respective basis distribution rules, from Example \ref{example:3}. Clearly, $f^T_{SV}$ is nonnegative for all $T\in\mathcal{T}$, whereas $f^T_{EQ}$ is not.}
\end{example}

\vspace{-0.05in}
\subsection{Consistency of basis distribution rules.}

It follows from Proposition \ref{prop:3}, that each budget-balanced distribution rule $f^W\in f^\mathbb{W}$ that guarantees the existence of an equilibrium in all games $G\in \mathcal{G}(N,f^\mathbb{W},\mathbb{W})$ is completely specified by a sequence of weight systems $\Omega^W=\left\{\omega^{W,T}\right\}_{T\in\mathcal{T}^W}$. But, these weight systems could be `inconsistent' across different coalitions and across different welfare functions. Thus, our next steps focus on proving that all the weight systems $\omega^{T,W}$ are consistent -- in other words, there exists a universal weight system $\omega^*$ that is equivalent to all the $\omega^{W,T}$ (replacing $\omega^{W,T}$ with $\omega^*$ for any coalition $T\in\mathcal{T}_W$ for any $W\in\mathbb{W}$ does not affect the distribution rule $f^{W,T}=f_{GWSV}^T[\omega^{W,T}]$).

To address the consistency of $\omega^{W,T}$ across different coalitions $T$ under the same welfare function $W$, it is sufficient to work with one welfare function at a time, just like in the previous module. However, to address the consistency across different welfare functions, it is necessary to work with more than one welfare function at a time -- for every subset of welfare functions $\mathbb{V}\subseteq\mathbb{W}$, we only focus on the corresponding distribution rules $f^\mathbb{V}$ that guarantee equilibrium existence for all games in the class $\mathcal{G}(N,f^\mathbb{V},\mathbb{V})$. The justification is similar -- $\mathcal{G}(N,f^\mathbb{V},\mathbb{V})\subseteq\mathcal{G}(N,f^\mathbb{W},\mathbb{W})$ for all $\mathbb{V}\subseteq\mathbb{W}$, and so, if $f^\mathbb{W}$ guarantees equilibrium existence for all games in $\mathcal{G}(N,f^\mathbb{W},\mathbb{W})$, then for every subset $\mathbb{V}\subseteq\mathbb{W}$, $f^\mathbb{V}\subseteq f^\mathbb{W}$ must guarantee equilibrium existence for all games in $\mathcal{G}(N,f^\mathbb{V},\mathbb{V})$.

In what follows, we work with $k>1$ welfare functions (not necessarily distinct) at a time, say $W_1,W_2,\ldots,W_k$, to address consistency across welfare functions, and then use the special case of $W_1=W_2=\ldots=W_k=W$ to address consistency across coalitions under the same welfare function $W$. In order to simplify notation, we drop $W$ from the superscripts. That is, for $1\leq j\leq k$, we write $f^j$ instead of $f^{W_j}$, $\mathcal{T}^j$ instead of $\mathcal{T}^{W_j}$, $q^j_T$ instead of $q^{W_j}_T$, $n^j_T(T')$ instead of $n^{W_j}_T(T')$, etc.

\vspace{-0.05in}
\subsubsection{Two consistency conditions.}\label{s.consistency}
Our goal in this section is to establish the following two important consistency properties that the basis distribution rules $f^{W,T}$ must satisfy, in order for the budget-balanced distribution rules $f^W=\sum_{T\in\mathcal{T}^W}q^W_T f^{W,T}$ to guarantee the existence of an equilibrium in all games $G\in \mathcal{G}(N,f^\mathbb{W},\mathbb{W})$. Recall that $\mathcal{T}^W_{ij}$, defined in (\ref{eq:notation:3}), refers to the set of coalitions in $\mathcal{T}^W$ containing both players $i$ and $j$. In addition, let $\mathcal{T}^W_{ij}(S)=\left\{T\in\mathcal{T}^W(S)\ |\ \{i,j\}\subseteq T\right\}$ denote the set of coalitions in $\mathcal{T}^W(S)$ containing both players $i$ and $j$.
\begin{enumerate}[label=(\alph*)]
\item \textit{Global consistency:} If there is a pair of players common to two coalitions (under the same or different welfare functions), then their shares from these two coalitions (given by the corresponding $f^{W,T}$ values) must be `consistent', as formalized in Lemma \ref{lemma:5}. Here, we deal with at most two welfare functions at a time.
\item \textit{Cyclic consistency:} If there is a sequence of $k\geq 3$ players, $(i_1,i_2,\ldots,i_k)$ such that for each of the $k$ neighbor-pairs $\left\{\left(i_1,i_2\right),\left(i_2,i_3\right),\ldots,\left(i_k,i_1\right)\right\}$, $\exists\ T_1 \in \left(\mathcal{T}^1_{i_1i_2}\right)^{\min}, T_2 \in \left(\mathcal{T}^2_{i_2i_3}\right)^{\min}, \ldots, T_k \in \left(\mathcal{T}^k_{i_ki_1}\right)^{\min}$ and in each $T_j$, at least one of the neighbors $i_j,i_{j+1}$ gets a nonzero share (given by the corresponding $f^{j,T_j}$ value), then the shares of these $k$ players from these $k$ coalitions must satisfy a `cyclic consistency' condition, as formalized in Lemma \ref{lemma:6}. Here, we deal with an arbitrary number of welfare functions at a time.
\end{enumerate}

\newpage

\begin{lemma}
\label{lemma:5}
Given any two local welfare functions $W_1,W_2$, if $f^1=\DS\sum_{T\in\mathcal{T}^1}q^1_T f^{1,T}$ and $f^2=\DS\sum_{T\in\mathcal{T}^2}q^2_T f^{2,T}$ are corresponding budget-balanced distribution rules that guarantee equilibrium existence in all games $G\in\mathcal{G}(N,\{f^1,f^2\},\{W_1,W_2\})$, then, for any two players $i,j\in N$, any two coalitions $T'\in\mathcal{T}^1_{ij}$ and $T\in\mathcal{T}^2_{ij}$,

\vspace{-0.125in}
\begin{equation}
\label{eq:lemma:5}
f^{1,T'}(i,T')f^{2,T}(j,T)=f^{2,T}(i,T)f^{1,T'}(j,T')
\end{equation}
\end{lemma}

\vspace{0.15in}
\proof{Proof.} Note that it is sufficient to show (\ref{eq:lemma:5}) for only those coalitions in $\mathcal{T}^1_{ij}$ and $\mathcal{T}^2_{ij}$ in which at least one among $i$ and $j$ get a nonzero share. Formally, define the collections $\mathcal{T}_{ij}^{1+}$ and $\mathcal{T}_{ij}^{2+}$ as,

\vspace{-0.125in}
\begin{equation}
\label{eq:notation:5}
\begin{split}
\mathcal{T}_{ij}^{1+} &= \left\{T\in\mathcal{T}^1_{ij}\ :\ f^{1,T}(i,T)>0\ \mbox{or}\ f^{1,T}(j,T)>0\right\}\\
\mathcal{T}_{ij}^{2+} &= \left\{T\in\mathcal{T}^2_{ij}\ :\ f^{2,T}(i,T)>0\ \mbox{or}\ f^{2,T}(j,T)>0\right\}
\end{split}
\end{equation}
Let $S$ be a minimal element (coalition) in the poset $\left(\mathcal{T}_{ij}^{1+},\subseteq\right)$, and without loss of generality, assume $f^{1,S}(i,S)>0$. Then, we need only show\footnote{It can be shown that (\ref{eq:lemma:5:1}) implies (\ref{eq:lemma:5}): For the special case when $W_2=W_1$, (\ref{eq:lemma:5:1}) implies that for all $T'\in\mathcal{T}_{ij}^{1+}$,

\vspace{-0.075in}
\begin{equation}
\label{eq:lemma:5:2}
f^{1,T'}(i,T')f^{1,S}(j,S)=f^{1,S}(i,S)f^{1,T'}(j,T')
\end{equation}
Let $T'\in\mathcal{T}^1_{ij}$ and $T\in\mathcal{T}^2_{ij}$. By assumption, $f^{1,S}(i,S)\not=0$. If $f^{1,S}(j,S)=0$, then (\ref{eq:lemma:5:1}) and (\ref{eq:lemma:5:2}) imply that $f^{2,T}(j,T)=0$ and $f^{1,T'}(j,T')=0$, in which case both sides of (\ref{eq:lemma:5}) are zero. If $f^{1,S}(j,S)\not=0$, then none of the four terms in equations (\ref{eq:lemma:5:1}) and (\ref{eq:lemma:5:2}) are zero, and therefore, by eliminating $f^{1,S}(i,S)$ and $f^{1,S}(j,S)$ between them, we get (\ref{eq:lemma:5}).} that for any coalition $T\in\mathcal{T}_{ij}^{2+}$,

\vspace{-0.05in}
\begin{equation}
\label{eq:lemma:5:1}
f^{2,T}(i,T)f^{1,S}(j,S)=f^{1,S}(i,S)f^{2,T}(j,T)
\end{equation}

The proof is by contradiction. Assume to the contrary, that for some $T\in\mathcal{T}_{ij}^{2+}$, $f^{2,T}(i,T)f^{1,S}(j,S)\not=f^{1,S}(i,S)f^{2,T}(j,T)$. We consider the following two cases:

\vspace{0.1in}
\noindent\textbf{Case 1:} $q^1_S q^2_T > 0$.

\vspace{-0.2in}
\begin{figure}[ht]
\FIGURE
{\includegraphics*[height=7cm,width=6.75cm]{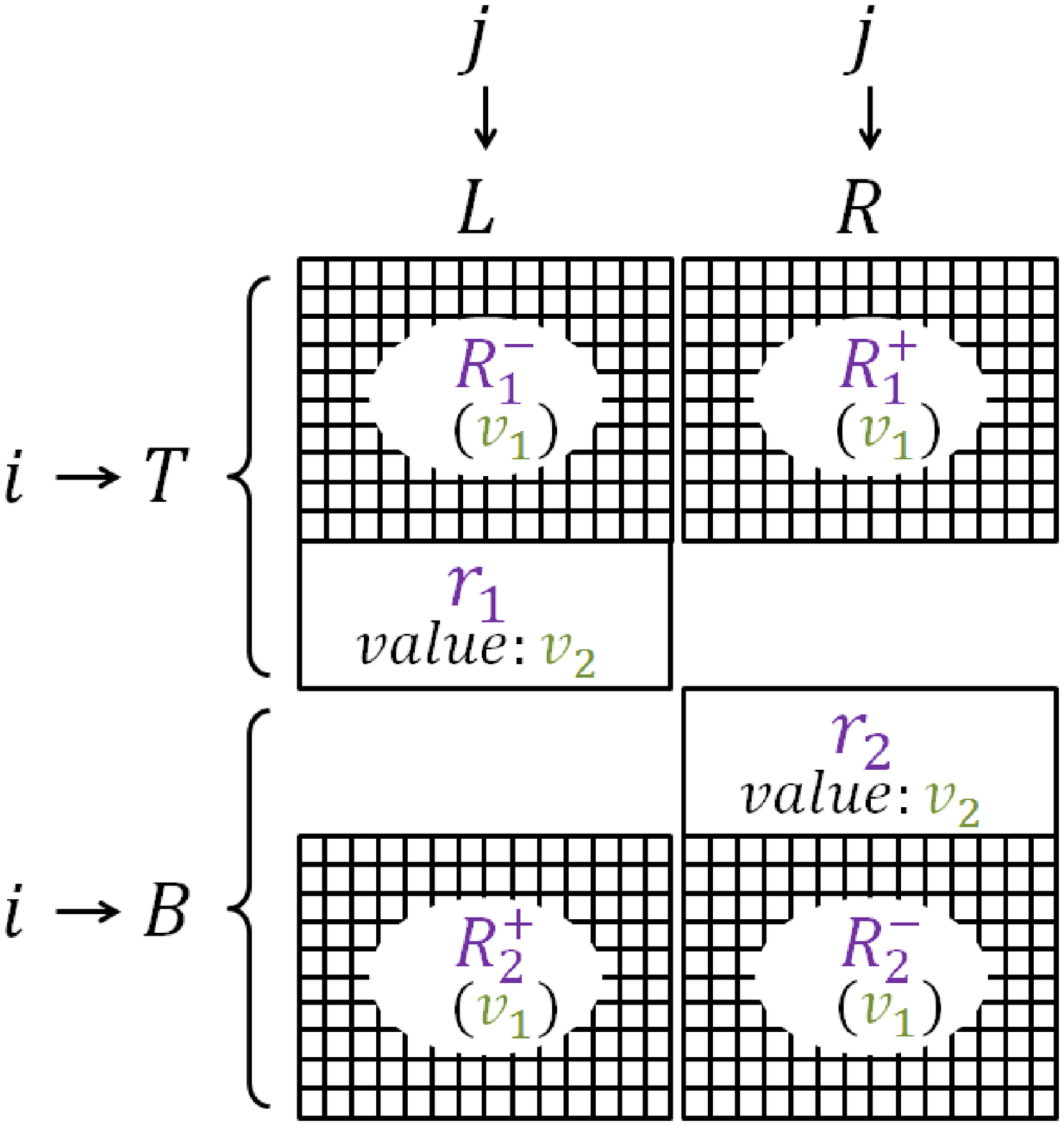}}
{Counterexample 5(a)\label{fig:5a}}
{}
\vspace{-0.15in}
\end{figure}

\textit{Counterexample 5(a).} Our goal is to build a counterexample that mimics \textit{Counterexample 4} illustrated in Figure \ref{fig:4.5}, leading to a similar best-response cycle involving just players $i$ and $j$. As before, we use the inclusion-exclusion principle (Corollary \ref{corollary:4.2.5}) to isolate just $f^{2,T}$, by appropriately adding resources and setting action sets. Consider the game in Figure \ref{fig:5a}, which is identical to \textit{Counterexample 4}, except for the following changes:

\newpage
\begin{enumerate}[label=(\roman*)]
\item There are two additional resources, $r_1$ and $r_2$, so the resource set is now
    \begin{equation*}
    R=\left\{r_1,r_2\right\}\cup R^+_1\cup R^-_1\cup R^+_2\cup R^-_1
    \end{equation*}
\item The welfare function at $r_1$ and $r_2$ is $W_1$. At all other resources, the welfare function is $W_2$.
\item The local resource coefficients are given by,
    \begin{equation*}
    v_{r_1}=v_{r_2}=v_2\quad\quad\text{and}\quad\quad (\forall T'\in\mathcal{T}^2(T))\quad v_{r^{T'}_1}=v_{r^{T'}_2}=v_1|n^2_T(T')|
    \end{equation*}
    where $v_1>0$ and $v_2>0$. We will discuss the specific choice of $v_1,v_2$ later.
\item In resources $r_1$ and $r_2$, we fix players in $S-\{i,j\}$.
\item The actions $T,B,L,R$ are modified to accommodate the two new resources:
    \begin{equation*}
    \begin{split}
    T &= \left\{r^{T'}_1 \in R^-_1 : i\in T'\right\}\cup \left\{r_1\right\}\cup \left\{r^{T'}_1 \in R^+_1 : i\in T'\right\}\mcr
    B &= \left\{r^{T'}_2 \in R^+_2 : i\in T'\right\}\cup \left\{r_2\right\}\cup \left\{r^{T'}_2 \in R^-_2 : i\in T'\right\}\mcr
    L &= \left\{r^{T'}_1 \in R^-_1 : j\in T'\right\}\cup \left\{r_1\right\}\cup \left\{r^{T'}_2 \in R^+_2 : j\in T'\right\}\mcr
    R &= \left\{r^{T'}_1 \in R^+_1 : j\in T'\right\}\cup \left\{r_2\right\}\cup \left\{r^{T'}_2 \in R^-_2 : j\in T'\right\}
    \end{split}
    \end{equation*}
\end{enumerate}

To complete the specification of \textit{Counterexample 5(a)}, we need to specify the values of $v_1>0$ and $v_2>0$. We now show that if $f^{2,T}(i,T)f^{1,S}(j,S)\not=f^{1,S}(i,S)f^{2,T}(j,T)$, then these values can be picked carefully in such a way that \textit{Counterexample 5(a)} does not possess an equilibrium, thereby contradicting the fact that $f^1$ and $f^2$ guarantee equilibrium existence in all games $G\in\mathcal{G}(N,\{f^1,f^2\},\{W_1,W_2\})$. Consider each of the four outcomes:
\begin{enumerate}[label=(\roman*)]
\item In action profiles $TL$ and $BR$, player $j$ has an incentive to deviate if $U_j(T,R)-U_j(T,L)=U_j(B,L)-U_j(B,R)>0$. This happens if,
    \begin{equation*}
    v_1\left(\sum_{T'\in\mathcal{T}^2(T)}n^2_T(T')f^2(j,T') - \sum_{T'\in\mathcal{T}^2(T)}n^2_T(T')f^2(j,T'-\{i\})\right) - v_2\left(f^1(j,S)-f^1(j,S-\{i\})\right) > 0
    \end{equation*}
    Using the inclusion-exclusion principle (Corollary \ref{corollary:4.2.5}) to simplify the terms in the first bracket, and the basis representation of $f^1$ to simplify the difference in the second bracket, this condition is equivalent to,
    \begin{equation*}
    v_1\left(q^2_T f^{2,T}(j,T)\right) - v_2\left(\sum_{T'\in\mathcal{T}^1_{ij}(S)}q^1_{T'}f^{1,T'}(j,T')\right) > 0
    \end{equation*}
    Since $S$ is minimal in $\mathcal{T}^{1+}_{ij}$, this reduces to,
    \begin{equation}
    \label{eq:lemma:5.1:cond1}
    v_1q^2_Tf^{2,T}(j,T)>v_2q^1_Sf^{1,S}(j,S)
    \end{equation}
\item Similarly, in action profiles $TR$ and $BL$, player $i$ has an incentive to deviate if $U_i(B,R)-U_i(T,R)=U_i(T,L)-U_i(B,L)>0$. This happens if,
    \begin{equation*}
    -v_1\left(\sum_{T'\in\mathcal{T}^2(T)}n^2_T(T')f^2(i,T') - \sum_{T'\in\mathcal{T}^2(T)}n^2_T(T')f^2(i,T'-\{j\})\right) + v_2\left(f^1(i,S)-f^1(i,S-\{j\})\right) > 0
    \end{equation*}
    By similar arguments as above, this condition reduces to,
    \begin{equation}
    \label{eq:lemma:5.1:cond2}
    v_1q^2_Tf^{2,T}(i,T)<v_2q^1_Sf^{1,S}(i,S)
    \end{equation}
\end{enumerate}
Without loss of generality, assume $q^2_T>0$ and $q^1_S>0$ (For the symmetric case when $q^2_T<0$ and $q^1_S<0$, the same arguments apply, but the deviations in the best-response cycle are reversed). By assumption, $f^{1,S}(i,S)>0$. Now we consider two cases for $f^{1,S}(j,S)$:
\begin{enumerate}[label=(\roman*)]
\item $f^{1,S}(j,S)=0$. In this case, $f^{2,T}(j,T)>0$ (for otherwise, (\ref{eq:lemma:5:1}) would be satisfied). It follows then, that (\ref{eq:lemma:5.1:cond1}) and (\ref{eq:lemma:5.1:cond2}) always have a solution in strictly positive integers $v_1$ and $v_2$.
\item $f^{1,S}(j,S)>0$. In this case, suppose $f^{2,T}(i,T)f^{1,S}(j,S) < f^{1,S}(i,S)f^{2,T}(j,T)$ (For the other case when $f^{2,T}(i,T)f^{1,S}(j,S) > f^{1,S}(i,S)f^{2,T}(j,T)$, the same arguments apply, but the deviations in the best-response cycle are reversed). Combining (\ref{eq:lemma:5.1:cond1}) and (\ref{eq:lemma:5.1:cond2}), we get,

    \vspace{-0.05in}
    \begin{equation*}
    \frac{q^2_T}{q^1_S}\frac{f^{2,T}(i,T)}{f^{1,S}(i,S)} < \frac{v_2}{v_1} < \frac{q^2_T}{q^1_S}\frac{f^{2,T}(j,T)}{f^{1,S}(j,S)}
    \end{equation*}

    \vspace{-0.025in}
    \noindent This inequality has a solution in strictly positive integers $v_1$ and $v_2$, if and only if,

    \vspace{-0.075in}
    \begin{equation*}
    \begin{split}
    &\frac{q^2_T}{q^1_S}\frac{f^{2,T}(i,T)}{f^{1,S}(i,S)} < \frac{q^2_T}{q^1_S}\frac{f^{2,T}(j,T)}{f^{1,S}(j,S)}\mcr
    \Longleftrightarrow\quad &f^{2,T}(i,T)f^{1,S}(j,S) < f^{1,S}(i,S)f^{2,T}(j,T)
    \end{split}
    \end{equation*}
    which is true by assumption.
\end{enumerate}

\vspace{0.05in}
\noindent\textbf{Case 2:} $q^1_S q^2_T<0$.

\vspace{-0.2in}
\begin{figure}[ht]
\FIGURE
{\includegraphics*[height=11.25cm,width=10cm]{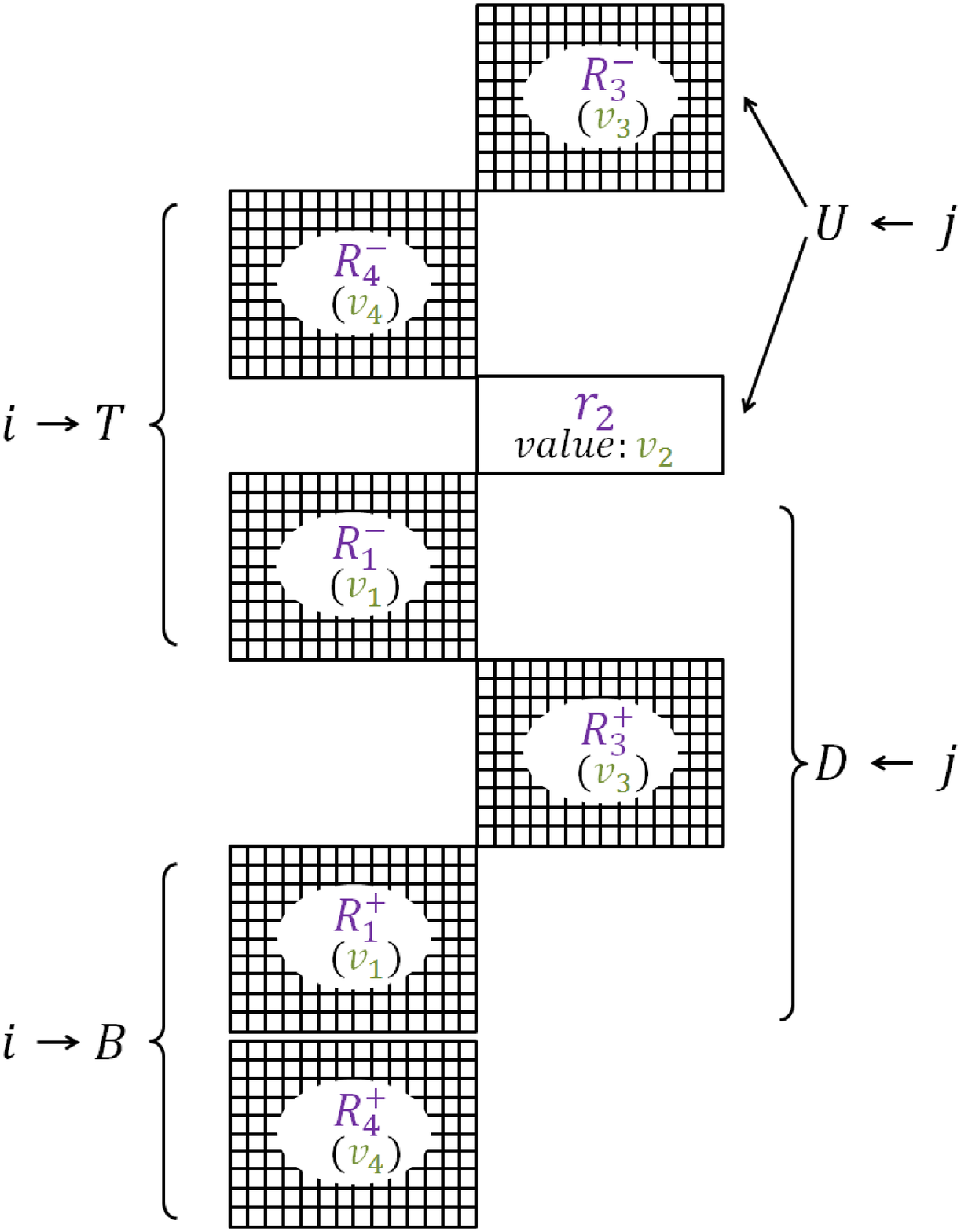}}
{Counterexample 5(b)\label{fig:5b}}
{}
\vspace{-0.15in}
\end{figure}

\textit{Counterexample 5(b).} Our goal remains the same -- to build a counterexample in which a best-response cycle involving just players $i$ and $j$ exists. This counterexample breaks from symmetry, and we use the inclusion-exclusion principle (Corollary \ref{corollary:4.2.5}) thrice here, to isolate two more basis distribution rules, in addition to $f^{2,T}$. We now present the formal details.

\newpage
Consider the game in Figure \ref{fig:5b}, where we have various boxes with labels on them indicating which resource or set of resources is present. Let $T_i$ and $T_j$ be some coalitions that contain $i$ and $j$ respectively. We will discuss the specific choice of $T_i,T_j$ later. As before, we use the resource sets $\left(R^+_1,R^-_1\right)$ (with $W_2$ the welfare function at all these resources) for isolating $f^{2,T}$. In addition, we use resource sets $\left(R^+_3,R^-_3\right)$ (with $W_x$ as the welfare function at all these resources) and $\left(R^+_4,R^-_4\right)$ (with $W_y$ as the welfare function at all these resources) to isolate two more basis distribution rules, $f^{x,T_j}$ and $f^{y,T_i}$, respectively, where the choice of $x,y\in\{1,2\}$ will be discussed later. In addition to these six sets, we also have a single resource $r_2$ whose welfare function is $W_1$. Formally,
\begin{equation*}
\begin{split}
R^+_1=\left\{r^{T'}_1:T'\in\mathcal{T}^2(T)\text{ and }n^2_T(T')>0\right\}\quad\quad\quad\quad\quad\quad\quad\quad\quad\quad\mcr
R^-_1=\left\{r^{T'}_1:T'\in\mathcal{T}^2(T)\text{ and }n^2_T(T')<0\right\}\quad\quad\quad\quad\quad\quad\quad\quad\quad\quad\mcr
R^+_3=\left\{r^{T'}_3:T'\in\mathcal{T}^x(T_j)\text{ and }n^x_{T_j}(T')>0\right\}\quad\quad R^+_4=\left\{r^{T'}_4:T'\in\mathcal{T}^y(T_i)\text{ and }n^y_{T_i}(T')>0\right\}\mcr
R^-_3=\left\{r^{T'}_3:T'\in\mathcal{T}^x(T_j)\text{ and }n^x_{T_j}(T')<0\right\}\quad\quad R^-_4=\left\{r^{T'}_4:T'\in\mathcal{T}^y(T_i)\text{ and }n^y_{T_i}(T')<0\right\}
\end{split}
\end{equation*}
The resource set $R$ is therefore given by,
\begin{equation*}
R=\left\{r_2\right\}\cup R^+_1\cup R^-_1\cup R^+_3\cup R^-_3\cup R^+_4\cup R^-_4
\end{equation*}
The local resource coefficients are given by,
\begin{equation*}
\begin{split}
v_{r_2}&=v_2\\
(\forall T'\in\mathcal{T}^2(T))\quad v_{r^{T'}_1}&=v_1|n^2_T(T')|\mcr
(\forall T'\in\mathcal{T}^x(T_j))\quad v_{r^{T'}_3}=v_3|n^x_{T_j}(T')|\quad\quad&\text{and}\quad\quad(\forall T'\in\mathcal{T}^y(T_i))\quad v_{r^{T'}_4}=v_4|n^y_{T_i}(T')|
\end{split}
\end{equation*}
where $v_1,v_2,v_3,v_4>0$. We will discuss the specific choice of $v_1, v_2, v_3, v_4$ later. In resource $r_2$, we fix players in $S-\{i,j\}$. For each $T'\in\mathcal{T}^2(T)$, in resource $r^{T'}_1$, we fix players in $T'-\{i,j\}$. For each $T'\in\mathcal{T}^x(T_j)$, in resource $r^{T'}_3$, we fix players in $T'-\{j\}$. For each $T'\in\mathcal{T}^y(T_i)$, in resource $r^{T'}_4$, we fix players in $T'-\{i\}$. Effectively, all players other than $i$ and $j$ have a fixed action in their action set, determined by these fixtures. In addition, these fixtures might also specify mandatory sets of resources $R_i$ and $R_j$ that players $i$ and $j$ must always be present in. The action sets of players $i$ and $j$ are given by $\mathcal{A}_i = \left\{T,B\right\}$ and $\mathcal{A}_j = \left\{U,D\right\}$, where,
\begin{equation*}
\begin{split}
T &= \left\{r^{T'}_4 \in R^-_4 : i\in T'\right\}\cup \left\{r_2\right\}\cup \left\{r^{T'}_1 \in R^-_1 : i\in T'\right\}\cup R_i\mcr
B &= \left\{r^{T'}_1 \in R^+_1 : i\in T'\right\}\cup \left\{r^{T'}_4 \in R^+_4 : i\in T'\right\}\cup R_i\mcr
U &= \left\{r^{T'}_1 \in R^-_3 : j\in T'\right\}\cup \left\{r_2\right\}\cup R_j\mcr
D &= \left\{r^{T'}_1 \in R^-_1 : j\in T'\right\}\cup \left\{r^{T'}_1 \in R^+_3 : j\in T'\right\}\cup \left\{r^{T'}_1 \in R^+_1 : j\in T'\right\}\cup R_j
\end{split}
\end{equation*}
This is essentially a game between players $i$ and $j$. The set of joint action profiles can therefore be represented as $\mathcal{A}=\left\{TU,TD,BU,BD\right\}$.

To complete the specification of \textit{Counterexample 5(b)}, we need to specify the values of $v_1,v_2,v_3,v_4>0$, $x,y\in\{1,2\}$, and $T_i,T_j$. We now show that if $f^{2,T}(i,T)f^{1,S}(j,S)\not=f^{1,S}(i,S)f^{2,T}(j,T)$, then these values can be picked carefully in such a way that \textit{Counterexample 5(b)} does not possess an equilibrium, thereby contradicting the fact that $f^1$ and $f^2$ guarantee equilibrium existence in all games $G\in\mathcal{G}(N,\{f^1,f^2\},\{W_1,W_2\})$. Consider each of the four outcomes:
\begin{enumerate}[label=(\roman*)]
\item In action profile $TU$, player $i$ has an incentive to deviate if $U_i(B,U)-U_i(T,U)>0$. This happens if,
    \begin{equation*}
    v_4 \left(\sum_{T'\in\mathcal{T}^y(T_i)}n^y_{T_i}(T')f^y(i,T')\right) + v_1 \left(\sum_{T'\in\mathcal{T}^2(T)}n^2_{T}(T')f^2(i,T'-\{j\})\right) - v_2 f^1(i,S) > 0
    \end{equation*}
    Note that $U_i(B,U)$ and $U_i(T,U)$ include utilities to player $i$ from resources in $R_i$, but while taking the difference, this cancels out, since $i$ is fixed in these resources, and between these two action profiles, all other players also have a fixed action. Now, using the inclusion-exclusion principle (Corollary \ref{corollary:4.2.5}) to simplify the terms in the first two brackets, we get,
    \begin{equation}
    \label{eq:lemma5b1}
    v_4 q^y_{T_i}f^{y,T_i}(i,T_i) >  v_2 f^1(i,S)
    \end{equation}
\item In action profile $BD$, player $i$ has an incentive to deviate if $U_i(T,D)-U_i(B,D)>0$. This happens if:
    \begin{equation*}
    - v_4 \left(\sum_{T'\in\mathcal{T}^y(T_i)}n^y_{T_i}(T')f^y(i,T')\right) - v_1 \left(\sum_{T'\in\mathcal{T}^2(T)}n^2_{T}(T')f^2(i,T')\right) + v_2 f^1(i,S-\{j\}) > 0
    \end{equation*}
    As before, the utility to player $i$ from resources in $R_i$ cancels out. Using the inclusion-exclusion principle to simplify the terms, we get,
    \begin{equation}
    \label{eq:lemma5b2}
    v_2 f^1(i,S-\{j\}) - v_1 q^2_T f^{2,T}(i,T) > v_4 q^y_{T_i}f^{y,T_i}(i,T_i)
    \end{equation}
\item In action profile $TD$, player $j$ has an incentive to deviate if $U_j(T,U)-U_j(T,D)>0$. This happens if:
    \begin{equation*}
    \begin{split}
    - v_3 &\left(\sum_{T'\in\mathcal{T}^x(T_j)}n^x_{T_j}(T')f^x(j,T')\right) + v_2 f^1(j,S)\mcr
    &- v_1\left(\sum_{\substack{T'\in\mathcal{T}^2(T)\mcr n^2_T(T')>0}}n^2_T(T')f^2(j,T'-\{i\})-\sum_{\substack{T'\in\mathcal{T}^2(T)\mcr n^2_T(T')<0}}n^2_T(T')f^2(j,T')\right) > 0
    \end{split}
    \end{equation*}
    The utility to player $j$ from resources in $R_j$ cancels out. Using the inclusion-exclusion principle to simplify the first term, we get,
    \begin{small}
    \begin{equation}
    \label{eq:lemma5b3}
    v_2 f^1(j,S) - v_1\left(\sum_{\substack{T'\in\mathcal{T}^2(T)\mcr n^2_T(T')>0}}n^2_T(T')f^2(j,T'-\{i\})-\sum_{\substack{T'\in\mathcal{T}^2(T)\mcr n^2_T(T')<0}}n^2_T(T')f^2(j,T')\right) > v_3 q^x_{T_j}f^{x,T_j}(j,T_j)
    \end{equation}
    \end{small}
\item In action profile $BU$, player $j$ has an incentive to deviate if $U_j(B,D)-U_j(B,U)>0$. This happens if:
    \begin{equation*}
    \begin{split}
    v_3 &\left(\sum_{T'\in\mathcal{T}^x(T_j)}n^x_{T_j}(T')f^x(j,T')\right) - v_2 f^1(j,S-\{i\})\mcr
    &+ v_1\left(\sum_{\substack{T'\in\mathcal{T}^2(T)\mcr n^2_T(T')>0}}n^2_T(T')f^2(j,T')-\sum_{\substack{T'\in\mathcal{T}^2(T)\mcr n^2_T(T')<0}}n^2_T(T')f^2(j,T'-\{i\})\right) > 0
    \end{split}
    \end{equation*}
    As before, the utility to player $j$ from resources in $R_j$ cancels out. Using the inclusion-exclusion principle to simplify the first term, we get,
    \begin{small}
    \begin{equation}
    \label{eq:lemma5b4}
    v_3 q^x_{T_j}f^{x,T_j}(j,T_j) > v_2 f^1(j,S-\{i\}) - v_1\left(\sum_{\substack{T'\in\mathcal{T}^2(T)\mcr n^2_T(T')>0}}n^2_T(T')f^2(j,T')-\sum_{\substack{T'\in\mathcal{T}^2(T)\mcr n^2_T(T')<0}}n^2_T(T')f^2(j,T'-\{i\})\right)
    \end{equation}
    \end{small}
\end{enumerate}
Combining inequalities (\ref{eq:lemma5b1}) and (\ref{eq:lemma5b2}), we get,
\begin{equation*}
v_1q^2_Tf^{2,T}(i,T) + v_2\left(f^1(i,S)-f^1(i,S-\{j\})\right) < 0
\end{equation*}
Using the basis representation of $f^1$ to simplify the difference in the bracket, this condition is equivalent to,
\begin{equation*}
v_1q^2_Tf^{2,T}(i,T) + v_2\left(\sum_{T'\in\mathcal{T}^1_{ij}(S)}q^1_{T'}f^{1,T'}(i,T')\right) < 0
\end{equation*}
Since $S$ is minimal in $\mathcal{T}^{1+}_{ij}$, this reduces to,
\begin{equation}
\label{eq:lemma:5.2:cond1}
v_1 q^2_T f^{2,T}(i,T) + v_2 q^1_S f^{1,S}(i,S) < 0
\end{equation}
Combining inequalities (\ref{eq:lemma5b3}) and (\ref{eq:lemma5b4}), we get,
\begin{small}
\begin{equation*}
v_1\left(\sum_{T'\in\mathcal{T}^2(T)}n^2_T(T')f^2(j,T') - \sum_{T'\in\mathcal{T}^2(T)}n^2_T(T')f^2(j,T'-\{i\})\right) + v_2\left(f^1(j,S)-f^1(j,S-\{i\})\right) > 0
\end{equation*}
\end{small}
Using the inclusion-exclusion principle to simplify the terms in the first bracket, and the basis representation of $f^1$ to simplify the difference in the second bracket, this condition is equivalent to,
\begin{equation*}
v_1q^2_Tf^{2,T}(j,T) + v_2\left(\sum_{T'\in\mathcal{T}^1_{ij}(S)}q^1_{T'}f^{1,T'}(j,T')\right) > 0
\end{equation*}
Since $S$ is minimal in $\mathcal{T}^{1+}_{ij}$, this reduces to,
\begin{equation}
\label{eq:lemma:5.2:cond2}
v_1 q^2_T f^{2,T}(j,T) + v_2 q^1_S f^{1,S}(j,S) > 0
\end{equation}
Without loss of generality, assume $q^2_T>0$ and $q^1_S<0$ (For the symmetric case when $q^2_T<0$ and $q^1_S>0$, the same arguments apply, but the deviations in the best-response cycle are reversed). By assumption, $f^{1,S}(i,S)>0$. Now we consider two cases for $f^{1,S}(j,S)$:
\begin{enumerate}[label=(\roman*)]
\item $f^{1,S}(j,S)= 0$. In this case, $f^{2,T}(j,T)>0$ (for otherwise, (\ref{eq:lemma:5:1}) would be satisfied). It follows then, that (\ref{eq:lemma:5.2:cond1}) and (\ref{eq:lemma:5.2:cond2}) always have a solution in strictly positive integers $v_1$ and $v_2$.
\item $f^{1,S}(j,S)> 0$. In this case, suppose $f^{2,T}(i,T)f^{1,S}(j,S) < f^{1,S}(i,S)f^{2,T}(j,T)$ (For the other case when $f^{2,T}(i,T)f^{1,S}(j,S) > f^{1,S}(i,S)f^{2,T}(j,T)$, the same arguments apply, but the deviations in the best-response cycle are reversed). Combining (\ref{eq:lemma:5.2:cond1}) and (\ref{eq:lemma:5.2:cond2}), we get,
    \begin{equation*}
    -\frac{q^2_T}{q^1_S}\frac{f^{2,T}(i,T)}{f^{1,S}(i,S)} < \frac{v_2}{v_1} < -\frac{q^2_T}{q^1_S}\frac{f^{2,T}(j,T)}{f^{1,S}(j,S)}
    \end{equation*}
    Therefore, this inequality has a solution in strictly positive integers $v_1$ and $v_2$, if and only if,
    \begin{equation*}
    \begin{split}
    &\frac{q^2_T}{q^1_S}\frac{f^{2,T}(i,T)}{f^{1,S}(i,S)} > \frac{q^2_T}{q^1_S}\frac{f^{2,T}(j,T)}{f^{1,S}(j,S)}\mcr
    \Longleftrightarrow\quad &f^{2,T}(i,T)f^{1,S}(j,S) < f^{1,S}(i,S)f^{2,T}(j,T)
    \end{split}
    \end{equation*}
    which is true by assumption.
\end{enumerate}
Finally, we need to show that given these carefully chosen values for $v_1$ and $v_2$, it is possible to find $v_3>0$, $v_4>0$, $x,y\in\{1,2\}$, $T_i$ and $T_j$ such that the inequalities (\ref{eq:lemma5b1})-(\ref{eq:lemma5b4}) are satisfied. These four inequalities can be consolidated as,
\begin{equation*}
\begin{split}
LHS_j<v_3q^x_{T_j}f^{x,T_j}(j,T_j)<RHS_j\mcr
LHS_i<v_4q^y_{T_i}f^{y,T_i}(i,T_i)<RHS_i
\end{split}
\end{equation*}
We describe the procedure to find $v_3>0$, $x\in\{1,2\}$, and $T_j$ here. Finding $v_4>0$, $y\in\{1,2\}$, and $T_i$ is analogous. Specifically, we consider the case where $RHS_j>0$ (we discuss the other case later). Consider the two quantities $f^{1,S}(j,S)$ and $f^{2,T}(j,T)$. They are not both zero (for otherwise, (\ref{eq:lemma:5:1}) would be satisfied). We consider two subcases:
\begin{enumerate}[label=(\roman*)]
\item If $f^{2,T}(j,T)>0$, choose $x=2$, $T_j=T$. Then, it is possible to find $v_3>0$ such that $LHS_j<v_3q^2_{T}f^{2,T}(j,T)<RHS_j$ because by assumption, $q^2_T>0$.
\item If $f^{2,T}(j,T)=0$, then $f^{1,S}(j,S)>0$. Here, we slightly alter \textit{Counterexample 5(b)} by modifying player $j$'s action set $\mathcal{A}_j=\{U,D\}$ as follows -- we simply switch the resources in $R^-_3$ and $R^+_3$ between action $U$ and action $D$. Formally,
    \begin{equation*}
    \begin{split}
    U' &= \left\{r_2\right\}\cup \left\{r^{T'}_1 \in R^+_3 : j\in T'\right\}\cup R_j\mcr
    D' &= \left\{r^{T'}_1 \in R^-_3 : j\in T'\right\}\cup \left\{r^{T'}_1 \in R^-_1 : j\in T'\right\}\cup \left\{r^{T'}_1 \in R^+_1 : j\in T'\right\}\cup R_j
    \end{split}
    \end{equation*}
    This alteration does not affect any of the arguments above, except that we now need to find $v_3>0$, $x\in\{1,2\}$, and $T_j$ such that $LHS_j<-v_3q_{T_j}f^{T_j}(j,T_j)<RHS_j$. (In particular, (\ref{eq:lemma:5.2:cond1}) and (\ref{eq:lemma:5.2:cond2}) remain unchanged.) Choose x=1, $T_j=S$. Then, it is possible to find $v_3>0$ such that $LHS_j<-v_3q^1_{S}f^{1,S}(j,S)<RHS_j$ because by assumption, $q^1_S<0$.
\end{enumerate}
The case when $RHS_j<0$ is symmetric -- we just interchange $S$ and $T$, and the choice of $x$, throughout in the two subcases. That is, if $f^{1,S}(j,S)>0$, we choose $x=1$, $T_j=S$. Otherwise, we choose $x=2$, $T_j=T$, and alter \textit{Counterexample 5(b)} the same way as above. This completes the proof.\footnote{Note that $v_1$ and $v_2$ can be scaled by an arbitrary positive constant if necessary, in order to ensure that an integral solution to $v_3,v_4$ exists.}\hfill\Halmos
\endproof

From Lemma \ref{lemma:4.3}, we inferred that each $f^{W,T}$ is a generalized weighted Shapley value for the corresponding unanimity game $W^T$, with weight system $\omega^{W,T}=\left(\boldsymbol{\lambda}^{W,T},\Sigma^{W,T}\right)$ that we constructed using $f^{W,T}$. As a result, the conditions on $\left\{\left\{f^{W,T}\right\}_{T\in\mathcal{T}^W}\right\}_{W\in\mathbb{W}}$ imposed by Lemma \ref{lemma:5} translate to equivalent conditions on the weight systems $\left\{\left\{\omega^{W,T}\right\}_{T\in\mathcal{T}^W}\right\}_{W\in\mathbb{W}}$. The following corollary restates Lemma \ref{lemma:5} in terms of the weight systems.

\begin{corollary}
\label{corollary:5}
Given any set of local welfare functions $\mathbb{W}$, if $f^\mathbb{W}$ are budget-balanced distribution rules that guarantee equilibrium existence in all games $G\in\mathcal{G}(N,f^\mathbb{W},\mathbb{W})$, where, for each $W\in\mathbb{W}$, $f^W:=\DS\sum_{T\in\mathcal{T}^W}q^W_T f^T_{GWSV}[\omega^{W,T}]$, where $\omega^{W,T}=\left(\boldsymbol{\lambda}^{W,T},\Sigma^{W,T}=\left(S_1^{W,T},S_2^{W,T}\right)\right)$, then, for any two players $i,j\in N$,
\begin{enumerate}[label=(\alph*)]
\item $\left(\exists\ W,T\right)\quad i\in S_1^{W,T},\ j\in S_2^{W,T}\ \Longrightarrow\ \left(\forall\ W',T'\right)\quad j\in S_2^{W',T'}$
\item $\left(\exists\ W,T\right)\quad i\in S_1^{W,T},\ j\in S_1^{W,T}\ \Longrightarrow\ \left(\forall\ W',T'\right)\quad
                      \left\{
                            \begin{array}{l l}
                            \mbox{(i)}\ \ i\in S_1^{W',T'}\Leftrightarrow j\in S_1^{W',T'}\\
                            \mbox{(ii)}\ \{i,j\}\subseteq S_1^{W',T'}\Rightarrow \frac{\lambda_i^{W,T}}{\lambda_i^{W',T'}} = \frac{\lambda_j^{W,T}}{\lambda_j^{W',T'}}
                            \end{array}
                      \right.$
\end{enumerate}
\end{corollary}

\proof{Explanation.} In essence, this corollary states consistency constraints that the various weight systems that define the distribution rules must satisfy, for every pair of players. It is obtained by applying Lemma \ref{lemma:5} for all pairs of welfare functions $\mathbb{W}\times\mathbb{W}$. Suppose $W,W'\in\mathbb{W}$, and let $T\in\mathcal{T}^W_{ij}$ and $T'\in\mathcal{T}^{W'}_{ij}$. Then, from Lemma \ref{lemma:5}, we have,
\begin{equation}
f^{W,T}(i,T)f^{W',T'}(j,T')=f^{W',T'}(i,T')f^{W,T}(j,T)
\end{equation}
The different parts of the corollary then follow directly from applying the definition of the generalized weighted Shapley value (see (\ref{eq:basisGWSV}) in Table \ref{table:basisdistr}) and simplifying the above equation for the corresponding cases.
\endproof

\begin{lemma}
\label{lemma:6}
For any integer $k\geq 3$, given any $k$ welfare functions $W_1,W_2,\ldots,W_k$, if $\left\{f^j:=\DS\sum_{T\in\mathcal{T}^j}q^j_T f^{j,T}\right\}_{j=1}^k$ are corresponding budget-balanced distribution rules that guarantee equilibrium existence in all games $G\in\mathcal{G}(N,\left\{f^1,f^2,\ldots,f^k\right\},\left\{W_1,W_2,\ldots,W_k\right\})$, and $i_1,i_2,\ldots,i_k\in N$ are any $k$ players such that $\exists\ T_1 \in \left(\mathcal{T}_{i_1i_2}^{1+}\right)^{\min}, T_2 \in \left(\mathcal{T}_{i_2i_3}^{2+}\right)^{\min}, \ldots, T_k \in \left(\mathcal{T}_{i_ki_1}^{k+}\right)^{\min}$, then,
\begin{equation}
\label{eq:lemma:6}
f^{1,T_1}(i_1,T_1)f^{2,T_2}(i_2,T_2)\cdots f^{k,T_k}(i_k,T_k) = f^{1,T_1}(i_2,T_1)f^{2,T_2}(i_3,T_2)\cdots f^{k,T_k}(i_1,T_k)
\end{equation}
\end{lemma}

\proof{Proof.} Recall that $\mathcal{T}_{xy}^{j+}$ denotes the collection of those coalitions from $\mathcal{T}_{xy}^j$ in which at least one of $x,y$ obtains a nonzero share (according to $f^j$). Refer to (\ref{eq:notation:5}) for the formal definition.

\vspace{0.05in}
\textit{Index arithmetic:} In the rest of this proof, the index set is $\{1,2,\ldots,k\}$, and when we add an integer $\ell$ to an index $j$, $j+\ell$ denotes the index that is $\ell$ positions away from index $j$ (cycling around if necessary). For example, suppose $k=3$. Then, for index $j=2$, $j+2=1$ and $j-2=3$.

\vspace{0.05in}
For simplicity, denote $q^j_{T_j}$ by $q_j$, $f^{j,T_j}(i_j,T_j)$ by $a_j$, and $f^{j,T_j}(i_{j+1},T_j)$ by $b_j$. Note that $a_j\geq 0$ and $b_j\geq 0$. Now, (\ref{eq:lemma:6}) can be written as:

\vspace{-0.1in}
\begin{equation}
\label{eq:lemma:6'}
\prod_{j=1}^k a_j = \prod_{j=1}^k b_j
\end{equation}
Our proof technique mirrors those in previous sections. Assuming (\ref{eq:lemma:6'}) is not satisfied, we present a game where no equilibrium exists. Without loss of generality, let
\begin{equation}
\label{eq:lemma:6''}
\prod_{j=1}^k a_j < \prod_{j=1}^k b_j
\end{equation}
We present a family of counterexamples, each corresponding to a specific sign profile of the coefficients $q_j$. (Recall that for a given choice of the local welfare functions, $q_j$ are fixed.) The proof is in three stages:
\begin{enumerate}[label=(\roman*)]
\item We present the details of the counterexample game (\textit{Counterexample 6}).
\item We present four \textit{validity conditions} on the action profiles of \textit{Counterexample 6}, and define those action profiles that satisfy at least one of them as \textit{valid action profiles}, observing that such action profiles are never equilibria.
\item We show that every action profile in \textit{Counterexample 6} is valid.
\end{enumerate}


\begin{figure}[ht]
\FIGURE
{\includegraphics*[height=9.5cm,width=9.5cm]{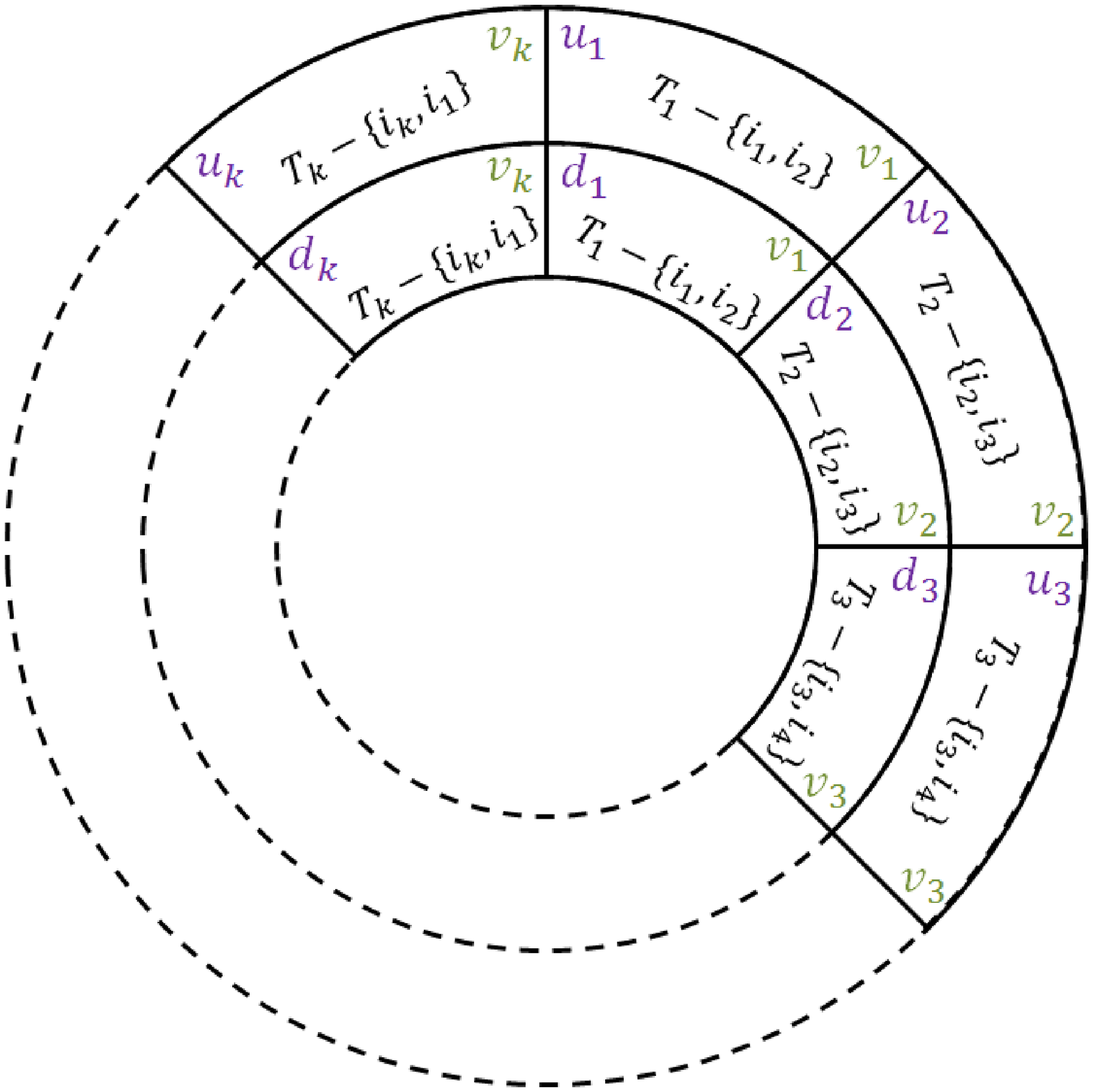}}
{Counterexample 6\label{fig:6}}
{}
\end{figure}

\textit{Counterexample 6.} Consider the game in Figure \ref{fig:6}, that involves only players $i_1,i_2,\ldots,i_k$. There are $2k$ resources, arranged in two circular rows of $k$ resources each. For each column $j$, both resources $u_j$ and $d_j$ share the same resource-specific coefficient $v_j>0$ and same local welfare function $W_j$, and in both these resources, we fix players in $T_j-\{i_j,i_{j+1}\}$. Effectively, all players other than $i_1,i_2,\ldots,i_k$ have a fixed action in their action set, determined by these fixtures. In addition, these fixtures might also specify mandatory sets of resources $R_1, R_2,\ldots,R_k$ that the players $i_1,i_2,\ldots,i_k$ must always be present in, respectively. However, for simplicity, we will not explicitly represent this, since such actions do not affect strategic behavior (utilities from these resources cancel out as far as unilateral deviations are concerned, just like they did in the proof of Lemma \ref{lemma:5}). Next, we need to specify the action sets of the players $i_1,i_2,\ldots,i_k$, which will depend intimately on properties regarding the sequence of signs of the coefficients $q_j$. We begin by relabeling the indices according to a cyclic transformation that is without loss of generality, to ensure that the following properties will be satisfied after the transformation:
\begin{enumerate}[label=(\roman*)]
\item The first coefficient, $q_1$, is negative, unless all coefficients are positive.
\item The last coefficient, $q_k$, is positive, unless all coefficients are negative.
\item The penultimate coefficient, $q_{k-1}$, is positive, unless no two adjacent coefficients are both positive.
\end{enumerate}
In essence, we cut down on the different sequences of signs of the coefficients that we need to consider. Formally, we define two special index sets, $J$ and $J^*$, as follows:
\begin{equation*}
J=\{j\ |\ 1\leq j\leq k \mbox{ and } q_{j-1}>0 \mbox{ and } q_j<0\}\quad\quad\quad J^*=\{j\ |\ j\in J \mbox{ and } q_{j-2}>0\}
\end{equation*}
Now, we define a special index $k^*$ as follows. If $J^*\not=\emptyset$, pick any $k^*\in J^*$. Otherwise, if $J\not=\emptyset$, then pick any $k^*\in J$. If $J=J^*=\emptyset$, set $k^*=1$. Now, we perform a cyclic transformation of the indices that resets $k^*=1$, by rotating Figure \ref{fig:6} counter-clockwise by $k^*-1$ columns. In other words, index $j$ becomes index $j-k^*+1$. In the rest of the proof, we assume that Figure \ref{fig:6} represents the counter-example after this transformation.

Next, we observe that given any profile of the signs of the coefficients $q_j$, the $k$ columns of resources in Figure \ref{fig:6} can be grouped into several segments, each of which can be classified as one the following three kinds:

\vfill
\begin{enumerate}[label=(\roman*)]
\item \textit{$\mathbb{P}_\ell$, a maximal plus segment of length $\ell$:} This segment consists of $\ell>1$ contiguous columns $i,i+1,\ldots,i+\ell-1$ such that $q_j>0$ for $i\leq j\leq i+\ell-1$. In addition, we require maximality, i.e., if $\ell\not=k$, then $q_{i-1}<0$ and $q_{i+\ell}<0$.

\vfill
\item \textit{$\mathbb{M}_\ell$, a maximal minus segment of length $\ell$:} This segment consists of $\ell>1$ contiguous columns $i,i+1,\ldots,i+\ell-1$ such that $q_j<0$ for $i\leq j\leq i+\ell-1$. In addition, we require maximality, i.e., if $\ell\not=k$, then $q_{i-1}>0$ and $q_{i+\ell}>0$.

\vfill
\item \textit{$\mathbb{Z}_\ell$, a maximal alternating minus-plus segment of length $\ell$:} This segment consists of $\ell>1$ contiguous columns ($\ell$ being even) $i,i+1,\ldots,i+\ell-1$ such that $q_j<0$ and $q_{j+1}>0$ for $j\in\{i,i+2,i+4,\ldots,i+\ell-2\}$. In addition, we require maximality, i.e., if $\ell\not=k$, then $q_{i-1}>0\Longrightarrow q_{i-2}>0$ and $q_{i+\ell}<0\Longrightarrow q_{i+\ell+1}<0$. Note that this kind of segment may share its first / last column with a preceding minus / succeeding plus segment.
\end{enumerate}

\vfill

\begin{example}
\label{example:8}
\textit{Consider the sign profile $(-,-,+,-,+,+,-)$. Our special index sets are given by $J=\{4,7\}$, $J^*=\{7\}$. Hence, $k^*=7$, and so, without loss of generality, we transform this sign profile to $(-,-,-,+,-,+,+)$. Now, the first three columns constitute an $\mathbb{M}_3$ segment, and the last two columns constitute a $\mathbb{P}_2$ segment. In between, we have a $(+,-)$ segment that doesn't fit any of our three definitions above. But, if the immediate neighbors on either side are taken into consideration, we have $(-,+,-,+)$, which is a $\mathbb{Z}_4$ segment. So, the (unique) decomposition of this sign profile is given by $\mathbb{M}_3\mathbb{Z}_4\mathbb{P}_2$, where columns $3$ and $6$ are shared between two neighboring segments.}
\end{example}

\vfill

Technically, the above definition of a $\mathbb{Z}_\ell$ segment allows for a spurious $\mathbb{Z}_2$ segment to be sandwiched between an $\mathbb{M}$ and an adjacent $\mathbb{P}$ segment. For example, the sign profile $(-,-,+,+)$ could be decomposed as either $\mathbb{M}_2\mathbb{P}_2$ or $\mathbb{M}_2\mathbb{Z}_2\mathbb{P}_2$. We exclude this possibility by requiring that every $\mathbb{Z}_\ell$ segment have at least one column that is not shared with a neighboring segment. Note that this requirement also guarantees that the above decomposition is always unique.

We now specify the action sets for the players $i_1,i_2,\ldots,i_k$. For any player $i_j$, his actions involve only the resources in adjacent columns $j-1$ and $j$, and specifically, in only one of two ways:

\vfill
\begin{enumerate}[label=(\roman*)]
\item \textit{Straight players} have the following action set:
\begin{equation}
\label{eq:straight}
\mathcal{A}_{i_{j}}=\{(u_{j-1},u_j),(d_{j-1},d_j)\}
\end{equation}
\item \textit{Diagonal players} have the following action set:
\begin{equation}
\label{eq:diagonal}
\mathcal{A}_{i_{j}}=\{(u_{j-1},d_j),(d_{j-1},u_j)\}
\end{equation}
\end{enumerate}

\newpage
\noindent Whether player $i_j$ is straight or diagonal is determined as follows. We consider two cases:

\vfill
\noindent\textbf{Case 1:} $j=1$. Player $i_1$ is straight if and only if one of the following conditions is satisfied:
\begin{enumerate}[label=(\roman*)]
\item $q_1>0$
\item column $1$ is at the beginning of an $\mathbb{M}$ segment and column $k$ is at the end of a $\mathbb{P}$ segment
\end{enumerate}

\vfill
\noindent\textbf{Case 2:} $j\not=1$. Player $i_j$ is straight if and only if one of the following conditions is satisfied:
\begin{enumerate}[label=(\roman*)]
\item column $j$ is at the end of a $\mathbb{P}_{2\ell+1}$, $\mathbb{M}_\ell$ or $\mathbb{Z}_\ell$ segment
\item $q_j<0$, and column $j$ is in a $\mathbb{Z}_\ell$ segment
\item $q_j>0$, $q_{j-1}<0$, and column $j$ is at the beginning of a $\mathbb{P}_\ell$ segment
\item $q_j<0$, $q_{j-1}>0$, $q_{j-2}<0$, and column $j$ is at the beginning of an $\mathbb{M}$ segment
\end{enumerate}

\vfill
\noindent A player is diagonal if and only if he is not straight.

\vfill

\begin{example}
\label{example:9}
\textit{Consider the $\mathbb{M}_3\mathbb{Z}_4\mathbb{P}_2$ sign profile of Example \ref{example:8}, namely, $(-,-,-,+,-,+,+)$. There are seven players, $i_1,\ldots,i_7$. Their action sets are represented pictorially (in blue) in Figure \ref{fig:8}. Formally, players $i_1,i_3,i_5,i_6$ are straight, whereas players $i_2,i_4,i_7$ are diagonal.}
\end{example}

\vspace{-0.125in}
\begin{figure}[ht]
\FIGURE
{\includegraphics*[height=8.5cm,width=8.5cm]{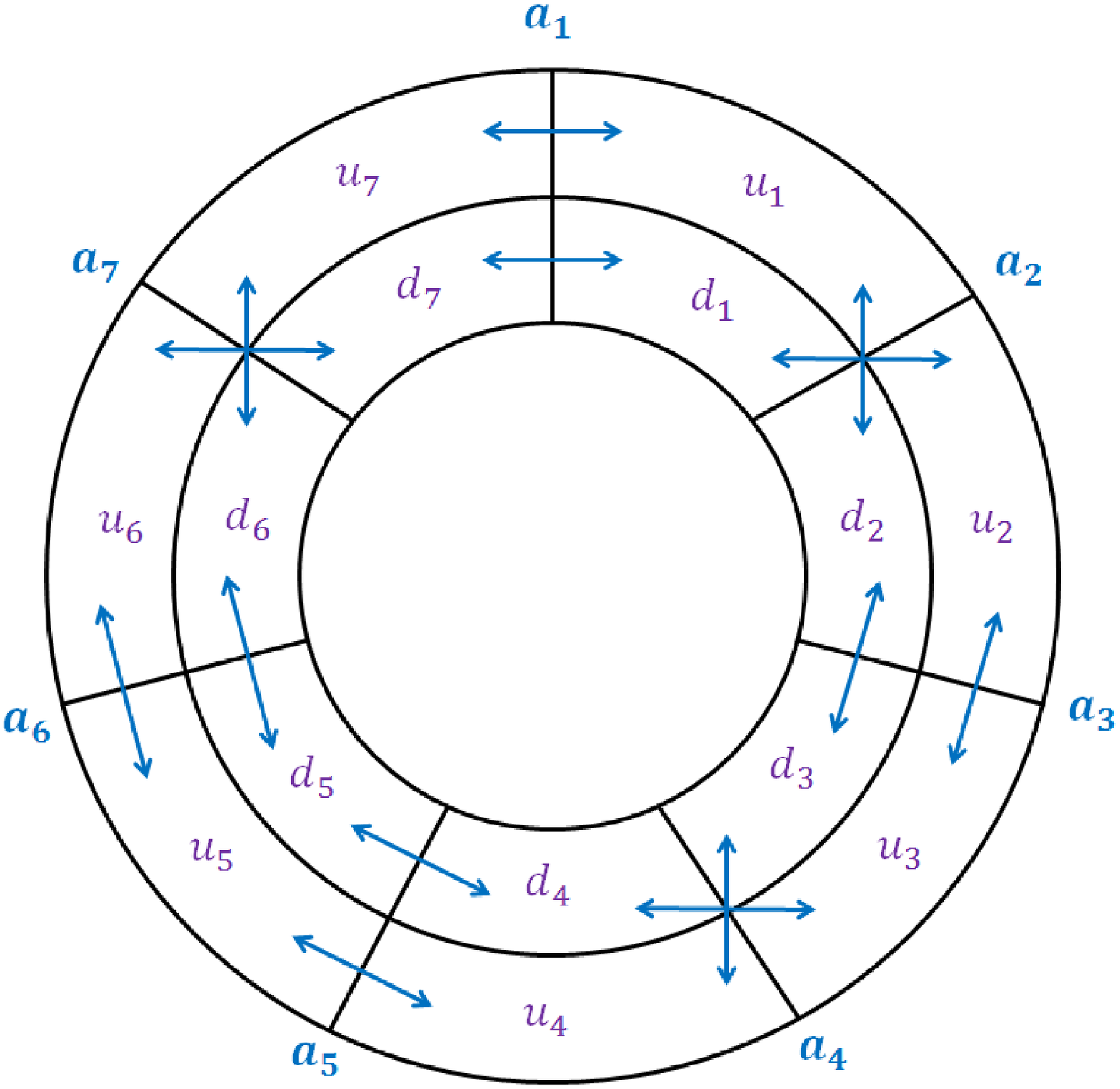}}
{Action sets of the players in Example \ref{example:9}\label{fig:8}}
{Each blue arrow connecting two resources in columns $j-1$ and $j$ denotes the action of player $j$ choosing those two resources.}
\vspace{-0.125in}
\end{figure}

\textit{Counterexample 6} is essentially a game between the players $i_1,i_2,\ldots,i_k$. To complete its specification, we need to specify the resource-specific coefficients $v_j$. For this, we pick any $v_j>0$ satisfying the following set of inequalities:

\vspace{-0.225in}
\begin{equation}
\label{eq:coeff_inequalities}
\begin{split}
v_1|q_1|b_1 &> v_2|q_2|a_2\mcr
v_2|q_2|b_2 &> v_3|q_3|a_3\mcr
&\vdots\mcr
v_k|q_k|b_k &> v_1|q_1|a_1
\end{split}
\end{equation}

\vspace{-0.125in}
\noindent We argue that if (\ref{eq:lemma:6''}) is satisfied, then this set of inequalities has a solution in strictly positive integers $v_j$. To see this, first observe that if (\ref{eq:lemma:6''}) is satisfied, then $b_j\not=0$ for all $1\leq j\leq k$. If $a_j=0$ for some $j$, say for $j=1$, then (\ref{eq:coeff_inequalities}) can be solved recursively as follows.

\newpage
First, pick any integer $v_k>0$; this satisfies the last inequality. Then, for $k-1\geq j\geq 1,$ pick any integer $v_j>0$ that satisfies the $j$th inequality, i.e.,
\begin{equation*}
v_j > v_{j+1}\frac{|q_{j+1}|}{|q_j|}\frac{a_{j+1}}{b_j}
\end{equation*}
Next, consider the case where $a_j\not=0$ for all $1\leq j\leq k$. For simplicity, denote $v_i|q_i|$ by $v'_i$.

For $i\not=1$, multiplying the first $i-1$ inequalities results in an upper bound for the ratio $\frac{v'_i}{v'_1}$, and multiplying the last $k-i+1$ inequalities results in a lower bound for the ratio $\frac{v'_i}{v'_1}$. These bounds are given by:

\vspace{-0.05in}
\begin{equation}
\label{eq:coeff_main_ineq}
(\forall 2\leq i\leq k)\quad\quad \frac{\displaystyle\prod_{j=i+1}^{k+1} a_j}{\displaystyle\prod_{j=i}^{k} b_j} < \frac{v'_i}{v'_1} < \frac{\displaystyle\prod_{j=1}^{i-1} b_j}{\displaystyle\prod_{j=2}^{i} a_j}
\end{equation}
It can be seen that (\ref{eq:coeff_main_ineq}) is feasible if and only if (\ref{eq:lemma:6''}) is satisfied. Algorithm \ref{alg:2} describes a procedure for obtaining $v'_i>0$ that solves (\ref{eq:coeff_inequalities}).

\vspace{-0.1in}
\begin{algorithm}
\caption{Solving (\ref{eq:coeff_inequalities})\label{alg:2}}
\begin{algorithmic}
    \State $v'_1 \gets 1$
    \State Pick $v'_2$ satisfying (\ref{eq:coeff_main_ineq}) for $i=2$
    \State $i \gets 3$
    \While{$i\leq k$}
        \State Pick $v'_i$ satisfying the inequality given by,
        \begin{equation}
        \label{eq:coeff_spec_ineq}
        \frac{\prod_{j=i}^{k+1} a_j}{\prod_{j=i}^{k} b_j} < a_i v'_i < b_{i-1}v'_{i-1}
        \end{equation}
        \State $i \gets i+1$
    \EndWhile
\end{algorithmic}
\end{algorithm}

\vspace{-0.1in}
\noindent Note that during the $i$th iteration, (\ref{eq:coeff_spec_ineq}) is feasible for $v'_i$, because, by using the inequalities in (\ref{eq:coeff_main_ineq}) for $i$ and $i-1$, it follows that,
\begin{equation*}
\frac{\displaystyle\prod_{j=i}^{k+1} a_j}{\displaystyle\prod_{j=i}^{k} b_j} < a_i v'_i\ ,\ b_{i-1}v'_{i-1} < \frac{\displaystyle\prod_{j=1}^{i-1} b_j}{\displaystyle\prod_{j=2}^{i-1} a_j}
\end{equation*}
and once again, this is feasible if and only if (\ref{eq:lemma:6''}) is satisfied.

Now, we verify that the $v'_i>0$ obtained through this procedure satisfy (\ref{eq:coeff_inequalities}):
\begin{enumerate}[label=(\roman*)]
\item From (\ref{eq:coeff_spec_ineq}), it is clear that the second through $(k-1)$th inequalities are satisfied.
\item The first inequality is satisfied, because $v'_1=1$ and we picked $v_2$ satisfying (\ref{eq:coeff_main_ineq}) for $i=2$, from which we get $v'_2 < \frac{b_1}{a_2}$.
\item The last inequality is satisfied, since $v'_1=1$, and the $v_k$ that we picked satisfying (\ref{eq:coeff_spec_ineq}) for $i=k$ also satisfies (\ref{eq:coeff_main_ineq}) for $i=k$, from which we get $v_k>\frac{a_1}{b_k}$.
\end{enumerate}
From these $v'_j$, we obtain $v_j$ by dividing out $|q_j|$. Note that it is always possible to choose $v'_j$ such that all $v_j$ are rational. If this is done, then these $v_j$ can all be scaled by a single positive constant to make them integers, while still satisfying (\ref{eq:coeff_inequalities}).

\newpage
We observe that our definition of the players' action sets in (\ref{eq:straight}) and (\ref{eq:diagonal}) ensures that in any action profile, every column $j$ (consisting of resources $u_j$ and $d_j$) must have both players $i_j$ and $i_{j+1}$. (There are four possible ways in which this can happen.) We call an action profile $a=\left(a_1,\ldots,a_k\right)$ a \textit{valid action profile} if there exists some $1\leq j\leq k$ (called a \textit{valid index}) such that one of four \textit{validity conditions} is true. Each validity condition involves a configuration consisting of two adjacent columns. We now present the four validity conditions: (The valid configurations that are referenced in these conditions are illustrated in Figure \ref{fig:9a}.)

\vfill
\begin{enumerate}[label=(\roman*)]
\item $q_j<0$, $q_{j+1}<0$, $a_j\cap a_{j+1}\not=\emptyset$. Visually, this corresponds to $V_1$ or $V_2$.

\vfill
\item $q_j>0$, $q_{j+1}>0$, $a_j\cap a_{j+1}=\emptyset$. Visually, this corresponds to $V_3$ or $V_4$.

\vfill
\item $q_j<0$, $q_{j+1}>0$, $a_j\cap a_{j+1}\not=\emptyset$, $a_{j+1}\cap a_{j+2}\not=\emptyset$. Visually, this corresponds to $V_2$.

\vfill
\item $q_j>0$, $q_{j+1}<0$, $a_j\cap a_{j+1}=\emptyset$, $a_{j+1}\cap a_{j+2}=\emptyset$. Visually, this corresponds to $V_4$.
\end{enumerate}

\begin{figure}[ht]
\FIGURE
{\includegraphics*[height=9cm,width=15cm]{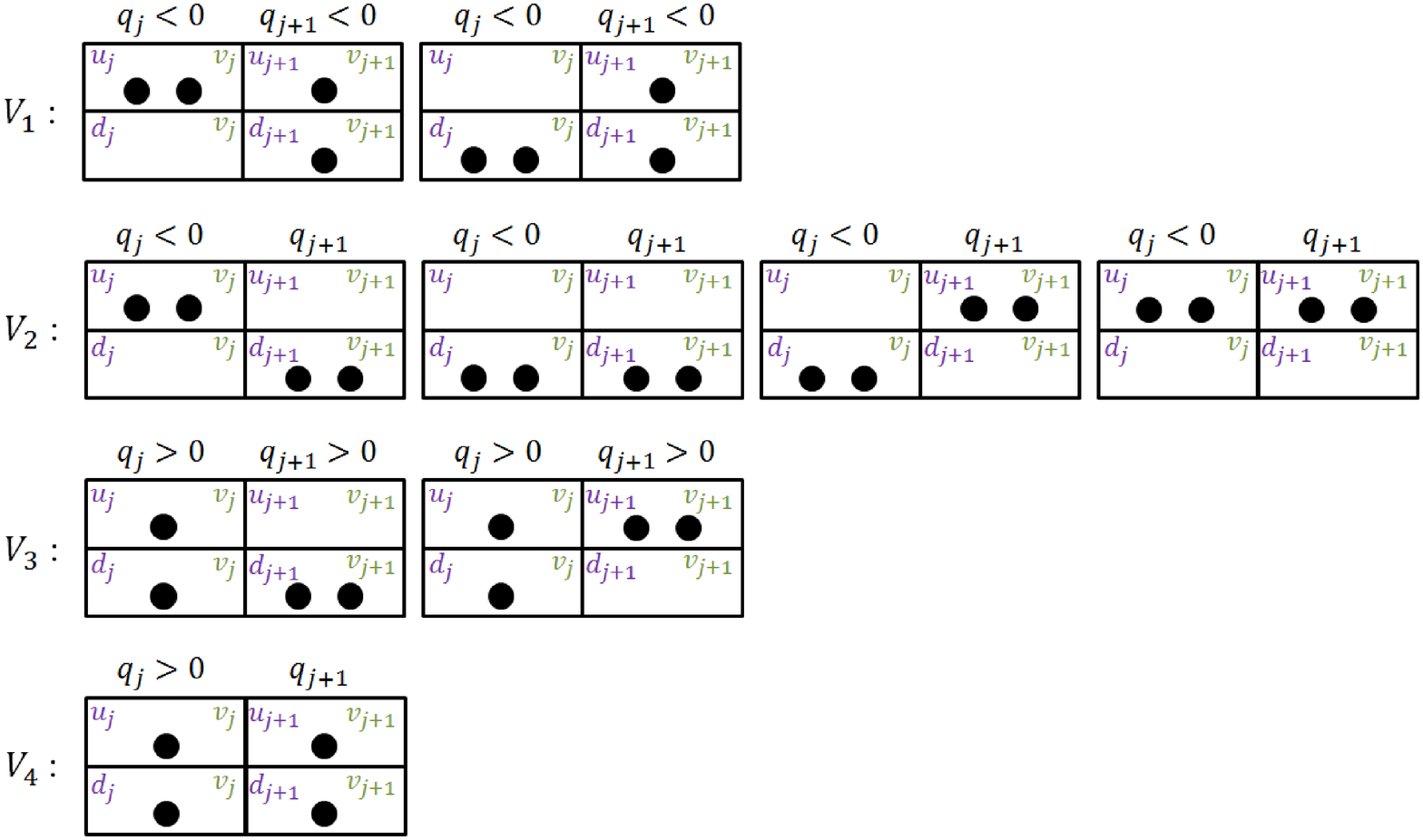}}
{Possible (anonymous) configurations for valid action profiles\label{fig:9a}}
{Each black circle denotes a player. Recall that each column $j$ must contain two players, $i_j$ and $i_{j+1}$. Hence, in every configuration, there are two black circles per column. In every configuration, at the top of each column, the sign of its coefficient is indicated. If there is no sign indicated, then it could be either positive or negative.}
\end{figure}

Figure \ref{fig:9b} is more detailed, where we enumerate all possible ways the black circles in Figure \ref{fig:9a} can correspond to players of their respective columns. For each set of valid configurations, we also show the utility to player $i_{j+1}$ for the action that he is shown taking ($a_{j+1}$), as well as for the action he could otherwise have chosen ($\overline{a_{j+1}}$).

\vfill
We now show that a valid action profile cannot be an equilibrium. From (\ref{eq:coeff_inequalities}), we get,
\begin{equation}
\label{eq:elaborateineq}
(\forall\ 1\leq j\leq k)\quad\quad v_j|q_j|f^{T_j}\left((i_{j+1},T_j\right) > v_{j+1}|q_{j+1}|f^{T_{j+1}}\left(i_{j+1},T_{j+1}\right)
\end{equation}Using (\ref{eq:elaborateineq}), it can be seen that in any valid action profile with a valid index $j$, i.e., action profiles containing any of the configurations of $V_1$ through $V_4$, player $i_{j+1}$ always has an incentive to deviate.

\newpage
\begin{figure}[ht]
\FIGURE
{\includegraphics*[height=15cm,width=16cm]{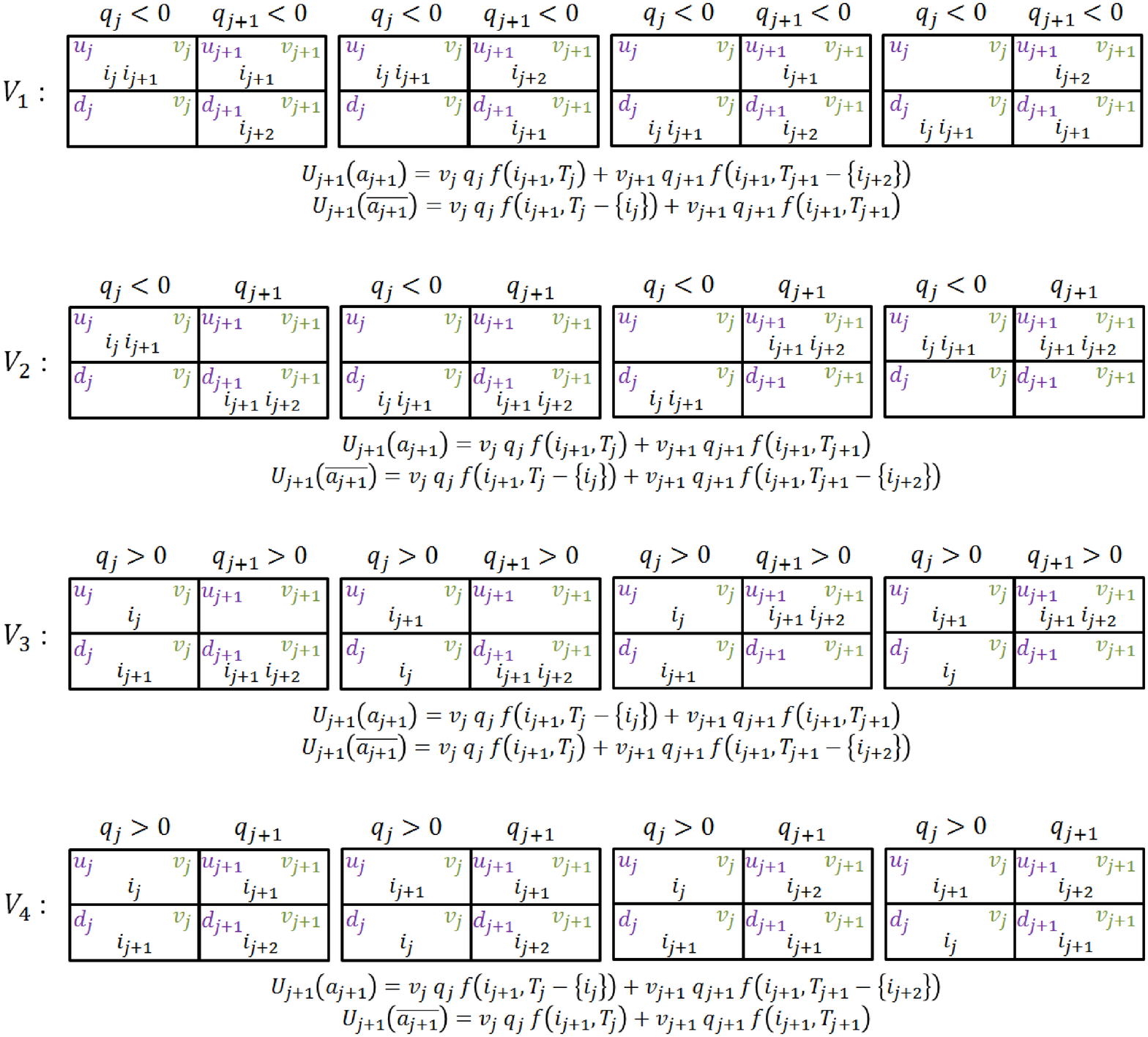}}
{Possible configurations for valid action profiles\label{fig:9b}}
{The action set of player $i_{j+1}$ is given by $\mathcal{A}_{i_{j+1}}=\left\{a_{j+1}, \overline{a_{j+1}}\right\}$. The specific resources that these two actions include will depend on whether the player is straight or diagonal, which is not important here. We do not explicitly highlight the actions of all players other than players $i_j,i_{j+1},i_{j+2}$. In each configuration, action $a_{j+1}$ corresponds to the action that player $i_{j+1}$ is shown taking. $U_{j+1}\left(a_{j+1}\right)$ denotes the utility to player $i_{j+1}$ in the configurations shown, and it only depends on whether it is $V_1$, $V_2$, $V_3$, or $V_4$. In every configuration, at the top of each column, the sign of its coefficient is indicated where relevant. If no sign is indicated, it could be either positive or negative.}
\end{figure}

\noindent For example, consider any of the four configurations of $V_1$, where $q_j<0$, $q_{j+1}<0$. The difference in the utilities to player $i_{j+1}$ between deviating to $\overline{a_{j+1}}$ and staying in $a_{j+1}$ is given by,
\begin{equation*}
\begin{split}
\Delta_{j+1} &= U_{j+1}(\overline{a_{j+1}})-U_{j+1}(a_{j+1})\mcr
&= -v_j\left(f(i_{j+1},T_j)-f(i_{j+1},T_j-\{i_j\})\right) + v_{j+1}\left(f(i_{j+1},T_{j+1})-f(i_{j+1},T_{j+1}-\{i_{j+2}\})\right)
\end{split}
\end{equation*}
Using the basis representation of $f$, this can be simplified as,
\begin{equation*}
\Delta_{j+1} = -v_j\sum_{T\in\mathcal{T}_{i_ji_{j+1}}(T_j)}q_Tf^{T}(i_{j+1},T) + v_{j+1}\sum_{T\in\mathcal{T}_{i_{j+1}i_{j+2}}(T_{j+1})}q_Tf^{T}(i_{j+1},T)
\end{equation*}

\newpage
\noindent But $T_j$ and $T_{j+1}$ are minimal in $\mathcal{T}^+_{i_ji_{j+1}}$ and $\mathcal{T}^+_{i_{j+1}i_{j+2}}$ respectively. Therefore, we get,

\vfill
\begin{equation*}
\begin{split}
\Delta_{j+1} &= -v_jq_jf^{T_j}(i_{j+1},T_j)+v_{j+1}q_{j+1}f^{T_{j+1}}(i_{j+1},T_{j+1})\mcr
&= v_j|q_j|f^{T_j}(i_{j+1},T_j)-v_{j+1}|q_{j+1}|f^{T_{j+1}}(i_{j+1},T_{j+1})
\end{split}
\end{equation*}

\vfill
\noindent which is strictly positive, from (\ref{eq:elaborateineq}). Hence, in configuration $V_1$, player $i_{j+1}$ has an incentive to deviate. Similar arguments can be constructed for configurations of $V_2$, $V_3$, and $V_4$.

\vfill
The final step is to show that no invalid action profile exists in \textit{Counterexample 6}. We do this by showing that any attempt to construct an invalid action profile by choosing actions from the players' action sets must fail. Before presenting the formal details, we return to Example \ref{example:9} to highlight the intuition behind our approach.

\vfill

\vfill

\vfill
\begin{example}
\label{example:10}
\textit{In Example \ref{example:9}, we specified the action sets of the seven players involved in an $\mathbb{M}_3\mathbb{Z}_4\mathbb{P}_2$ sign profile, namely, $(-,-,-,+,-,+,+)$. Here, we show that for this sign profile, every admissible action profile is valid, i.e., it satisfies one of the four validity properties. We do this by showing that any attempt to construct an invalid action profile must fail:}

\vfill
\begin{enumerate}[label=(\roman*)]
\item \textit{First, consider the $\mathbb{M}_3$ segment. Recall that player $i_2$ is diagonal and player $i_3$ is straight. It can be seen that in any invalid action profile, there are only four possible ways in which this segment can be filled up -- configurations $M_1$, $M_2$, or their symmetric counterparts, $M'_1$, $M'_2$, as illustrated in Figure \ref{fig:ex:10a}. To see this, take $M_1$ for example:}

    \vfill
    \begin{itemize}
    \item \textit{If $i_2$ switches, the first two columns form a valid configuration of $V_2$.}
    \item \textit{If $i_3$ switches, the second and third columns form a valid configuration of $V_2$.}
    \item \textit{If both $i_2$ and $i_3$ switch, the first two columns form a valid configuration of $V_1$.}
    \end{itemize}

\vfill
\item \textit{Next, consider the $\mathbb{Z}_4$ segment. Recall that player $i_4$ is diagonal, and players $i_5$ and $i_6$ are straight. It can be seen that in any invalid action profile, there are only ten possible ways in which this segment can be filled up -- configurations $Z_1$ through $Z_4$, or their symmetric counterparts, $Z'_1$ through $Z'_4$, as illustrated in Figure \ref{fig:ex:10b}. Here, if any configuration other than these ten occurs, there will be adjacent columns what would form valid configurations of either $V_2$ or $V_4$.}

\vfill
\item \textit{Finally, consider the $\mathbb{P}_2$ segment. Recall that player $i_7$ is diagonal. It can be seen that in any invalid action profile, there are only four possible ways in which this segment can be filled up -- configurations $P_1$, $P_2$, or their symmetric counterparts, $P'_1$, $P'_2$, as illustrated in Figure \ref{fig:ex:10c}. Here, if any configuration other than these four occurs, the two columns would form valid configurations of either $V_3$ or $V_4$.}
\end{enumerate}

\vfill
\noindent\textit{It follows that any invalid action profile must be constructed by picking one configuration from each of Figures \ref{fig:ex:10a}-\ref{fig:ex:10c} and `stringing' them together. Note that in doing so, two columns shaded with the same color must be identical to be strung together, since they correspond to overlapping columns. Therefore, it can be seen that there are only four ways of gluing together such configurations: $M_1Z_4P_1$, $M_1Z_4P_2$, and their symmetric counterparts, $M'_1Z'_4P'_1$, $M'_1Z'_4P'_2$. None of these four action profiles are invalid:}

\vfill
\begin{enumerate}[label=(\roman*)]
\item \textit{$M_1Z_4P_2$ and $M'_1Z'_4P'_2$ are illegal action profiles, since player $i_1$ is straight, whereas in these two action profiles, he chooses a diagonal action.}
\item \textit{$M_1Z_4P_1$ and $M'_1Z'_4P'_1$ are legal action profiles, but are valid, because the last column and the first column (when wrapped around) form a valid configuration of $V_4$.}
\end{enumerate}

\vfill
\noindent\textit{Hence, all action profiles are valid.}
\end{example}

\newpage
\begin{figure}[ht]
    \centering
    \subfloat[][Invalid configurations for an $\mathbb{M}_3$ segment.\\ \begin{scriptsize}\textit{Note.} The vertical green shade indicates that the third column of this $\mathbb{M}_3$ segment must match the first column of the succeeding $\mathbb{Z}_4$ segment, since they overlap.\end{scriptsize}\label{fig:ex:10a}]{
        \includegraphics*[height=3.5cm,width=10.5cm]{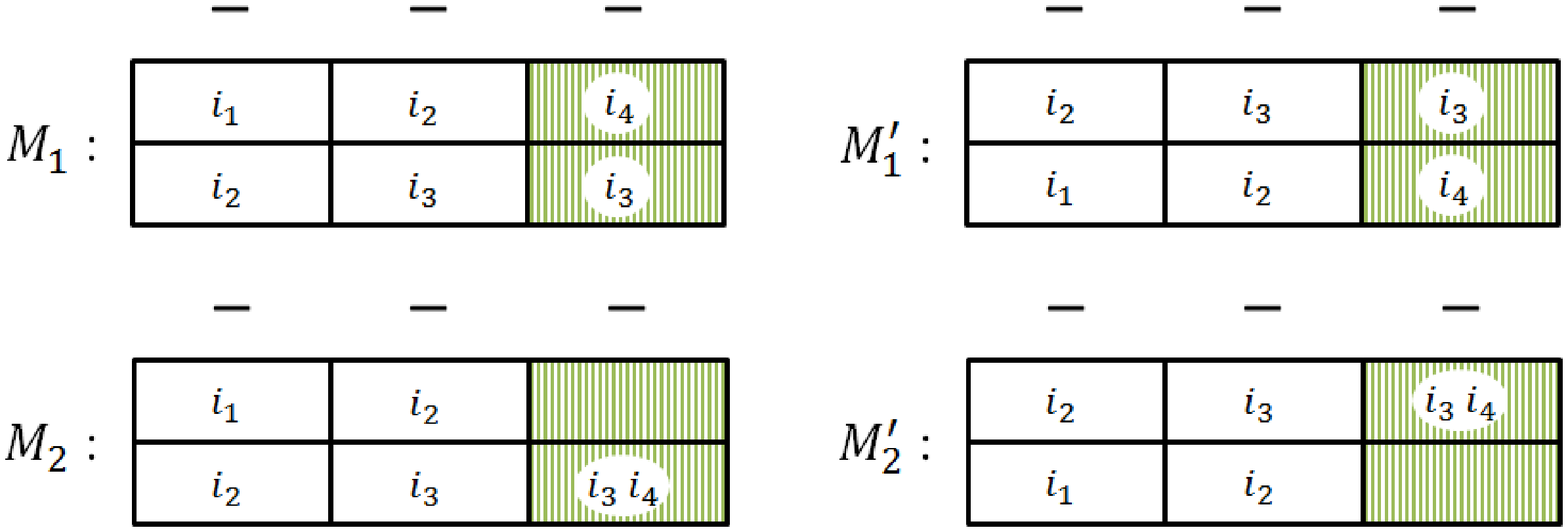}}\\
    \subfloat[][Invalid configurations for a $\mathbb{Z}_4$ segment.\\ \begin{scriptsize}\textit{Note.} We have clubbed together two configurations as $Z_3$ (and similarly, $Z'_3$), since they have identical boundaries (first and last columns) -- only boundary compatibility matters when gluing together different segments. The vertical green shade indicates that the first column of this $\mathbb{Z}_4$ segment must match the third column of the preceding $\mathbb{M}_3$ segment, since they overlap. The horizontal blue shade indicates that the fourth column of this $\mathbb{Z}_4$ segment must match the first column of the succeeding $\mathbb{P}_2$ segment, since they overlap.\end{scriptsize}\label{fig:ex:10b}]{
        \includegraphics*[height=8cm,width=12.5cm]{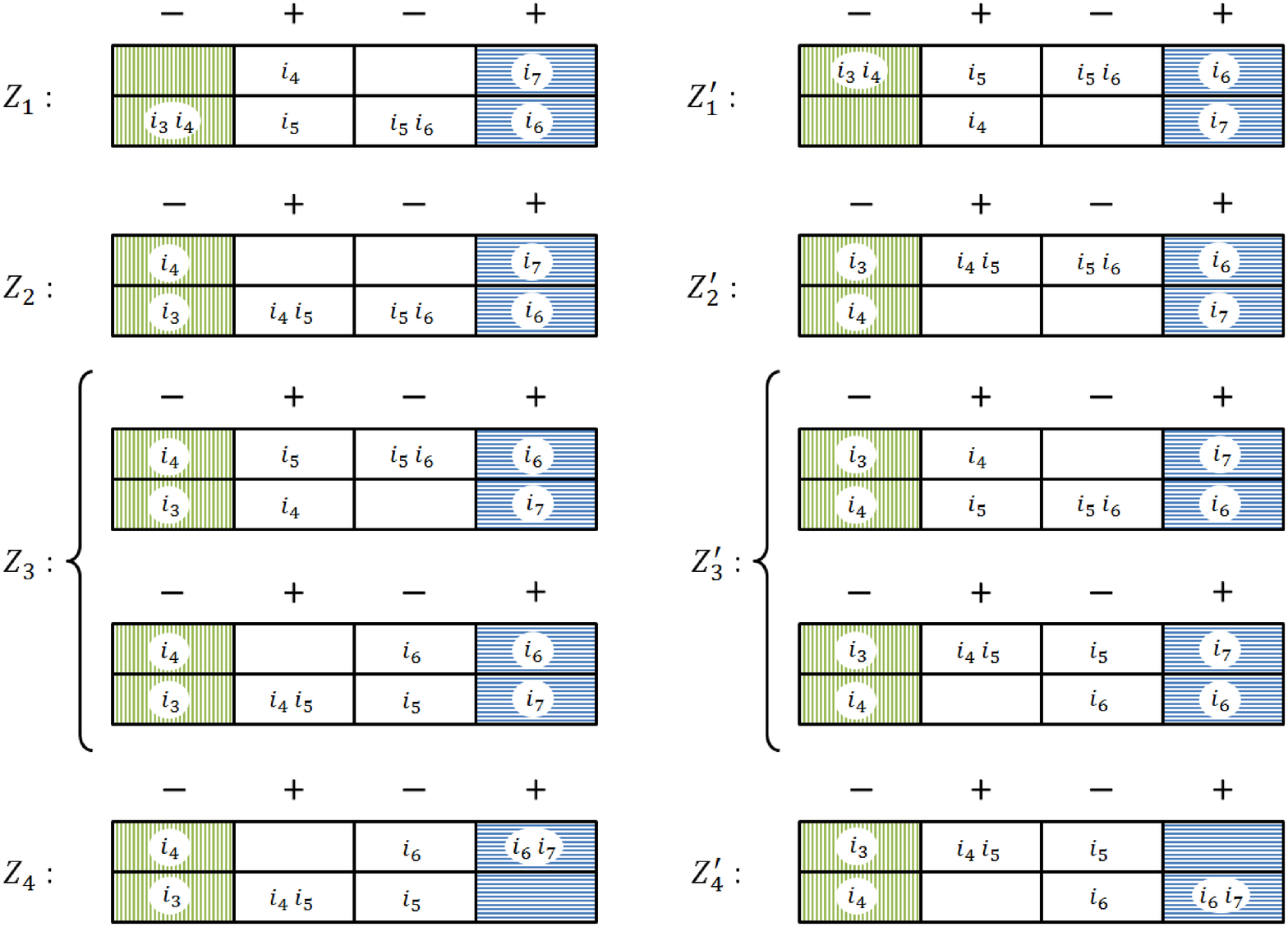}}\\
    \subfloat[][Invalid configurations for a $\mathbb{P}_2$ segment.\\ \begin{scriptsize}\textit{Note.} The horizontal blue shade indicates that the first column of this $\mathbb{P}_2$ segment must match the fourth column of the preceding $\mathbb{Z}_4$ segment, since they overlap.\end{scriptsize}\label{fig:ex:10c}]{
        \includegraphics*[height=3.5cm,width=7.5cm]{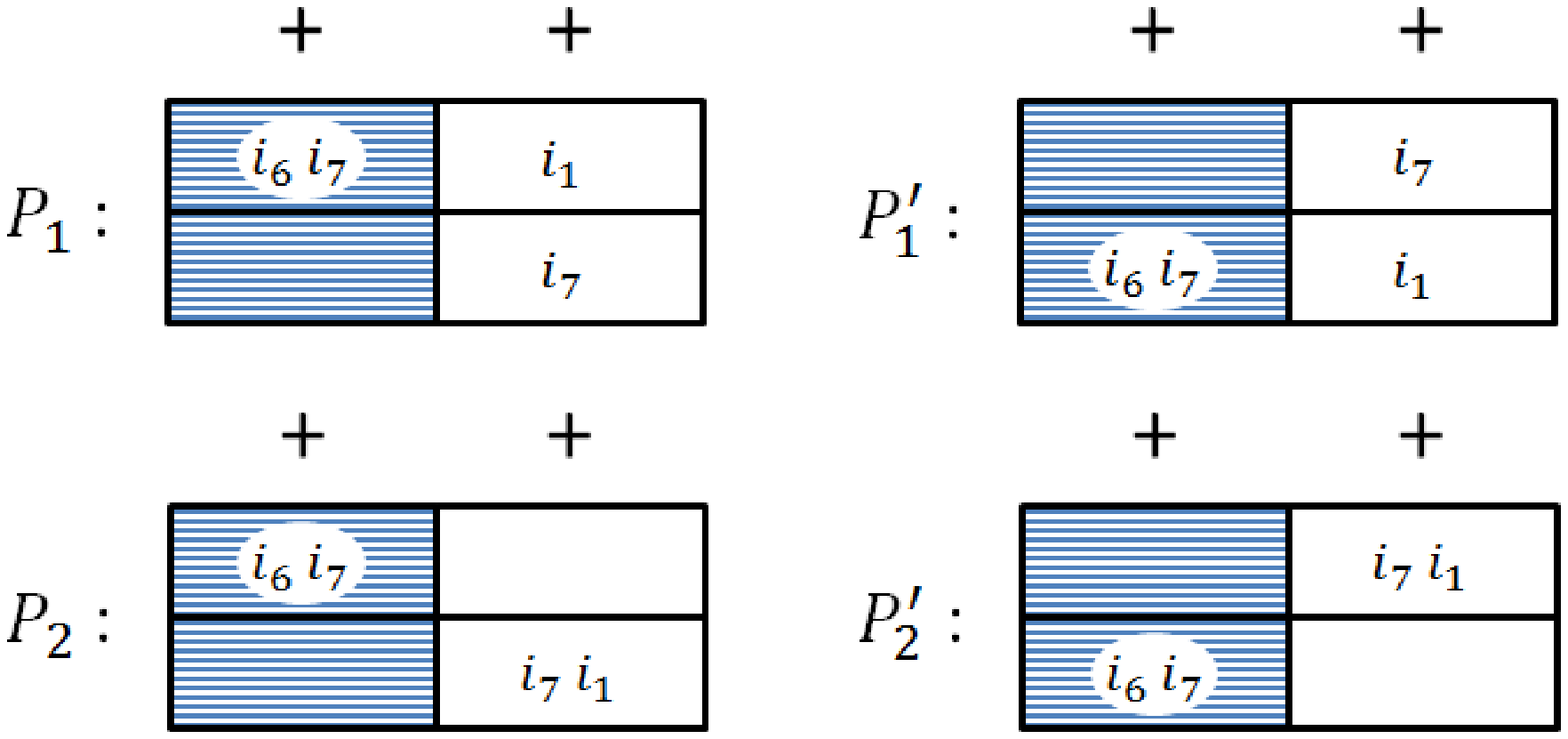}}
    \caption{{\footnotesize Possible configurations for $\mathbb{M}_3$, $\mathbb{Z}_4$, $\mathbb{P}_2$ segments within an invalid action profile for Example \ref{example:10}}\label{fig:ex:10}}
\end{figure}

We now build on the general intuition that was demonstrated in the example above to provide a complete proof. First, we state some necessary conditions that any invalid action profile must satisfy within a $\mathbb{P}_\ell$, $\mathbb{M}_\ell$, and $\mathbb{Z}_\ell$ segment. The proofs are by induction, and involve arguing that in order to avoid any of the valid configurations $V_1$ through $V_4$, while still respecting how the action sets are defined, such segments must satisfy these necessary conditions. The visual configurations that are referenced in these conditions are illustrated in Figure \ref{fig:7}.
\begin{enumerate}[label=(\roman*)]
\item Let columns $j,j+1,\ldots,j+\ell-1$ form an $\mathbb{M}_\ell$ segment. Then, in any invalid action profile, $a_j\cap a_{j+1}=\emptyset$. In addition,
    \begin{itemize}
    \item $u_j\in a_j \Longrightarrow d_{j+\ell-1}\in a_{j+\ell-1}$
    \item $d_j\in a_j \Longrightarrow u_{j+\ell-1}\in a_{j+\ell-1}$
    \end{itemize}
    Visually, any $\mathbb{M}_\ell$ segment of an invalid action profile must match configurations $M_1,M_2$ or their symmetric counterparts $M'_1,M'_2$. If not, there will be adjacent columns that would form valid configurations of either $V_1$ or $V_2$.

    The proof is by induction on $\ell$, the length of the segment. For the base case, when $\ell=2$, the arguments are similar to those for the $\mathbb{M}_3$ segment in Example \ref{example:10}. Our induction hypothesis is that every invalid $\mathbb{M}_\ell$ segment must match one of the four configurations $M_1,M_2,M'_1,M'_2$, for some $\ell>2$. Assuming this is true, now consider an invalid $\mathbb{M}_{\ell+1}$ segment. Keep in mind that from the definition of action sets, player $i_{j+\ell}$ is required to be straight, since he is at the end of this segment. Let $\widetilde{\mathbb{M}}_\ell$ denote the subsegment formed by its first $\ell$ columns. From the induction hypothesis, $\widetilde{\mathbb{M}}_\ell$ must match one of the four configurations $M_1,M_2,M'_1,M'_2$:
    \begin{itemize}
    \item $\widetilde{\mathbb{M}}_\ell$ cannot match $M_2$ or $M'_2$, because in either case, in $\mathbb{M}_{\ell+1}$, columns $\ell$ and $\ell+1$ together will form valid configurations of either $V_1$ or $V_2$ (depending on how column $\ell+1$ is occupied by the players $i_{j+\ell}$ and $i_{j+\ell+1}$).
    \item If $\widetilde{\mathbb{M}}_\ell$ matches $M_1$ or $M'_1$, then $\mathbb{M}_{\ell+1}$ will match one of the four configurations $M_1$, $M_2$, $M'_1$, $M'_2$ (depending on how column $\ell+1$ is occupied by the players $i_{j+\ell}$ and $i_{j+\ell+1}$).
    \end{itemize}
\item Let columns $j,j+1,\ldots,j+\ell-1$ form a $\mathbb{P}_\ell$ segment. Then, in any invalid action profile, $a_j\cap a_{j+1}\not=\emptyset$. In addition,
    \begin{itemize}
    \item $u_j\in a_j \Longrightarrow d_{j+\ell-1}\in a_{j+\ell-1}$
    \item $d_j\in a_j \Longrightarrow u_{j+\ell-1}\in a_{j+\ell-1}$
    \end{itemize}
    Visually, any $\mathbb{P}_\ell$ segment of an invalid action profile must match configurations $P_1,P_2$ or their symmetric counterparts $P'_1,P'_2$. If not, there will be adjacent columns that would form valid configurations of either $V_3$ or $V_4$. The proof is by a similar inductive argument as the $\mathbb{M}_\ell$ case above, except that it is more complicated -- we need to consider segments of odd and even lengths separately, because whether player $i_{j+\ell-1}$ is straight or diagonal in a $\mathbb{P}_\ell$ segment depends on whether $\ell$ is even or odd. We omit the proof for brevity.
\item Let columns $j,j+1,\ldots,j+\ell-1$ form a $\mathbb{Z}_\ell$ segment. Then, in any invalid action profile, one of the following three statements must hold:
    \begin{itemize}
    \item $a_j\cap a_{j+1}\not=\emptyset$ and $a_{j+\ell-1}\cap a_{j+\ell}=\emptyset$. In addition,
        \begin{itemize}
        \item $u_j\in a_j \Longrightarrow \left(u_{j+\ell-1}\in a_{j+\ell-1}\text{ AND }d_{j+\ell-1}\in a_{j+\ell}\right)$
        \item $d_j\in a_j \Longrightarrow \left(d_{j+\ell-1}\in a_{j+\ell-1}\text{ AND }u_{j+\ell-1}\in a_{j+\ell}\right)$
        \end{itemize}
        Visually, this corresponds to configuration $Z_1$ or its symmetric counterpart $Z'_1$.
    \item $a_j\cap a_{j+1}=\emptyset$ and $a_{j+\ell-1}\cap a_{j+\ell}=\emptyset$. Visually, this corresponds to configurations $Z_2, Z_3$ or their symmetric counterparts $Z'_2, Z'_3$.
    \item $a_j\cap a_{j+1}=\emptyset$ and $a_{j+\ell-1}\cap a_{j+\ell}\not=\emptyset$. In addition,
        \begin{itemize}
        \item $u_j\in a_j \Longrightarrow d_{j+\ell-1}\in a_{j+\ell-1}\cap a_{j+\ell}$
        \item $d_j\in a_j \Longrightarrow u_{j+\ell-1}\in a_{j+\ell-1}\cap a_{j+\ell}$
        \end{itemize}
        Visually, this corresponds to configuration $Z_4$ or its symmetric counterpart $Z'_4$.
    \end{itemize}
    Note that if none of these conditions are satisfied, then there will be adjacent columns that would form valid configurations of either $V_2$ or $V_4$. Once again, the proof is by a similar inductive argument, and is omitted for brevity.
\end{enumerate}

\begin{figure}[ht]
    \FIGURE
    {\includegraphics*[height=17.5cm,width=12cm]{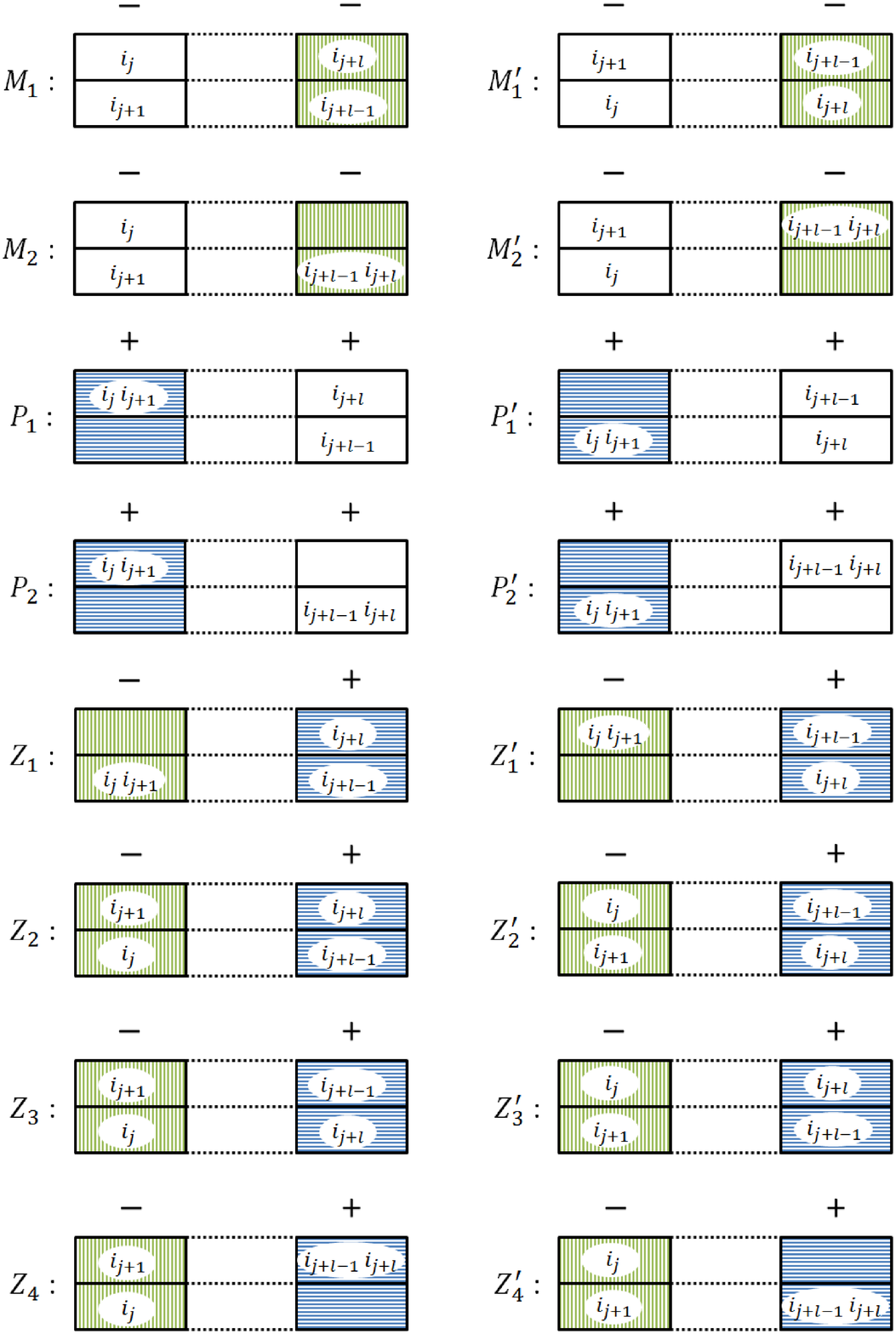}}
    {Possible configurations for $\mathbb{M}_\ell$, $\mathbb{P}_\ell$, and $\mathbb{Z}_\ell$ segments (spanning columns $j$ through $j+\ell-1$) within an invalid action profile\label{fig:7}}
    {An action profile can be constructed by performing one of two operations repeatedly: (1) Two segments can be ``hooked'' together to form a larger segment if the last column of the first segment and the first column of the second segment are identical in both occupancy and shading scheme. In this case, these two columns overlap to become one column. For example, $M_1$ and $Z_2$ can be hooked together. (2) Two segments can potentially be ``glued'' together to form a larger segment if the last column of the first segment and the first column of the second segment do not have matching shading schemes. In this case, these two columns would not overlap, but sit next to each other. Such a gluing is permitted only if the resulting action of the boundary player $i_{j+\ell}$ is legal (permissible according to the definition of the action sets). For example, $Z_2$ and $M_1$ can be glued together, since it results in the boundary player $i_{j+\ell}$ being straight, which respects the definition of a straight player.}
\end{figure}

It follows that any invalid action profile must somehow be constructed by `stringing' together different-length instantiations of these sixteen configurations. Figure \ref{fig:7} illustrates how this is done. There are four possible configurations for an $\mathbb{M}_\ell$ segment, four possible configurations for a $\mathbb{P}_\ell$ segment, and eight possible configurations for a $\mathbb{Z}_\ell$ segment. Note that, by definition, these are maximal segments, so when stringing together two configurations, they cannot be of the same type of segment. We discuss all possible ways of putting together an invalid action profile below:

\vfill
\begin{enumerate}[label=(\roman*)]

\vfill
\item First, observe that an invalid action profile cannot be constructed using exactly one of these sixteen configurations ($j=1,\ell=k$ in this case), because, when wrapped around, at the boundary that is formed by the last column and the first column, either of the following two scenarios occur:

    \vfill
    \begin{itemize}
    \item The boundary player, $i_1$, ends up making an illegal choice, rendering the whole action profile illegal.

    \vfill
    \item The boundary configuration (the configuration formed by the last column and the first column) ends up being a valid configuration.
    \end{itemize}

\vfill
\item Now, we need at least two configurations to be strung together to create an invalid action profile. We investigate possible ways of stringing together two configurations. Observe that configurations $Z_1$, $Z_2$, $Z_3$ and their symmetric counterparts $Z'_1$, $Z'_2$, $Z'_3$ cannot be hooked to any configurations of a $\mathbb{P}_\ell$ segment (since the overlapping columns do not match), and cannot be glued to any configurations of an $\mathbb{M}_\ell$ segment (since this results in a valid boundary configuration of $V_4$). So, these six configurations cannot be used to construct an invalid action profile, and can be eliminated.

\vfill
\item Among the remaining ten configurations, observe that $P_1$ and its symmetric counterpart $P'_1$ cannot be glued to any configurations of an $\mathbb{M}_\ell$ or any remaining configurations of a $\mathbb{Z}_\ell$ segment (since this results in a valid boundary configuration of $V_4$). Also, $M_2$ and its symmetric counterpart $M'_2$ cannot be hooked to any remaining configurations of a $\mathbb{Z}_\ell$ segment (since the overlapping columns do not match), and cannot be glued to any configurations of a $\mathbb{P}_\ell$ segment (since this results in a valid boundary configuration of $V_2$). So, these four configurations cannot be used to construct an invalid action profile, and can be eliminated.

\vfill
\item We are left with $M_1$, $P_2$, $Z_4$, which can all be hooked or glued with each other, and their symmetric counterparts, $M'_1$, $P'_2$, $Z'_4$, which can also all be hooked or glued with each other. Also, none of $M_1$, $P_2$, $Z_4$ can be hooked or glued with any of $M'_1$, $P'_2$, $Z'_4$ and vice versa (since either overlapping columns (if any) do not match, or the boundary player $i_{j+\ell}$ ends up making an illegal choice).

\vfill
\item Therefore, an invalid action profile must be constructed by using only $M_1,P_2,Z_4$ or only $M'_1, P'_2, Z'_4$. Also, such invalid action profiles must begin with a minus sign and end with a plus sign (due to our cyclic transformation at the beginning). There are now only three cases to be considered:

    \vfill
    \begin{itemize}
    \item Begin with $M_1$ (respectively, $M'_1$) and end with $Z_4$ (respectively, $Z'_4$)

    \vfill
    \item Begin with $M_1$ (respectively, $M'_1$) and end with $P_2$ (respectively, $P'_2$)

    \vfill
    \item Begin with $Z_4$ (respectively, $Z'_4$) and end with $P_2$ (respectively, $P'_2$)
    \end{itemize}

    \vfill
    \noindent But, in each of these cases, when wrapped around, the boundary player $i_1$ ends up making an illegal choice, rendering the whole action profile illegal.
\end{enumerate}

\vfill
\noindent Hence, there exist no invalid action profiles. This concludes the proof.\hfill\Halmos
\endproof

\vfill
As before, we now present a useful inference from Lemma \ref{lemma:6} in terms of the weight systems:

\newpage
\begin{corollary}
\label{corollary:6}
Given any set of local welfare functions $\mathbb{W}$, let $f^\mathbb{W}$ be budget-balanced distribution rules that guarantee equilibrium existence in all games $G\in\mathcal{G}(N,f^\mathbb{W},\mathbb{W})$, where, for each $W\in\mathbb{W}$, $f^W:=\DS\sum_{T\in\mathcal{T}^W}q^W_T f^T_{GWSV}[\omega^{W,T}]$, where $\omega^{W,T}=\left(\boldsymbol{\lambda}^{W,T},\Sigma^{W,T}=\left(S_1^{W,T},S_2^{W,T}\right)\right)$. Let $i_1,i_2,\ldots,i_k\in N$ be any $k$ players ($k\geq 3$) such that $\exists\ T_1 \in \left(\mathcal{T}_{i_1i_2}^{1+}\right)^{\min}, T_2 \in \left(\mathcal{T}_{i_2i_3}^{2+}\right)^{\min}, \ldots, T_k \in \left(\mathcal{T}_{i_ki_1}^{k+}\right)^{\min}$, for any $k$ welfare functions $W_1,W_2,\ldots,W_k\in\mathbb{W}$. Then,

\vfill
\begin{enumerate}[label=(\alph*)]
\item $(\forall\ 1\leq j \leq k)\ \{i_j,i_{j+1}\}\subseteq S_1^{j,T_j}$

\vfill
\item $\DS\prod_{j=1}^k \frac{\lambda_{i_j}^{j,T_j}}{\lambda_{i_{j+1}}^{j,T_j}}=1$
\end{enumerate}
\end{corollary}

\vfill
\subsubsection{Existence of a universal weight system.}\label{s.universal}
In order to establish the global consistency of the sequence of weight systems $\Omega=\left\{\left\{\omega^{W,T}\right\}_{T\in\mathcal{T}^W}\right\}_{W\in\mathbb{W}}$, we need to show that there exists a universal weight system $\omega^*=\left(\boldsymbol{\lambda}^*,\Sigma^*\right)$ that is equivalent to all the weight systems in $\Omega$, i.e., replacing $\omega^{W,T}$ with $\omega^*$ for any coalition $T\in\mathcal{T}_W$ for any $W\in\mathbb{W}$ does not affect the distribution rule $f^{W,T}=f_{GWSV}^T[\omega^{W,T}]$. We show this by explicitly constructing $\Sigma^*$ and $\boldsymbol{\lambda}^*$. Before doing so, we use $\Omega$ to define two useful relations $\succeq_\Omega$ and $=_\Omega$ on the set $N$, as follows. For any two elements $i,j\in N$,

\begin{equation}
\label{eq:partial_order}
\begin{split}
i \succeq_\Omega j &\Longleftrightarrow \Big((\exists\ W\in\mathbb{W})\ (\exists\ T\in\mathcal{T}_{ij}^{W+})\ s.t.\ i\in S_1^{W,T}\Big)\ \mbox{OR}\ \Big(i=j\Big)\mcr
i =_\Omega j &\Longleftrightarrow (i \succeq_\Omega j)\ \mbox{AND}\ (j \succeq_\Omega i)
\end{split}
\end{equation}

\vfill
\noindent Using Corollary \ref{corollary:5}(a) and \ref{corollary:5}(b)(i), we can write down an equivalent set of definitions for these relations:
\begin{equation}
\label{eq:partial_order_alternate}
\begin{split}
i \succeq_\Omega j &\Longleftrightarrow \Big((\forall\ W\in\mathbb{W})\ (\forall\ T\in\mathcal{T}_{ij}^{W+})\ i\in S_1^{W,T}\Big)\ \mbox{OR}\ \Big(i=j\Big)\mcr
i =_\Omega j &\Longleftrightarrow \Big((\forall\ W\in\mathbb{W})\ (\forall\ T\in\mathcal{T}_{ij}^{W+})\ \{i,j\}\subseteq S_1^{W,T}\Big)\ \mbox{OR}\ \Big(i=j\Big)
\end{split}
\end{equation}

\vfill
\noindent We denote the transitive closures of these relations by $\succeq_\Omega^+$ and $=_\Omega^+$ respectively.

\vfill
\begin{lemma}
\label{lemma:7}
Given any set of local welfare functions $\mathbb{W}$, if $f^\mathbb{W}$ are budget-balanced distribution rules that guarantee equilibrium existence in all games $G\in\mathcal{G}(N,f^\mathbb{W},\mathbb{W})$, described completely by the sequence of weight systems $\Omega$, then, $\succeq_\Omega^+$ constitutes a partial order on $N$.
\end{lemma}

\proof{Proof.} By definition, $\succeq_\Omega^+$ is both \textit{reflexive} and \textit{transitive}. To prove that it is a partial order on $N$, we need only show \textit{antisymmetry}, i.e., we need to show that for any $i,j\in N$, if $i\succeq_\Omega^+ j$ and $j\succeq_\Omega^+ i$, then $i =_\Omega^+ j$. This is equivalent to showing that if there is a cycle in $\succeq_\Omega$, i.e., if there exists a sequence of $k$ distinct players $i_1,\ldots,i_k\in N$ such that $i_1\succeq_\Omega i_2\succeq_\Omega\ldots\succeq_\Omega i_k\succeq_\Omega i_1$, then it must be that $i_1=_\Omega i_2=_\Omega\ldots=_\Omega i_k=_\Omega i_1$. The case where $k=1$ is trivial. For $k=2$, the proof is vacuous by definition of $=_\Omega$. For $k\geq 3$, suppose there is a cycle in $\succeq_\Omega$. Then, using the definition of $\succeq_\Omega$ from (\ref{eq:partial_order_alternate}), $\forall\ 1\leq j\leq k\ \forall\ W\in\mathbb{W}\ \forall\ T_j\in\mathcal{T}_{i_ji_{j+1}}^{W+},\ i_j\in S_1^{W,T_j}$. Then, from Corollary \ref{corollary:6}(a), $\forall\ 1\leq j\leq k\ \exists W\in\mathbb{W}\ \exists\ T_j\in\mathcal{T}_{i_ji_{j+1}}^{W+}\ s.t. \ \{i_j,i_{j+1}\}\subseteq S_1^{W,T_j}$. The conclusion then follows by using the definitions from (\ref{eq:partial_order}).\hfill\Halmos
\endproof

\newpage
We now present the construction of a universal weight system, $\omega^*=\left(\boldsymbol{\lambda}^*,\Sigma^*\right)$.

\vfill
\begin{itemize}
\item \textbf{Construction of $\Sigma^*$.} From Lemma \ref{lemma:7}, the relation $\succeq_\Omega^+$ constitutes a partial order on $N$. And the corresponding relation $=_\Omega^+$ is an equivalence relation. We let $\Sigma^*=\left(S_1^*, S_2^*, \ldots, S_k^*\right)$ be an ordered partition of $N$ into its equivalence classes according to $=_\Omega^+$, ordered in any manner that does not violate $\succeq_\Omega^+$, i.e., for any $1\leq j < \ell\leq k$, any $i_j\in S_j^*$ and $i_\ell\in S_\ell^*$, $i_\ell \nsucceq_\Omega^+ i_j$.

\vfill
\item \textbf{Construction of $\boldsymbol{\lambda}^*$.} We construct $\boldsymbol{\lambda}^*=\left(\lambda_i^*\right)_{i\in N}$ in a piecewise fashion as follows. For each equivalence class $S_r^*\in\Sigma^*$, consider the following two cases:

    \vfill
    \begin{enumerate}[label=(\arabic*)]
    \item $|S_r^*|=1$. In this case, for $i\in S_r^*$, set $\lambda_i^*$ to an arbitrary strictly positive number.

    \vfill
    \item $|S_r^*|=k>1$. Let $S_r^*=\{i_1, i_2, \ldots, i_k\}$. $S_r^*$ is an equivalence class determined by the relation $=_\Omega^+$, the transitive closure of $=_\Omega$. So, by definition, it must be that for some permutation of its elements, without loss of generality the identity permutation, $i_1=_\Omega i_2=_\Omega\ldots=_\Omega i_k$. Using the definition of $=_\Omega$ from (\ref{eq:partial_order_alternate}), this means,
        \begin{equation*}
        \left(\forall\ 1\leq j < k\right)\quad(\forall\ W\in\mathbb{W})\ (\forall\ T_j\in\mathcal{T}_{i_ji_{j+1}}^{W+})\quad\left\{i_j,i_{j+1}\right\}\subseteq S_1^{W,T_j}
        \end{equation*}
        For each $1\leq j < k$, pick any\footnote{At least one such $(W_j,T_j)$ pair is guaranteed by the definitions in (\ref{eq:partial_order}).} $W_j\in\mathbb{W}$ and $T_j\in \left(\mathcal{T}_{i_ji_{j+1}}^{j+}\right)^{\min}$. To begin with, set $\left(\lambda^*_{i_1}, \lambda^*_{i_2}\right)=\left(\lambda^{1,T_1}_{i_1},\lambda^{1,T_1}_{i_2}\right)$. If $k=2$, we are done. Otherwise, for $3\leq j \leq k$, recursively set:
        \begin{equation}
        \label{eq:recursive_construction}
        \lambda^*_{i_j}=\frac{\lambda^{j-1,T_{j-1}}_{i_j}}{\lambda^{j-1,T_{j-1}}_{i_{j-1}}}\lambda^*_{i_{j-1}}
        \end{equation}
    \end{enumerate}
\end{itemize}

\vfill

\vfill

\vfill

\begin{example}
\textit{Let $N=\{i,j,k,\ell,m,n\}$ be the set of players, and let there be just one local welfare function $W$, $\mathcal{T}^W=\left\{T_1^W=\{i,j\},T_2^W=\{j,k,\ell\},T_3^W=\{m,n\},T_4^W=\{i,m,n\}\right\}$. Also, let $f$ be a distribution rule that guarantees the existence of an equilibrium in all games $G\in\mathcal{G}(N,f,W)$, described by the sequence of weight systems $\Omega=\left\{\omega^{W,T_i}=(\boldsymbol{\lambda}^{W,T_i},\Sigma^{W,T_i})\right\}_{i=1}^4$, where,}

\vfill
\begin{equation*}
\begin{split}
\boldsymbol{\lambda}^{W,T_1}=\boldsymbol{\lambda}^{W,T_3}=(1,2)\quad&\quad\boldsymbol{\lambda}^{W,T_2}=\boldsymbol{\lambda}^{W,T_4}=(1,2,3)\mcr
\Sigma^{W,T_1}=\left\{\{i,j\},\{\}\right\}\quad\Sigma^{W,T_2}=\left\{\{j,k\},\{\ell\}\right\}\;\;\;&\;\;\;\Sigma^{W,T_3}=\left\{\{m,n\},\{\}\right\}\quad\Sigma^{W,T_4}=\left\{\{i\},\{m,n\}\right\}
\end{split}
\end{equation*}

\vfill
\noindent\textit{Using the definitions in (\ref{eq:partial_order}) or (\ref{eq:partial_order_alternate}), it can be seen that the players are related as follows:}

\vfill
\begin{equation*}
i\ =_\Omega^+\ j\ =_\Omega^+\ k\ \succeq_\Omega^+\ \ell \quad\text{and}\quad i\ \succeq_\Omega^+\ m\ =_\Omega^+\ n
\end{equation*}

\vfill
\noindent\textit{Using the construction above, it can be seen that for $\Sigma^*$, both $\left\{\{i,j,k\},\{m,n\},\{\ell\}\right\}$ and $\left\{\{i,j,k\},\{\ell\},\{m,n\}\right\}$ are admissible orderings of the three equivalence classes of $=_\Omega^+$ (they do not violate $\succeq_\Omega^+$). As for the weights, we get $\boldsymbol{\lambda}^*=(1,2,4,a,1,2)$, where $a$ can be any strictly positive number. Any strictly positive scaling of $\boldsymbol{\lambda}^*$ would also be admissible.}
\end{example}

\newpage
Before proceeding to show that $\omega^*$ as constructed above is equivalent to all the weight systems in $\Omega$, we prove an important property of $\boldsymbol{\lambda}^*$ in a quick lemma:

\vfill
\begin{lemma}
\label{lemma:7.1}
With $\boldsymbol{\lambda}^*$ as derived above, for any $S_r^*\in\Sigma^*$ with $|S_r^*|>1$, for any two players $i,j\in S_r^*$, for any $W\in\mathbb{W}$, for any coalition $T\in\mathcal{T}_{ij}^{W+}$ with $\{i,j\}\subseteq S_1^{W,T}$,

\vfill
\begin{equation}
\label{eq:lemma:7.1}
\frac{\lambda^{W,T}_i}{\lambda^{W,T}_j}=\frac{\lambda^*_i}{\lambda^*_j}
\end{equation}
\end{lemma}

\vfill
\proof{Proof.} Let $|S_r^*|=k>1$. Equivalently, we show that for all $m\in\{1,2,\ldots,k-1\}$, for all $\ell\in\{1,2,\ldots,k-m\}$, (\ref{eq:lemma:7.1}) holds for players $i=i_\ell$ and $j=i_{\ell+m}$. The base case, where $m=1$ follows by construction (\ref{eq:recursive_construction}), and by using Corollary \ref{corollary:5}(b)(ii). For $m\geq 2$, suppose there exists a welfare function $W\in\mathbb{W}$ and a coalition $T\in\mathcal{T}_{ij}^{W+}$, with $\{i,j\}\subseteq S_1^{W,T}$. Recall the welfare functions and coalitions $(W_j,T_j)$, $1\leq j< k$, that were picked for constructing $\boldsymbol{\lambda}^*$. From Corollary \ref{corollary:5}(b), it is sufficient to prove this lemma for $T\in\left(\mathcal{T}_{ij}^{W+}\right)^{\min}$. Now, using the definitions in (\ref{eq:partial_order}), $j=_\Omega i$. Therefore, it follows that the players $i=i_\ell, i_{\ell+1}, \ldots, i_{\ell+m}=j$ form a cycle in $=_\Omega$, i.e., $i_\ell=_\Omega i_{\ell+1}=_\Omega\ldots=_\Omega i_{\ell+m}=_\Omega i_\ell$. This means that, $i_\ell\in S_1^{\ell,T_\ell}, i_{\ell+1}\in S_1^{\ell+1,T_{\ell+1}}, \ldots, i_{\ell+m-1}\in S_1^{\ell+m-1,T_{\ell+m-1}}, i_{\ell+m}\in S_1^{W,T}$. Applying Corollary \ref{corollary:6}(b) and (\ref{eq:recursive_construction}), we have:

\vfill
\begin{small}
\begin{equation*}
\left(\prod_{j=\ell}^{\ell+m-1} \frac{\lambda_{i_j}^{j,T_j}}{\lambda_{i_{j+1}}^{j,T_j}}\right)\frac{\lambda_{i_{\ell+m}}^{W,T}}{\lambda_{i_{\ell}}^{W,T}} = 1\quad
\Longrightarrow\quad\frac{\lambda_i^{W,T}}{\lambda_j^{W,T}} = \prod_{j=\ell}^{\ell+m-1} \frac{\lambda_{i_j}^{j,T_j}}{\lambda_{i_{j+1}}^{j,T_j}}\quad
\Longrightarrow\quad \frac{\lambda_i^{W,T}}{\lambda_j^{W,T}}= \prod_{j=\ell}^{\ell+m-1} \frac{\lambda_{i_j}^*}{\lambda_{i_{j+1}}^*}\quad
\Longrightarrow\quad\frac{\lambda_i^{W,T}}{\lambda_j^{W,T}}= \frac{\lambda_i^*}{\lambda_j^*}
\end{equation*}
\end{small}

\vfill
\noindent This concludes the proof.\hfill\Halmos
\endproof

\vfill
Now we present the final lemma that establishes the global consistency of $\Omega$.

\vfill
\begin{lemma}
\label{lemma:8}
Given any set of local welfare functions $\mathbb{W}$, if $f^\mathbb{W}$ are budget-balanced distribution rules that guarantee equilibrium existence in all games $G\in\mathcal{G}(N,f^\mathbb{W},\mathbb{W})$, where, for each $W\in\mathbb{W}$, $f^W:=\DS\sum_{T\in\mathcal{T}^W}q^W_T f^{W,T}_{GWSV}[\omega^{W,T}]$, then, there exists a weight system $\omega^*$, such that,

\vfill
\begin{equation}
\label{eq:lemma:8}
(\forall\ W\in\mathbb{W})\ (\forall T\in\mathcal{T}^W)\quad f^{W,T}_{GWSV}[\omega^{W,T}]=f^T_{GWSV}[\omega^*]
\end{equation}
\end{lemma}

\vfill
\proof{Proof.} We prove that $\omega^*$ as constructed above satisfies (\ref{eq:lemma:8}). Consider any welfare function $W\in\mathbb{W}$ and any coalition $T\in\mathcal{T}^W$. Let $k=\min\left\{r|S^*_r\cap T\not=\emptyset\right\}$. Then, we need only show the following:

\vfill
\begin{enumerate}[label=(\roman*)]
\item $S_1^{W,T}\subseteq S^*_k$

\vfill
\item $S_2^{W,T}\cap S^*_k=\emptyset$

\vfill
\item $\left(\forall i,j\in S_1^{W,T}\right)\ \frac{\lambda_i^{W,T}}{\lambda_j^{W,T}}=\frac{\lambda_i^*}{\lambda_j^*}$
\end{enumerate}

\vfill
\noindent Of these, the first two are immediate from the construction of $\Sigma^*$, and the third follows from Lemma \ref{lemma:7.1}. This completes the proof.\hfill\Halmos
\endproof

\newpage
\section{Proof of Proposition \ref{prop:svmcequiv}.}
\label{appendix:prop:svmcequiv}

First, note that we only need to prove one direction, since from (\ref{eq:basisGWSV}) and (\ref{eq:basisGWMC}) in Table \ref{table:basisdistr}, it follows that,

\vfill
\begin{equation}\label{eq:prop:svmcequiv}
q'_T=\left(\sum_{j\in\overline{T}}\lambda_j\right)q''_T\quad\Longleftrightarrow\quad q'_Tf^T_{GWSV}[\omega]=q''_Tf^{T}_{GWMC}[\omega]
\end{equation}

\vfill
\noindent To prove the other direction, it suffices to show that,

\vfill
\begin{equation*}
f^{W'}_{GWSV}[\omega]=f^{W''}_{GWMC}[\omega]\quad\Longrightarrow\quad \left(\forall\ T\subseteq N\right)\;q'_T=\left(\sum_{j\in\overline{T}}\lambda_j\right)q''_T
\end{equation*}

\vfill
\noindent with the understanding that $q_T = 0$ whenever $T\notin\mathcal{T}$, for $\mathcal{T}=\mathcal{T}',\mathcal{T}''$. The proof is by contradiction. Suppose $f^{W'}_{GWSV}[\omega]=f^{W''}_{GWMC}[\omega]$, and let $T$ be a smallest subset for which $q'_T\not=(\sum_{j\in\overline{T}}\lambda_j)q''_T$, and $i\in\overline{T}$. Then, we have,

\vfill
\begin{equation*}
\begin{split}
f^{W'}_{GWSV}[\omega]=f^{W''}_{GWMC}[\omega]\quad&\Longrightarrow\quad f^{W'}_{GWSV}[\omega](i,T)=f^{W''}_{GWMC}[\omega](i,T)\mcr
&\Longrightarrow\quad \sum_{S\subseteq T}q'_S f^S_{GWSV}[\omega](i,S) = \sum_{S\subseteq T}q''_S f^S_{GWMC}[\omega](i,S)\mcr
&\Longrightarrow\quad q'_T f^T_{GWSV}[\omega](i,T)+\sum_{S\subsetneq T}q'_S f^S_{GWSV}[\omega](i,S)\mcr
&\qquad\qquad\qquad= q''_T f^T_{GWMC}[\omega](i,T) + \sum_{S\subsetneq T}q''_S f^S_{GWMC}[\omega](i,S)\mcr
&\Longrightarrow\quad q'_T=\left(\sum_{j\in\overline{T}}\lambda_j\right)q''_T\qquad\qquad\qquad\qquad\qquad\quad\,\text{\big(from (\ref{eq:prop:svmcequiv})\big)}
\end{split}
\end{equation*}

\vfill
\noindent which contradicts our assumption. This completes the proof.\hfill\Halmos

\vfill
\section{Generalized weighted potential games.}\label{appendix:potentialfunction}

In Hart and Mas-Colell \cite{Hart89}, the authors show that weighted Shapley values result in weighted potential games, by explicitly constructing a potential function (in a recursive form, almost identical to the one in (\ref{eq:GWSVpotential})). The authors claim that their result extends to generalized weighted Shapley values, but do not provide a proof. In this section, we fill this gap by showing that generalized weighted Shapley values result in a slight variant of weighted potential games, which we call \textit{generalized weighted potential games}, defined as follows:

\vfill
\begin{definition}\label{defn:piecewiseWPG}
\textit{A finite game $G=\left(N,\left\{\mathcal{A}_i\right\}_{i\in N},\left\{U_i\right\}_{i\in N}\right)$ is a \textbf{generalized weighted potential game} if there exists a potential function $\boldsymbol{\Phi}\ :\ \mathcal{A}\rightarrow\mathbb{R}^m$ (where $m$ is some positive integer), and a positive weight $w_i > 0$ for each agent such that for every agent $i\in N$, for every $a_{-i}\in\mathcal{A}_{-i}$, and for every $a'_i,a''_i\in \mathcal{A}_i$,}

\vfill
\begin{equation}\label{eq:piecewiseWPG}
U_i\left(a'_i,a_{-i}\right) - U_i\left(a''_i,a_{-i}\right) = w_i\left(\Phi_{k(i)}\left(a'_i,a_{-i}\right) - \Phi_{k(i)}\left(a''_i,a_{-i}\right)\right),
\end{equation}

\vfill
\noindent\textit{where $k(i)$ denotes the index of the first nonzero term of $\boldsymbol{\Phi}\left(a'_i,a_{-i}\right) - \boldsymbol{\Phi}\left(a''_i,a_{-i}\right)$.}
\end{definition}

\newpage
Generalized weighted potential games are a special subclass of the weaker, more general class of ordinal potential games, where the difference in the potential function (right hand side of (\ref{eq:piecewiseWPG})) is merely required to be of the same sign as the difference in the utility function (left hand side of (\ref{eq:piecewiseWPG})). Note that weighted potential games are simply generalized weighted potential games with a one-dimensional potential function ($m=1$).

Definition \ref{defn:piecewiseWPG} applies to any finite noncooperative game in normal form. However, recall from Section \ref{s.model} that, in our model, the agent utility functions are separable, given by
\begin{equation}\label{eq:utilityseparable}
U_i(a) = \sum_{r\in a_i}f^r(i,\{a\}_r)
\end{equation}
Hence, in searching for a potential function for $G$, it is natural to seek a separable potential function $\boldsymbol{\Phi}\ :\ \mathcal{A}\rightarrow\mathbb{R}^m$ (where $m$ is some positive integer), given by
\begin{equation}\label{eq:potentialseparable}
\boldsymbol{\Phi}(a) = \sum_{r\in R}\boldsymbol{\phi}_r(\{a\}_r)
\end{equation}
where $\boldsymbol{\phi}_r\ :\ 2^N\rightarrow\mathbb{R}^m$ is the `local' potential function at resource $r$. Therefore, to show that $\boldsymbol{\Phi}$ is a potential function for $G$, it is sufficient to show that for every agent $i\in N$, there exists a positive weight $w_i > 0$ such that, for every resource $r\in R$, for every player subset $S\subseteq N$ containing $i$,
\begin{equation}\label{eq:piecewiseWPGseparable}
f^r(i,S) = w_i\left(\left(\phi_r\right)_{k(i)}(S) - \left(\phi_r\right)_{k(i)}(S-\{i\})\right),
\end{equation}
where $k(i)$ denotes the index of the first nonzero term of $\boldsymbol{\phi}_r(S) - \boldsymbol{\phi}_r(S-\{i\})$. Verifying this is quite straightforward; use (\ref{eq:utilityseparable})-(\ref{eq:piecewiseWPGseparable}) to check that (\ref{eq:piecewiseWPG}) is satisfied.

We now state our formal result. Recall that a weight system $\omega=\left(\boldsymbol{\lambda},\Sigma\right)$ consists of a strictly positive vector of player weights $\boldsymbol{\lambda}\in\mathbb{R}^N_{++}$, and an ordered partition $\Sigma=(S_1,S_2,\ldots,S_m)$ of the set of players $N$.
\begin{theorem}
For any welfare sharing game $G = \left(N, R, \left\{\mathcal{A}_i\right\}_{i\in N}, \left\{f^r\right\}_{r\in R}, \left\{W_r\right\}_{r\in R}\right)$ and any weight system $\omega$, if for every $r\in R$, the distribution rule $f^r=f^{W'_r}_{GWSV}[\omega]=f^{W''_r}_{GWMC}[\omega]$ with $W'_r,W''_r$ being any two ground welfare functions related according to (\ref{eq:svmcequiv}), then, $G$ is a generalized weighted potential game, with player weights $\boldsymbol{\lambda}$, and the local potential function at resource $r$, $\boldsymbol{\phi}_r[\omega](S)=\left(\left(\phi_r[\omega]\right)_1(S),\left(\phi_r[\omega]\right)_2(S),\ldots,\left(\phi_r[\omega]\right)_m(S)\right)$, where, for all $1\leq k\leq m$, $\left(\phi_r[\omega]\right)_k(S)$ is given in terms of $W'_r$ in the following recursive form:
\begin{equation}\label{eq:GWSVpotentialappendix}
\left(\phi_r[\omega]\right)_k(S)=\frac{1}{\sum_{i\in S}\lambda_i}\left(W'_r(\overline{S}_{m-k+1}) + \sum_{i\in S}\lambda_i\left(\phi_r[\omega]\right)_k(S-\{i\})\right),
\end{equation}
and in terms of $W''_r$ in the following closed form:
\begin{equation}\label{eq:GWMCpotentialappendix}
\left(\phi_r[\omega]\right)_k(S) = W''_r(\overline{S}_{m-k+1}),
\end{equation}
where $\overline{S}_k=S-\cup_{\ell=1}^{k-1}S_\ell$.
\end{theorem}

\proof{Proof.} First, we use the closed form expression in (\ref{eq:GWMCpotentialappendix}) to show that $\boldsymbol{\phi}_r[\omega]$ satisfies (\ref{eq:piecewiseWPGseparable}). This involves proving, for any $1\leq k\leq m$, for any subset $S$ containing a player $i\in S_k$, the following two steps:
\begin{enumerate}[label=(\alph*)]
\item $k(i)=m-k+1$, i.e., $W''_r(\overline{S}_\ell) - W''_r\left(\big(\overline{S-\{i\}}\big)_\ell\right) = 0$ for all $k+1\leq\ell\leq m$
\item $f^{W''_r}_{GWMC}[\omega](i,S) = \lambda_i\left(W''_r(\overline{S}_k) - W''_r\left(\big(\overline{S-\{i\}}\big)_k\right)\right)$
\end{enumerate}

\newpage
\noindent Observe that $\big(\overline{S-\{i\}}\big)_\ell = \overline{S}_\ell-\{i\}$ for all $1\leq\ell\leq m$. Part (a) is straightforward, since if $i\in S_k$, then by definition, $i\notin\overline{S}_\ell$ for all $k+1\leq\ell\leq m$. We now focus on part (b), which is exactly the definition of $f^{W''_r}_{GWMC}[\omega]$ in Table \ref{table:distr}. Hence, the following is simply an exercise in verifying the equivalence of the two definitions of $f^{W''_r}_{GWMC}[\omega]$ from Tables \ref{table:distr} and \ref{table:basisdistr}, using the basis representation discussed in Section \ref{s.basisD}. Evaluating the left hand side, we get,
\begin{equation*}
\lambda_i\left(W''_r(\overline{S}_k) - W''_r\left(\big(\overline{S-\{i\}}\big)_k\right)\right)=\lambda_i\left(\sum_{T\in\mathcal{T}''(\overline{S}_k)}q''_T-\sum_{T\in\mathcal{T}''(\overline{S}_k-\{i\})}q''_T\right)
\end{equation*}
where, for any player subset $S\subseteq N$, $\mathcal{T}''(S)$ denotes the set of all coalitions $T\in\mathcal{T}''$ that are contained in $S$ ($T\subseteq S$). Notice that $\overline{S}_k$ does not contain any players in $S_1\cup S_2\cup\ldots S_{k-1}$. Therefore, $\mathcal{T}''(\overline{S}_k)$ consists of those coalitions contained in $S$ that do not contain any player in $S_1\cup S_2\cup\ldots S_{k-1}$. Similarly, $\mathcal{T}''(\overline{S}_k-\{i\})$ consists of those coalitions contained in $S-\{i\}$ that do not contain any player in $S_1\cup S_2\cup\ldots S_{k-1}$. Therefore, the collection $\mathcal{T}''(\overline{S}_k)-\mathcal{T}''(\overline{S}_k-\{i\})$ consists precisely of those coalitions $T\in\mathcal{T}''(S)$ that do not contain any player in $S_1\cup S_2\cup\ldots S_{k-1}$, but contain player $i$. Since $i\in S_k$, this is the same as saying that the collection $\mathcal{T}''(\overline{S}_k)-\mathcal{T}''(\overline{S}_k-\{i\})$ contains precisely those coalitions $T\in\mathcal{T}''(S)$ for which $i\in\overline{T}$. So, we get,
\begin{equation*}
\begin{split}
\lambda_i\left(W''_r(\overline{S}_k) - W''_r\left(\big(\overline{S-\{i\}}\big)_k\right)\right)&=\lambda_i\left(\sum_{T\in\mathcal{T}''(\overline{S}_k)}q''_T-\sum_{T\in\mathcal{T}''(\overline{S}_k-\{i\})}q''_T\right)\mcr
&=\sum_{T\in\mathcal{T}''(S)\ :\ i\in\overline{T}}q''_T\lambda_i\mcr
&=\sum_{T\in\mathcal{T}''}q''_T f^T_{GWMC}[\omega](i,S)\qquad\qquad\qquad\big(\text{from (\ref{eq:basisGWMC})}\big)\mcr
&= f^{W''_r}_{GWMC}[\omega](i,S)
\end{split}
\end{equation*}
To complete the proof, observe that when $W'_r$ and $W''_r$ are related according to (\ref{eq:svmcequiv}), then, for all $1\leq k\leq m$, and all $S\subseteq N$, the expression for $\left(\phi_r[\omega]\right)_k(S)$ (\ref{eq:GWMCpotentialappendix}) satisfies the recursion (\ref{eq:GWSVpotentialappendix}).\hfill\Halmos
\endproof
\end{APPENDICES}

\section*{Acknowledgments.}
This research was supported by AFOSR grants \#FA9550-09-1-0538 and \#FA9550-12-1-0359, ONR grant \#N00014-12-1-0643, and NSF grants \#CNS-0846025 and \#CCF-1101470.


\bibliographystyle{ormsv080} 
\bibliography{cost_sharing_journal} 

\begin{thebibliography}{51}
\expandafter\ifx\csname natexlab\endcsname\relax\def\natexlab#1{#1}\fi
\expandafter\ifx\csname url\endcsname\relax
  \def\url#1{{\tt #1}}\fi
\expandafter\ifx\csname urlprefix\endcsname\relax\def\urlprefix{URL }\fi
\expandafter\ifx\csname urlstyle\endcsname\relax
  \expandafter\ifx\csname doi\endcsname\relax
  \def\doi#1{doi:\discretionary{}{}{}#1}\fi \else
  \expandafter\ifx\csname doi\endcsname\relax
  \def\doi{doi:\discretionary{}{}{}\begingroup \urlstyle{rm}\Url}\fi \fi

\bibitem[{Aadithya et~al.(2010)Aadithya, Ravindran, Michalak, and
  Jennings}]{Aadithya10}
Aadithya, K.~V., B.~Ravindran, T.~P. Michalak, N.~R. Jennings. 2010.
\newblock Efficient computation of the shapley value for centrality in
  networks.
\newblock {\it {WINE}\/}. 1--13.

\bibitem[{Adlakha et~al.(2009)Adlakha, Johari, Weintraub, and
  Goldsmith}]{Adlakha09}
Adlakha, S., R.~Johari, G.~Weintraub, A.~Goldsmith. 2009.
\newblock Oblivious equilibrium: an approximation to large population dynamic
  games with concave utility.
\newblock {\it GameNets\/}. 68--69.

\bibitem[{Anshelevich et~al.(2004)Anshelevich, Dasgupta, Kleinberg, Tardos,
  Wexler, and Roughgarden}]{Anshelevich04}
Anshelevich, E., A.~Dasgupta, J.~Kleinberg, {\'E}.~Tardos, T.~Wexler,
  T.~Roughgarden. 2004.
\newblock The price of stability for network design with fair cost allocation.
\newblock {\it {FOCS}\/}. 295--304.

\bibitem[{Anshelevich et~al.(2003)Anshelevich, Dasgupta, Tardos, and
  Wexler}]{Anshelevich03}
Anshelevich, E., A.~Dasgupta, \'{E}. Tardos, T.~Wexler. 2003.
\newblock Near-optimal network design with selfish agents.
\newblock {\it {STOC}\/}. 511--520.

\bibitem[{Blume(1993)}]{Blume93}
Blume, L.~E. 1993.
\newblock The statistical mechanics of strategic interaction.
\newblock {\it {Game. and Econ. Behav.}\/} {\bf 5}(3) 387--424.

\bibitem[{Chekuri et~al.(2007)Chekuri, Chuzhoy, Lewin-Eytan, Naor, and
  Orda}]{Chekuri07}
Chekuri, C., J.~Chuzhoy, L.~Lewin-Eytan, J.~Naor, A.~Orda. 2007.
\newblock Non-cooperative multicast and facility location games.
\newblock {\it {IEEE J. Sel. Areas Commun.}\/} {\bf 25}(6) 1193--1206.

\bibitem[{Chen et~al.(2008)Chen, Roughgarden, and Valiant}]{Chen08}
Chen, H-L., T.~Roughgarden, G.~Valiant. 2008.
\newblock Designing networks with good equilibria.
\newblock {\it {SODA}\/}. 854--863.

\bibitem[{Chen et~al.(2010)Chen, Roughgarden, and Valiant}]{Chen10}
Chen, H.-L., T.~Roughgarden, G.~Valiant. 2010.
\newblock Designing network protocols for good equilibria.
\newblock {\it {SIAM J. Comput.}\/} {\bf 39}(5) 1799--1832.

\bibitem[{Christodoulou et~al.(2010)Christodoulou, Chung, Ligett, Pyrga, and
  Stee}]{Christodoulou10}
Christodoulou, G., C.~Chung, K.~Ligett, E.~Pyrga, R.~Stee. 2010.
\newblock On the price of stability for undirected network design.
\newblock {\it {WAOA}\/}. 86--97.

\bibitem[{Conitzer and Sandholm(2004)}]{Conitzer04}
Conitzer, V., T.~Sandholm. 2004.
\newblock {Computing Shapley values, manipulating value division schemes, and
  checking core membership in multi-issue domains}.
\newblock {\it {AAAI}\/}. 219--225.

\bibitem[{Corbo and Parkes(2005)}]{Corbo05}
Corbo, J., D.~C. Parkes. 2005.
\newblock The price of selfish behavior in bilateral network formation.
\newblock {\it {PODC}\/}. 99--107.

\bibitem[{Deng and Papadimitriou(1994)}]{Deng94}
Deng, X., C.~H. Papadimitriou. 1994.
\newblock On the complexity of cooperative solution concepts.
\newblock {\it {Math. Oper. Res.}\/} {\bf 19}(2) 257--266.

\bibitem[{Dobzinski et~al.(2008)Dobzinski, Mehta, Roughgarden, and
  Sundararajan}]{Dobzinski08}
Dobzinski, S., A.~Mehta, T.~Roughgarden, M.~Sundararajan. 2008.
\newblock {Is Shapley cost sharing optimal?}
\newblock {\it {SAGT}\/}. 327--336.

\bibitem[{Feigenbaum et~al.(2001)Feigenbaum, Papadimitriou, and
  Shenker}]{Feigenbaum01}
Feigenbaum, J., C.~H. Papadimitriou, S.~Shenker. 2001.
\newblock Sharing the cost of multicast transmissions.
\newblock {\it Comput. and Syst. Sciences\/} {\bf 63}(1) 21--41.

\bibitem[{Fiat et~al.(2006)Fiat, Kaplan, Levy, Olonetsky, and Shabo}]{Fiat06}
Fiat, A., H.~Kaplan, M.~Levy, S.~Olonetsky, R.~Shabo. 2006.
\newblock On the price of stability for designing undirected networks with fair
  cost allocations.
\newblock {\it {ICALP}\/}. 608--618.

\bibitem[{Gopalakrishnan et~al.(2011)Gopalakrishnan, Marden, and
  Wierman}]{Gopalakrishnan11}
Gopalakrishnan, R., J.~R. Marden, A.~Wierman. 2011.
\newblock An architectural view of game theoretic control.
\newblock {\it {SIGMETRICS Perf. Eval. Rev.}\/} {\bf 38}(3) 31--36.

\bibitem[{Granot and Huberman(1981)}]{Granot81}
Granot, D., G.~Huberman. 1981.
\newblock Minimum cost spanning tree games.
\newblock {\it {Math. Program.}\/} {\bf 21} 1--18.

\bibitem[{Hart and Mas-Colell(1989)}]{Hart89}
Hart, S., A.~Mas-Colell. 1989.
\newblock Potential, value, and consistency.
\newblock {\it Econometrica\/} {\bf 57}(3) 589--614.

\bibitem[{Hoefer and Krysta(2005)}]{Hoefer05}
Hoefer, M., P.~Krysta. 2005.
\newblock Geometric network design with selfish agents.
\newblock {\it {COCOON}\/}. 167--178.

\bibitem[{Immorlica and Pountourakis(2012)}]{Immorlica12}
Immorlica, N., E.~Pountourakis. 2012.
\newblock On budget-balanced group-strategyproof cost sharing mechanisms.
\newblock {\it {WINE}\/}.
\newblock To appear.

\bibitem[{Jain and Mahdian(2007)}]{Jain07}
Jain, K., M.~Mahdian. 2007.
\newblock Cost sharing.
\newblock N.~Nisan, T.~Roughgarden, {\'E}.~Tardos, V.~V. Vazirani, eds., {\it
  {Algorithmic Game Theory}\/}, chap.~15. {Cambridge University Press},
  385--410.

\bibitem[{Johari and Tsitsiklis(2004)}]{Johari04}
Johari, R., J.~N. Tsitsiklis. 2004.
\newblock Efficiency loss in a network resource allocation game.
\newblock {\it {Math. Oper. Res.}\/} {\bf 29}(3) 407--435.

\bibitem[{Liben-Nowell et~al.(2012)Liben-Nowell, Sharp, Wexler, and
  Woods}]{Liben-Nowell12}
Liben-Nowell, D., A.~Sharp, T.~Wexler, K.~Woods. 2012.
\newblock {Computing Shapley value in supermodular coalitional games}.
\newblock J.~Gudmundsson, J.~Mestre, T.~Viglas, eds., {\it {Computing and
  Combinatorics}\/}, {\it {LNCS}\/}, vol. 7434. Springer, 568--579.

\bibitem[{Marden(2012)}]{Marden12b}
Marden, J.~R. 2012.
\newblock State based potential games.
\newblock {\it Automatica\/} {\bf 48}(12) 3075--3088.

\bibitem[{Marden et~al.(2009)Marden, Arslan, and Shamma}]{Marden09a}
Marden, J.~R., G.~Arslan, J.~S. Shamma. 2009.
\newblock Joint strategy fictitious play with inertia for potential games.
\newblock {\it {Trans. on Autom. Control}\/} {\bf 54}(2) 208--220.

\bibitem[{Marden and Effros(2012)}]{Marden12}
Marden, J.~R., M.~Effros. 2012.
\newblock The price of selfishness in network coding.
\newblock {\it {IEEE Trans. Inf. Theory}\/} {\bf 58}(4) 2349--2361.

\bibitem[{Marden and Shamma(2012)}]{Marden12a}
Marden, J.~R., J.~S. Shamma. 2012.
\newblock {Revisiting log-linear learning: Asynchrony, completeness and
  payoff-based implementation}.
\newblock {\it {Game. Econ. Behav.}\/} {\bf 75}(2) 788--808.

\bibitem[{Marden and Wierman(2008)}]{Marden08}
Marden, J.~R., A.~Wierman. 2008.
\newblock Distributed welfare games with applications to sensor coverage.
\newblock {\it {CDC}\/}. 1708--1713.

\bibitem[{Marden and Wierman(2013{\natexlab{a}})}]{Marden13a}
Marden, J.~R., A.~Wierman. 2013{\natexlab{a}}.
\newblock Distributed welfare games.
\newblock {\it {Oper. Res.}\/} To appear.

\bibitem[{Marden and Wierman(2013{\natexlab{b}})}]{Marden13b}
Marden, J.~R., A.~Wierman. 2013{\natexlab{b}}.
\newblock Overcoming the limitations of utility design for multiagent systems.
\newblock {\it {IEEE Trans. Autom. Control}\/} To appear.

\bibitem[{Matsui and Matsui(2000)}]{Matsui00}
Matsui, T., Y.~Matsui. 2000.
\newblock A survey of algorithms for calculating power indices of weighted
  majority games.
\newblock {\it {J. Oper. Res. Soc. Japan}\/} {\bf 43} 71--86.

\bibitem[{Mishra and Rangarajan(2007)}]{Mishra07}
Mishra, D., B.~Rangarajan. 2007.
\newblock Cost sharing in a job scheduling problem.
\newblock {\it {Soc. Choice Welfare}\/} {\bf 29} 369--382.

\bibitem[{Moulin(2002)}]{Moulin02}
Moulin, H. 2002.
\newblock Axiomatic cost and surplus sharing.
\newblock K.~J. Arrow, A.~K. Sen, K.~Suzumura, eds., {\it {Handbook of Social
  Choice and Welfare}\/}, vol.~1, chap.~6. Elsevier, 289--357.

\bibitem[{Moulin(2010)}]{Moulin10}
Moulin, H. 2010.
\newblock An efficient and almost budget balanced cost sharing method.
\newblock {\it {Games and Economic Behavior}\/} {\bf 70}(1) 107--131.

\bibitem[{Moulin(2013)}]{Moulin13}
Moulin, H. 2013.
\newblock {Cost sharing in networks: Some open questions}.
\newblock {\it {Int. Game Theory Rev.}\/} To appear.

\bibitem[{Moulin and Shenker(2001)}]{Moulin01}
Moulin, H., S.~Shenker. 2001.
\newblock {Strategyproof sharing of submodular costs: Budget balance versus
  efficiency}.
\newblock {\it Econ. Theory\/} {\bf 18}(3) 511--533.

\bibitem[{Panagopoulou and Spirakis(2008)}]{Panagopoulou08}
Panagopoulou, P.~N., P.~G. Spirakis. 2008.
\newblock A game theoretic approach for efficient graph coloring.
\newblock {\it {ISAAC}\/}. 183--195.

\bibitem[{Rosenthal(1973)}]{Rosenthal73}
Rosenthal, R.~W. 1973.
\newblock A class of games possessing pure-strategy nash equilibria.
\newblock {\it {Int. J. Game Theory}\/} {\bf 2}(1) 65--67.

\bibitem[{Roughgarden and Sundararajan(2009)}]{Roughgarden09}
Roughgarden, T., M.~Sundararajan. 2009.
\newblock Quantifying inefficiency in cost-sharing mechanisms.
\newblock {\it {J. ACM}\/} {\bf 56}(4) 23:1--23:33.

\bibitem[{Roughgarden and Tardos(2002)}]{Roughgarden02}
Roughgarden, T., \'{E}. Tardos. 2002.
\newblock How bad is selfish routing?
\newblock {\it {J. ACM}\/} {\bf 49}(2) 236--259.

\bibitem[{Sandholm(2010)}]{Sandholm10}
Sandholm, W.~H. 2010.
\newblock Decompositions and potentials for normal form games.
\newblock {\it {Game. Econ. Behav.}\/} {\bf 70}(2) 446--456.

\bibitem[{Shapley(1953{\natexlab{a}})}]{Shapley53a}
Shapley, L.~S. 1953{\natexlab{a}}.
\newblock Additive and non-additive set functions.
\newblock Ph.D. thesis, {Department of Mathematics, Princeton University}.

\bibitem[{Shapley(1953{\natexlab{b}})}]{Shapley53b}
Shapley, L.~S. 1953{\natexlab{b}}.
\newblock A value for n-person games.
\newblock {\it {Contributions to the theory of games -- II}\/}. Princeton
  University Press.

\bibitem[{Su and van~der Schaar(2009)}]{Su09}
Su, Y., M.~van~der Schaar. 2009.
\newblock Conjectural equilibrium in multiuser power control games.
\newblock {\it {IEEE Trans. Signal Process.}\/} {\bf 57}(9) 3638--3650.

\bibitem[{Tardos and Wexler(2007)}]{Tardos07}
Tardos, {\'E}., T.~Wexler. 2007.
\newblock Network formation games and the potential function method.
\newblock N.~Nisan, T.~Roughgarden, {\'E}.~Tardos, V.~V. Vazirani, eds., {\it
  {Algorithmic Game Theory}\/}, chap.~19. {Cambridge University Press},
  487--516.

\bibitem[{Ui(2000)}]{Ui00}
Ui, T. 2000.
\newblock A shapley value representation of potential games.
\newblock {\it {Game. Econ. Behav.}\/} {\bf 31}(1) 121--135.

\bibitem[{von Falkenhausen and Harks(2013)}]{vonFalkenhausen13}
von Falkenhausen, P., T.~Harks. 2013.
\newblock Optimal cost sharing for resource selection games.
\newblock {\it {Math. Oper. Res.}\/} To appear.

\bibitem[{Wolpert and Tumer(1999)}]{Wolpert99}
Wolpert, D.~H., K.~Tumer. 1999.
\newblock An introduction to collective intelligence.
\newblock Tech. rep., Handbook of Agent technology. AAAI.

\bibitem[{Yang and Hajek(2007)}]{Yang07}
Yang, S., B.~Hajek. 2007.
\newblock {VCG-Kelly mechanisms for allocation of divisible goods: Adapting VCG
  mechanisms to one-dimensional signals}.
\newblock {\it {IEEE J. Sel. Areas Commun.}\/} {\bf 25}(6) 1237--1243.

\bibitem[{Young(1994{\natexlab{a}})}]{Young94a}
Young, H.~P. 1994{\natexlab{a}}.
\newblock Cost allocation.
\newblock R.~Aumann, S.~Hart, eds., {\it {Handbook of Game Theory with Economic
  Applications}\/}, vol.~2. Elsevier, 1193--1235.

\bibitem[{Young(1994{\natexlab{b}})}]{Young94b}
Young, H.~P. 1994{\natexlab{b}}.
\newblock {\it Cost sharing\/}, chap.~5.
\newblock {Equity: In theory and practice}, {Princeton University Press},
  81--96.

\end{thebibliography}


\end{document}